\documentclass[useAMS,usenatbib]{mn2e}

\newcommand{\be}{\begin{equation}}
\newcommand{\ee}{\end{equation}}
\newcommand{\bea}{\begin{eqnarray}}
\newcommand{\eea}{\end{eqnarray}}
\newcommand{\bfig}{\begin{figure}}
\newcommand{\efig}{\end{figure}}
\newcommand{\bfigs}{\begin{figure*}}
\newcommand{\efigs}{\end{figure*}}
\newcommand{\bt}{\begin{table}}
\newcommand{\et}{\end{table}}
\renewcommand{\vec}[1]{ {\bf #1} }

\usepackage[dvips]{graphicx}

\title[Radiative transfer in cosmological hydrodynamic
  simulations] {A novel approach for
  accurate radiative transfer in cosmological hydrodynamic
  simulations} \author[M. Petkova and V. Springel]{Margarita
  Petkova$^1$\thanks{E-mail: mpetkova@mpa-garching.mpg.de} and Volker
  Springel$^{2,3,1}$\thanks{E-mail:
    volker.springel@h-its.org}\vspace{0.1cm}\\ $^1$Max-Planck-Institut
  f\"ur Astrophysik, Karl-Schwarzschild-Strasse 1, 85748 Garching,
  Germany\\ $^2$Heidelberg Institute for Theoretical Studies,
  Schloss-Wolfsbrunnenweg 35, 69118 Heidelberg, Germany\\ $^3$Zentrum
  f\"ur Astronomie der Universit\"at Heidelberg, ARI,
  M\"{o}nchhofstr. 12-14, 69120 Heidelberg, Germany}

\begin{document}

\pagerange{\pageref{firstpage}--\pageref{lastpage}} \pubyear{2009}

\maketitle

\label{firstpage}


\begin{abstract}
  We present a numerical implementation of radiative transfer based on
  an explicitly photon-conserving advection scheme, where radiative
  fluxes over the cell interfaces of a structured or unstructured mesh
  are calculated with a second-order reconstruction of the intensity
  field. The approach employs a direct discretisation of the radiative
  transfer equation in Boltzmann form with adjustable angular
  resolution that in principle works equally well in the optically
  thin and optically thick regimes. In our most general formulation of
  the scheme, the local radiation field is decomposed into a linear
  sum of directional bins of equal solid-angle, tessellating the unit
  sphere. Each of these ``cone-fields'' is transported independently,
  with constant intensity as a function of direction within the
  cone. Photons propagate at the speed of light (or optionally using a
  reduced speed of light approximation to allow larger timesteps),
  yielding a fully time-dependent solution of the radiative transfer
  equation that can naturally cope with an arbitrary number of
  sources, as well as with scattering. The method casts sharp shadows,
  subject to the limitations induced by the adopted angular
  resolution. If the number of point sources is small and scattering
  is unimportant, our implementation can alternatively treat each
  source exactly in angular space, producing shadows whose sharpness
  is only limited by the grid resolution. A third hybrid alternative
  is to treat only a small number of the locally most luminous point
  sources explicitly, with the rest of the radiation intensity
  followed in a radiative diffusion approximation.  We have
  implemented the method in the moving-mesh code {\small AREPO}, where
  it is coupled to the hydrodynamics in an operator splitting approach
  that subcycles the radiative transfer alternatingly with the
  hydrodynamical evolution steps. We also discuss our treatment of
  basic photon sink processes relevant for cosmological reionisation,
  with a chemical network that can accurately deal with
  non-equilibrium effects.  We discuss several tests of the new
  method, including shadowing configurations in two and three
  dimensions, ionised sphere expansion in static and dynamic density
  field and the ionisation of a cosmological density
  field. The tests agree favourably with analytic expectations and
  results based on other numerical radiative transfer approximations.

\end{abstract}



\section{Introduction}

The transport of radiation and its interaction with matter is of
fundamental importance in astrophysics, playing a crucial role in the
formation and evolution of objects as diverse as stars, black holes,
or galaxies. It would therefore be highly desirable to be able to
calculate radiative transfer (RT) processes with equal accuracy and ease as
ordinary hydrodynamical and gravitational dynamics.  Unfortunately,
the difficult mathematical structure of the radiative transfer
equation, which takes the form of a partial differential equation in
six dimensions (3 spatial dimensions, 2 angular dimensions, 1
frequency dimension) makes this an extremely challenging goal.  In
fact, the RT problem is so hard, even in isolation,
that coupled radiation hydrodynamics methods are still in their
infancy in cosmology thus far.

However, a large array of different approximations to the RT problem
have been developed over the years, which are often specifically tuned
to the requirements and characteristics of particular types of
problems, and in many cases are applied to static density fields only.
In this study, we are primarily concerned with RT in
calculations of cosmological reionisation and in star formation,
leaving aside other important areas such as stellar spectra and
atmospheres.  Especially for the reionisation problem, recent years
have seen a flurry of activity in the development of new RT solvers
that are well suited to this problem. These numerical methods include
long and short characteristics schemes, ray-tracing, 
moment methods and direct solvers, and other particle- or Monte-Carlo-based
transport methods. In the following we briefly describe these schemes.

In the long characteristics method \citep{MM, ANM1999,
  Sokasian2001, Cen2002, Abel2002b, Razoumov2005, Susa2006}, each
source cell in the computational volume is connected to all other
relevant cells. Then the RT equation
is integrated individually from that cell to
each of the selected cells. While this method is relatively simple and
straight-forward, it is also very time consuming, since it requires
${\cal O}(N^2)$ interactions between the cells. Moreover, parallelisation of
this approach is cumbersome and requires large amounts of data
exchange between the different processors. 

Short characteristics methods \citep{KA1988, NUS2001,
  Mellema1998, MIAS2006, Shapiro2004,
  Whalen2006, Alvarez2006a, Ahn2007, Altay2008, CFMR2001, MFC2003,
  iVINE, Cantalupo2010arXiv, Hasegawa2010, Baek2009}
try to gain efficiency by integrating the equation of
radiative transfer only along lines that connect nearby cells, and not
to all other cells in the computational domain. This reduces the
redundancy of the computations and makes the scheme easier to
parallelise.

A widely used incarnation of the long-characteristics method are so-called
ray-tracing schemes.
Here a discrete number of rays is traced from
each source, along which  the RT equation is integrated in 1D,
considering absorptions and recombinations. As
the angular resolution decreases with increasing distance from the
source, rays may be split into subrays
\citep[e.g.][]{Abel2002b, Trac2007} for higher efficiency. The ray-tracing itself 
can be performed either
on grids \citep{MIAS2006,Whalen2006} or using particles
 as interpolation points \citep{Baek2009,iVINE,Altay2008}. Other innovative
methods trace photons on unstructured grids, for example Delaunay tessellations,
that are adapted to the mean photon optical depth of the gas
\citep{Rijkhorst2006, Paardekooper2010, Ritzerveld2006}. 

Stochastic integration methods, specifically Monte Carlo methods,
employ a ray-casting strategy where the rays are discretised into
photon packets \citep{MFC2003,Baek2009} or particles
\citep{Nayakshin2009}. For each photon packet, its frequency and its
direction of propagation are determined by sampling the appropriate
distribution function of the emitters that have been assigned in the
initial conditions. A particular advantage of this approach is that
comparatively few approximations to the radiative transfer equations
need to be made, so that the quality of the results is primarily a
function of the number of photon packets employed, which can be made
larger in proportion to the CPU time spent. A disadvantage of these
schemes is the comparatively high computational cost and the sizable
level of noise in the simulated radiation field, which only slowly
diminishes as more photon packets are used. The `cone' transport
scheme of \citep{Pawlik2008}, where radiation is directly transferred
between particles, tries to improve on these limitations. If needed,
this method can also create further sampling points dynamically to
improve the resolution locally.

Using moments of the radiative transfer equations instead of the full
set of equations can lead to very substantial simplifications that can
drastically speed up the calculations. In this approach, the radiation
is represented by its mean intensity field throughout the
computational domain, which is evolved either in a diffusion
approximation or based on a suitably estimated local Eddington tensor
\citep{GA2001,AT2007,Petkova2009, Finlator2009}. 
Instead of following rays, the moment equations are solved directly on
the grid, or in a mesh-less fashion on a set of sampling
particles. Due to its local nature, the moment approach is
comparatively easy to parallelise, but its accuracy is highly problem
dependent, making it difficult to judge whether the simplifications
employed still provide sufficient accuracy. 
The simplest and most popular moment method is radiative
diffusion
\citep[e.g.][]{Whitehouse2005,Reynolds2009}, where the RT equation is approximated in terms of an integrated
energy density in each discretised mass or volume element, and this radiation energy
 density
is then evolved through the flux-limited diffusion approximation,
where the  flux limiter is introduced to prevent the occurrence of transfer speeds larger than the speed of light.
While the diffusion approximation works very well in the optically thick regime, its accuracy is hard to judge in general situations.

A comparison of the relative accuracy of these established RT methods
has been carried out in the ``cosmological radiative transfer comparison
project'' \citep{Iliev2006comparison,Iliev2009comparison}, where a
subset of the above implementations has been compared on a variety of
simple test problems. Such a comparison is particularly useful as the
lack of known analytic solutions for the RT problem, except for a
small number of simple situations, makes the validation of a RT method
quite difficult. Reassuringly, the comparison project found in most
cases reasonably good agreement between the different methods, but it
also highlighted that each of the different techniques shows
individual strengths and weaknesses, providing ample motivation to
search for still better methods.

It is the goal of this study to propose a new numerical scheme for
RT that is competitive with the best of the known
methods in terms of accuracy and general applicability, but is also
fast enough to allow self-consistent radiation-hydrodynamic
simulations in the context of cosmological reionisation and star
formation problems. We also aim  to couple the method
to the new moving-mesh code {\small AREPO} \citep{Arepo}, which solves
the equations of hydrodynamics on an unstructured Voronoi mesh that
moves with the flow and automatically adapts its resolution to the
gravitational clustering of matter. This mesh-based code computes
hydrodynamics similar to high-accuracy Eulerian codes on Cartesian
grids, but it features reduced advection errors when the flow velocity
is large.

Our new method is based on a radiation advection technique where a
second-order accurate, piece-wise linear reconstruction of the photon
intensity field is used to estimate upwind photon fluxes for each face
of the mesh. If there is only a single point source, such a scheme can
exploit the fact that the local streaming direction of the photons is
known everywhere -- it is along the ray from the source's position to
the local coordinate. If there are multiple sources, the radiation
field can be treated equally accurately by decomposing it into a
linear sum for each source, and treating each component
independently. Alternatively, we introduce a direct discretisation
of angular space, allowing a description of arbitrary source fields,
albeit at the cost of a finite angular resolution. We note that in all
these variants the conservation of photon number is manifest in the
transport step.  We treat the source terms and  the coupling
to the hydrodynamics in an operator split approach, where the
emission, advection, and absorption of radiation are calculated in
separate steps. This makes our approach fully photon conserving, which
is especially useful for the cosmic reionisation problem, as it ensures
that all photons emitted by an ionising source are really used up in
exactly one ionisation event.

We note that the advection scheme discussed in this paper normally
propagates the photons at their physical speed of light, based on an
explicit time integration scheme. While this has the advantage of
allowing general, fully time-dependent radiative transfer simulations,
it can make them computationally very expensive due to the required
small Courant time steps. This can however be greatly alleviated by
using a reduced speed of light approximation \citep{GA2001}, which
allows much larger timesteps while still preserving the speed of
cosmological ionisation fronts (I-fronts). With this approximation, it then
becomes possible to calculate high-resolution cosmological radiation
hydrodynamics simulations of structure formation that
simultaneously account for cosmic reionisation, with no restriction on
the number of sources.

In Section~\ref{sec:method} of this paper, we present our methodology
in detail. We first give a brief introduction to the RT equation in
Section~\ref{sec:rt}. Then we discuss three variants of our solution
method for the radiation advection equation in
Sections~\ref{sec:summation}, \ref{sec:summback}, and
\ref{sec:cones}. In Section~\ref{sec:sources}, we briefly describe our
treatment of emission and absorption processes, with an emphasis on
the hydrogen chemistry relevant for the cosmic reionisation problem,
and in Section~\ref{sec:heating} we specify our formulation of
photo-heating and radiative cooling. Issues of time stepping and code
implementation are discussed in Sections~\ref{sec:timestep} and
\ref{sec:imple}.  We move on to a presentation of basic test results
in Section~\ref{sec:tests}, starting with a variety of shadowing
(Section~\ref{sec:shadow}) and Str\"omgren sphere tests
(Section~\ref{sec:isphere}).  We then consider the more demanding
tests of I-front trapping in Section~\ref{sec:trap}, the ionisation of
a cosmological density field in Section~\ref{sec:cosmo}, and an
ionisation problem with dynamic density field in
Section~\ref{sec:tsphere}.  Finally, we present our conclusions in
Section~\ref{sec:end}.

\section{An advection solver for the radiative transfer problem} \label{sec:method}

\subsection{The radiative transfer equation} \label{sec:rt}

Let us briefly discuss different forms of the RT
equation, which is helpful to clarify how our new method differs from
other approaches, and for specifying our notation.  Let $f_\gamma \equiv
f_\gamma(t, \vec{x}, \vec{p})$ be the photon distribution function for
comoving coordinate $\vec{x}$ and photon momentum \be \vec{p} =
a\frac{h\nu}{c}\vec{\hat{n}} \, , \ee where $a \equiv a(t)$ is the
cosmological scale factor, $h$ is the Planck constant, $\nu$ is the
frequency of the photons, and $\vec{\hat{n}}$ is the unit vector in the
direction of photon propagation. Then the number of photons in some part
of the Universe is \be N_{\gamma} = \int {\rm d}\vec{x}\, {\rm d}\vec{p}
\, f_\gamma(t, \vec{x}, \vec{p}) \, .  \ee

We can quite generally write the phase-space continuity equation for the
distribution function $f_\gamma$ of photons as \be \frac{\partial
  f_\gamma}{\partial t} + \frac{\partial}{\partial
  \vec{x}}(\dot{\vec{x}} f_\gamma) + \frac{\partial}{\partial
  \vec{p}}(\dot{\vec{p}} f_\gamma) = \left . \frac{\partial
  f_\gamma}{\partial t} \right | _{\rm sources} - \left . \frac{\partial
  f_\gamma}{\partial t} \right | _{\rm sinks} .
\label{eqn:ph_dista}
\ee In this Boltzmann-like transport equation, the source and sink terms
on the right-hand side of the equation represent photon emission and
absorption processes, respectively.  If we neglect gravitational lensing
effects, individual photons propagate along straight lines with conserved
momenta, i.e. we have $\vec{\dot{x}}=(c/a)\vec{\hat n}$ and
$\vec{\dot{p}} =0$. The transport equation hence simplifies to
\be \frac{\partial
  f_\gamma}{\partial t} + \frac{c}{a} \frac{\partial}{\partial
  \vec{x}}(\vec{\hat{n}} f_\gamma) = \left . \frac{\partial
  f_\gamma}{\partial t} \right | _{\rm sources} - \left . \frac{\partial
  f_\gamma}{\partial t} \right | _{\rm sinks} .
\label{eqn:ph_dist}
\ee

Normally, a direct use of equation (\ref{eqn:ph_dist}) through a
discretisation of phase-space is considered prohibitively expensive due
to the high-dimensionality of the problem. However, if only
monochromatic radiation is considered, which is often sufficient, the
momentum-space dimensions reduce to just two angular coordinates. If
furthermore only a relatively coarse angular resolution for the photon
transport is sufficient, then the $4\pi$ solid angle described by these
angular dimensions may be discretised into a limited set of cones, say
up to 10-100, at which point a brute-force solution of equation
(\ref{eqn:ph_dist}) on a 3D mesh becomes computationally feasible and
attractive, as we shall argue here.

Before we discuss this in more detail, let us first briefly recall for
clarity how the specific intensity $I_\nu$ that is normally used in RT
studies relates to equation (\ref{eqn:ph_dist}).  We can define the
specific radiation intensity $I_\nu$ in a certain direction $\vec{\hat
  n}$ through the energy $\Delta E_\nu = I_\nu \Delta \nu \Delta A
\Delta \Omega \Delta t$ of photons that pass through a physical area
$\Delta A$ normal to $\vec{\hat n}$ and within solid angle $\Delta
\Omega$ around $\vec{\hat n}$, over a time interval $\Delta t$ and in a
frequency bin $\Delta \nu$.  With this definition, the specific
intensity $I_\nu $ is then related to the photon distribution function
$f_\gamma$ as  \be I_\nu = h\nu f_\gamma \frac{{\rm d}^3 x \,{\rm
    d}^3 p} {{\rm d}\nu \, {\rm d} \Omega \, {\rm d}A\, {\rm d}t} =
\frac{h^4 \nu^3}{c^2}f_\gamma \, .  \ee

Substituting into equation~(\ref{eqn:ph_dist}), and writing  the
absorption and emission terms in their conventional form, one obtains the 
cosmological RT equation in the form
\be
\frac{1}{c}\frac{\partial I_\nu}{\partial t}  +
\frac{\vec{\hat n}}{a}\frac{\partial I_\nu}{\partial \vec{x}} 
- \frac{H(a)}{c}\left(\nu \frac{\partial I_\nu}{\partial \nu} - 3 I_\nu \right)
=
-\kappa_\nu I_\nu + j_\nu \, ,
\label{eqn:int_cont}
\ee where $\kappa_\nu$ is the absorption coefficient, $j_\nu$ is the
emission coefficient, and $H(a)$ is the Hubble rate.  Defining the solid
angle averaged intensity as \be J_\nu = \frac{1}{4\pi}\int {\rm
  d}\Omega\, I_\nu , \ee we can calculate the physical photon number
density from the specific intensity as \be n_{\gamma}^{\rm phys} =
\frac{1}{c} \int \frac{4 \pi J_{\nu}}{h\nu}\,{\rm d}\nu . \ee Another
equivalent way to obtain the photon number density is simply to
integrate the distribution function, \be n_{\gamma} = \int {\rm d}\vec{p}
\, f_\gamma(t, \vec{x}, \vec{p}), \ee which yields the comoving number
density of photons, $n_{\gamma} =a^3 n_{\gamma}^{\rm phys}$. This
highlights again that describing the radiation field with the arguably
more familiar RT equation (\ref{eqn:int_cont}), or with the distribution
function and the Boltzmann-like equation (\ref{eqn:ph_dist}), is fully
equivalent.  In this paper, we will mostly work in the latter
formulation.

In general, to solve the RT problem on some discretised
mesh, we can split off the source and sink terms and treat them
separately in the time integration. In such an operator splitting
approach, known as Strang splitting, we are basically left with two
separate problems that are interleaved in the time integration, one is
to follow the conservative transport of photons on the mesh, the other
is the local updating of the photon density field through the source and
sink terms. In the following, we first focus on the conservative transport
problem, which is where the primary computational challenge lies.

\bfig
\begin{center}
\includegraphics[width=0.4\textwidth]{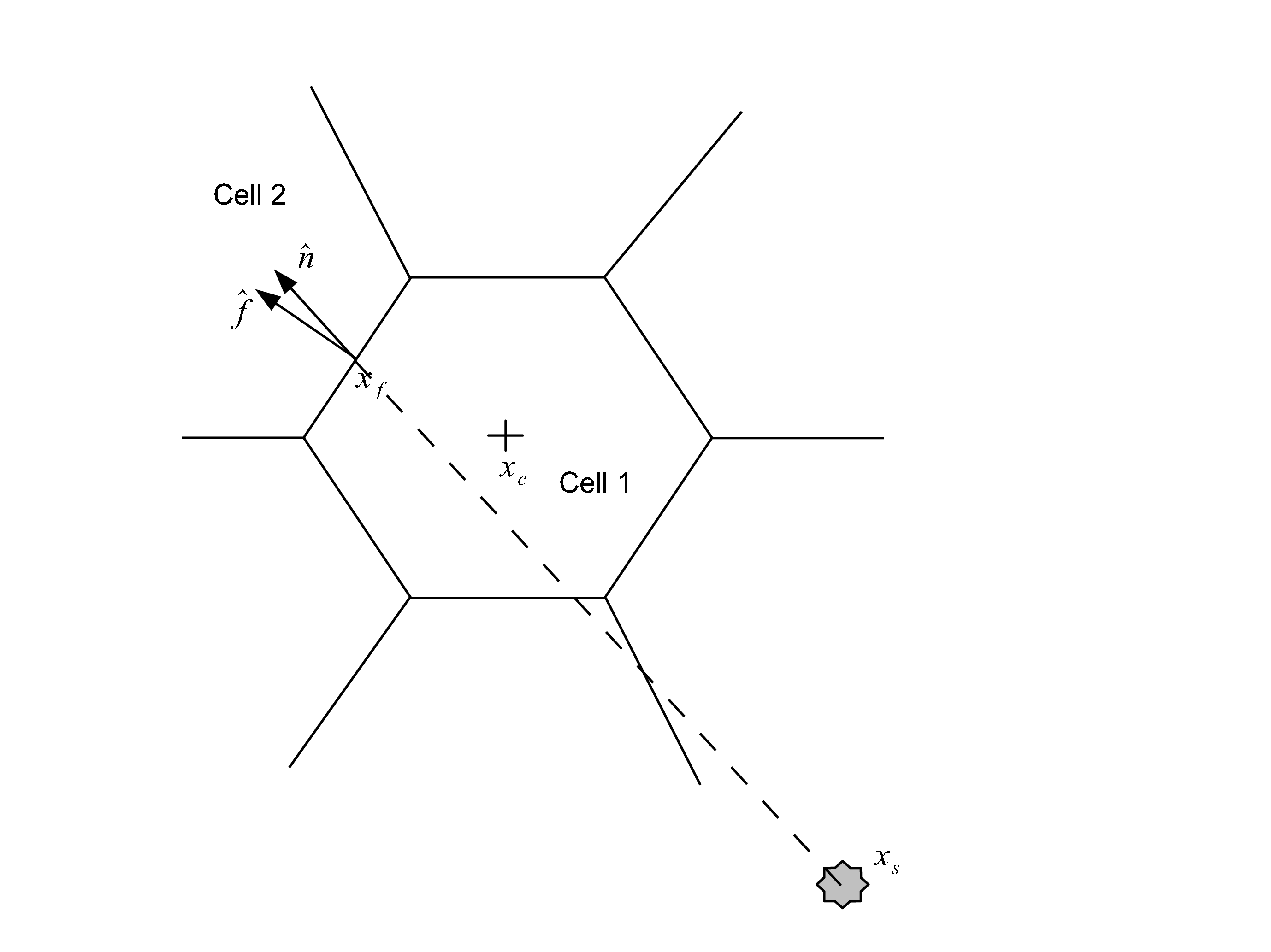}
\end{center}
\caption{A simple sketch showing the geometry involved in our
  advection scheme for a single point source at coordinate
  $\vec{x}_s$.  Here $\vec{\hat n}$ is the photon propagation
  direction, $\vec{\hat{f}}$ is the normal vector of a face of a cell,
  $\vec{x}_c$ is the center of mass of the corresponding cell and
  $\vec{x}_f$ is the center of mass of the face for which the photon
  flux is calculated. \label{fig:Voronoi_s}} \efig

\subsection{Transferring radiation by advection for point sources}
\label{sec:summation}

Suppose for the moment that we know at a given point in space that all
photons stream in the same direction $\vec{\hat{n}}$.  This is for
example the case if there is a single point source at coordinate
$\vec{x}_s$ (i.e.~no other sources and no scattering are present). For
simplicity, we shall also restrict ourselves to a spatially invariant
photon momentum spectrum.  One then obtains a simple advection equation
for the comoving photon density $n_\gamma$: \be \frac{\partial n_\gamma}{\partial
  t} + \frac{c\, \vec{\hat n}}{a} \cdot \nabla n_\gamma =
0, \label{eqn:advection} \ee where the local advection direction
$\vec{\hat n}$ is known at every point $\vec{x}$ and is simply given by
\be \vec{\hat n}(\vec{x}) = \frac{\vec{x} - \vec{x}_s}{|\vec{x} -
  \vec{x}_s|}.\label{eqn:vel}\ee This advection is conservative and may
be solved with the techniques commonly employed to treat the hyperbolic
conservation laws of ideal fluid dynamics on spatial meshes. Indeed,
this is the approach we are going to employ: we shall use a conservative
transport scheme based on a second-order accurate upwind method that is
inlined with the hydrodynamic calculations of our unstructured
moving-mesh hydrodynamics code AREPO, which is described in some more
detail below.  It is important to note that knowledge of the local
number density field of photons combined with the source location
$\vec{x}_s$ is sufficient to accurately solve the radiative transport,
simply because this information suffices to specify the photon
streaming direction at every point in space. Apart from the spatial
discretisation, no approximations need to be made for the case of a
single monochromatic point source in this treatment.

In practice, we use a second-order accurate spatial reconstruction
technique to convert photon numbers $N_i$ stored for each cell $i$ of
a given mesh into a photon density field. For every cell, we first
obtain an estimate $\left<\vec{\nabla} n_\gamma\right>_i$ for the
gradient of $n_\gamma$, which allows a piece-wise linear conservative
reconstruction of the photon density field, in the form \be
n_\gamma(\vec{x}) = \left< n_\gamma \right>_i + \left<\vec{\nabla}
n_\gamma\right>_i(\vec{x} - \vec{x}^c_i), \;\;\;\; {\rm for}\; \vec{x}
\in {\rm cell}\; i. \label{eqn:interp}\ee Here $\left< n_\gamma
\right>_i = N_i/V_i$ is the mean photon number density in the cell
with center-of-mass $\vec{x}^c_i$ and volume $V_i$.  As illustrated in
the sketch of Figure~\ref{fig:Voronoi_s}, for every face centroid
$\vec{x}_{f}$ of the mesh, we can then identify the upwind side of the
photon flow, based on the sign of the dot product between face normal
$\vec{\hat f}$ and the photon streaming direction $\vec{\hat n} =
(\vec{x}_f-\vec{x}_s)/|\vec{x}_f-\vec{x}_s|$.  This allows us to
estimate the photon flux $F_\gamma$ over the face as \be F_\gamma =
\frac{c}{a} \, (\vec{f} \cdot \vec{\hat{n}}) \,
n_\gamma(\vec{x}_f), \label{eqn:flux} \ee where the photon density
$n_\gamma(\vec{x}_f)$ at the face centroid is estimated based on the
linear reconstruction of the cell on the upwind side.  If the face has
comoving area $A$, the number of photons exchanged during time $\Delta
t$ between the cells that share the face is then given by \be \Delta
N_\gamma = F_\gamma A \Delta t .\ee Due to the pairwise exchange of
photons, the conservation of total photon number is manifest, which is
important for guaranteeing that I-fronts propagate at their physical
speeds.  We note that in our code the mesh is composed of Voronoi
cells (of which a Cartesian mesh is a special case), but this is not
important for the general approach.

There are two important caveats with this transport scheme, which need
to be pointed out. One is that this explicit transport scheme requires a
time step that is given by a local Courant criterion for the photons,
which can become very small due to the high speed of light. For
reionisation problems, this can however be circumvented with the reduced
speed of light approximation, which we will discuss in more detail later
on.  The other caveat is that close to a point source the mesh
resolution will always be coarse, so that our use of a single Gauss
point per mesh face may introduce sizable errors in the discretised
advection fluxes.  This can happen when the opening angle under which a
mesh face is seen by the point source is large, so that adopting a
single propagation direction for the entire face is inaccurate. As a
result, isophots of the radiation field produced by the point source may
then deviate from sphericity with distortions that reflect the local
geometry of the mesh around the point source. We have found however that
this problem can be cured quite effectively by injecting the photons of
the source in a kernel-weighted fashion over 2-3 mesh cells or so.  With
such slightly extended sources, the above scheme is able to quite
accurately treat single point sources.

\bfig
\begin{center}
\includegraphics[width=0.4\textwidth]{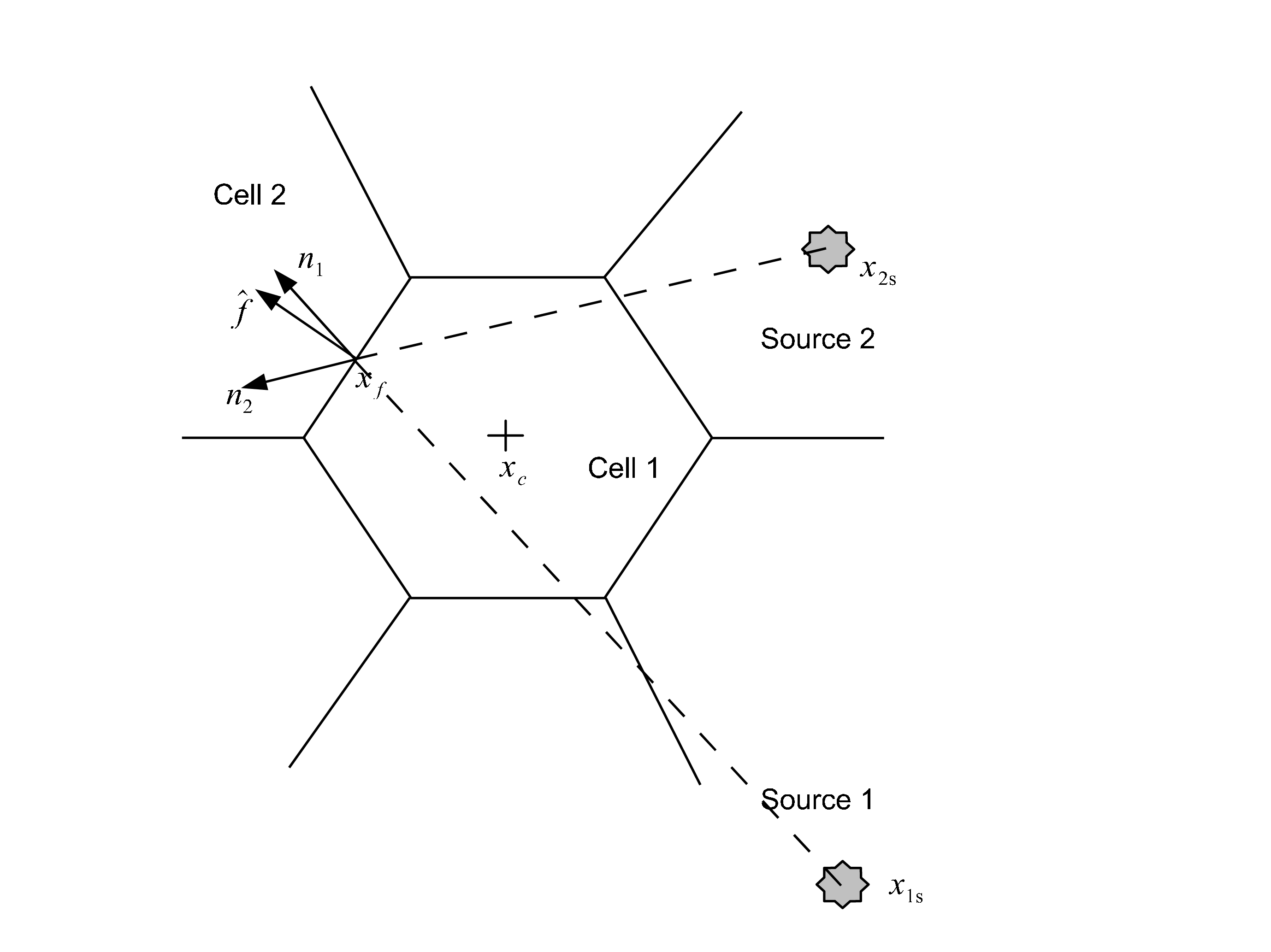}
\end{center}
\caption{Sketch that illustrates the linear summation principle used to
  treat the radiative transfer for multiple sources. Here $\vec{\hat n}_1$
  and $\vec{\hat n}_2$ are the photon propagation directions from the two
  sources, as seen from the center $\vec{x}_f$ of a face with normal
  vector $\vec{\hat{f}}$. The total flux passing through the face is
  then computed as a linear sum of the contributions from the partial
  fields created by each source. \label{fig:Voronoi_2s}} \efig

The approach can also be straightforwardly extended to multiple point
sources simply by linear superposition of the radiation fields
produced by each of the individual sources, as sketched in
Figure~\ref{fig:Voronoi_2s}. This means that the advection equation is
solved for the radiation field of each source separately. This
obviously involves a computational and storage cost that scales with
the number of sources, but if the number of sources is small, this is
an interesting technique for certain applications due to its high
accuracy. As we show in our test problems, the method in particular is
able to accurately cast shadows, and unlike for example in the
optically thin variable Eddington tensor approximation (OTVET), there
is no accuracy-degrading mutual influence of multiple sources on each
other.

However, for a large number of ionising sources, the linear
superposition approach will quickly become infeasible. For example, in
large cosmological simulations, we would like to allow every star
particle to act as a source of ionising radiation. Here we obviously
cannot decompose the radiation field into all its single point
sources, instead, we need to employ another decomposition.  We have
actually developed two possible schemes for this, which we describe in the
following.

\subsection{A hybrid between point-source treatment and local
  diffusion}
\label{sec:summback}

One possibility to address the multiple point sources problem is to only
retain a finite number $N_{\rm br}$ of locally brightest sources in an
explicit treatment, while all the remaining sources are lumped together
into a background radiation field that is treated with radiative
diffusion. The idea here is that especially in cosmic reionisation
problems the local ionisation ``bubble'' is expected to be driven
primarily by one or a few sources, and only at very late stages,
multiple sources may become visible at a given point, but then 
reionisation has largely completed already anyway. By making the set of sources
that are treated exactly as point sources spatially
variable, we should then get a quite accurate approximation of the
reionisation phenomenon even for moderate values of $N_{\rm br}$. Since
in the limit of large $N_{\rm br}$, the scheme will become essentially
exact (apart from spatial discretisation errors), the degree to which
imposing a limiting value for $N_{\rm br}$ affects the results can be
readily tested.

We will discuss some results obtained with this approach later on, but
we note that it clearly involves several complication when applied in
practice. First of all, the need to allow a local change of the list
of bright sources requires that one keeps track of all locally
incoming fluxes of radiation, sorting them appropriately, and matching
them through the use of unique source identifiers to the already
stored radiation intensities from the previous step.  Also, since
neighboring cells may have different source lists, a matching
procedure is required for gradient estimates, with the additional
complication that the accuracy of the gradients will be reduced at
``domain boundaries'', i.e.~regions of the mesh that differ in their
assessment what the locally most important $N_{\rm br}$ point sources
are. Furthermore, if the number of sources is very large and spread
out in space (e.g.~the individual stars in galaxies), the injection of
photons needs to be treated in some sort of clustered fashion,
otherwise faint individual sources may not be able to compete with the
$N_{\rm br}$ bright sources already stored locally, so that they are
channeled into the radiatively treated flux reservoir right away
without having a chance to build up to a significant source when
combined with the potentially many nearby sources that are equally
faint. Finally, one also needs a separate radiative diffusion solver,
which requires a small timestep for stability when integrated
explicitly in time, as we do here. For all these reasons, we actually
favour in most applications our second approach for treating a
large number of sources, which is facilitated by discretising the
solid angle explicitly, as we describe next.

\subsection{Full angular discretisation and cone transport}\label{sec:cones}

For general radiation fields we seek a method that can directly
represent the angular distribution of the local radiation field. This
can, for example, be done in terms of moments of the radiation
field. However, we here want to propose a more flexible approach that
is based on a direct angular discretisation of the photon space.  To
this extent, we can decompose the full solid angle into a set of cones
of equal size, for example based on the well-known {\small HEALPIX}
\citep{healpix} tessellation of the unit-sphere, which we shall use in
the following. Our strategy could however be straightforwardly
generalised also for other discretisations of angular space. In
{\small HEALPIX}, the unit sphere is decomposed into $N_{\rm pix}= 12
\, N_{\rm side}^2$ patches of equal solid angle (which we call
``cones'' for simplicity, even though they are not exactly
axi-symmetric), each centred around a central direction $\vec{\hat
  n}_j$, where $j=1\ldots N_{\rm pix}$. We now linearly decompose the
radiation field $f_\gamma$ into $N_{\rm pix}$ components, each
containing the photons that propagate along a direction within the
corresponding cone: \be f_\gamma(\vec{x}, \vec{\hat n}) = \sum_j
f_\gamma^j(\vec{x}, \vec{\hat n}), \ee where $f_\gamma^j(\vec{x},
\vec{\hat n})=0$ if the photon direction $\vec{\hat n}$ lies outside
of $\Delta\Omega_j$ around $\vec{\hat n}_j$. The basic simplification
we now make is that we assume that each of the partial radiation
fields, $f_\gamma^j(\vec{x}, \vec{\hat n})$, can be taken to be
constant as a function of direction within the corresponding cone. Or
in other words, each of the partial fields
$f_\gamma^j(\vec{x},\vec{\hat n})$ describes the intensity of a
homogeneously illuminated beam of opening angle $\Delta\Omega_j$
around direction $\vec{\hat n}_j$, emanating from the local coordinate
$\vec{x}$. Our goal is now to generalise the radiation advection
scheme for point sources outlined above such that it can accurately
transport the radiation cones occurring in this discretisation.

If we simply transport one of the partial radiation fields
$f_\gamma^j$ locally always along the primary direction of its cone,
i.e.  \be \frac{\partial f_\gamma^{j}}{\partial t} + \frac{c\,
  \vec{\hat n}_j} {a} \cdot \nabla f_\gamma^j =
0 \label{eqn:advectionparallel} , \ee we will invariably observe a
central ``focusing effect'', i.e.~the radiation emanating from a point
will not illuminate the finite solid angle $\Delta\Omega_j$
homogeneously, but rather tend to concentrate along the primary axis of
the cone. It is clear that this ``focusing effect'' arises from the
parallel transport described by
equation~(\ref{eqn:advectionparallel}); instead of transporting the
photon field over different directions that are uniformly spread over
the finite solid angle, all of the photons are transported along the
single direction $\vec{\hat n}_j$, with any residual angular spread
around $\vec{\hat n}_j$ arising only from numerical diffusion due to
the finite mesh resolution.

\bfig
\begin{center}
\includegraphics[width=0.4\textwidth]{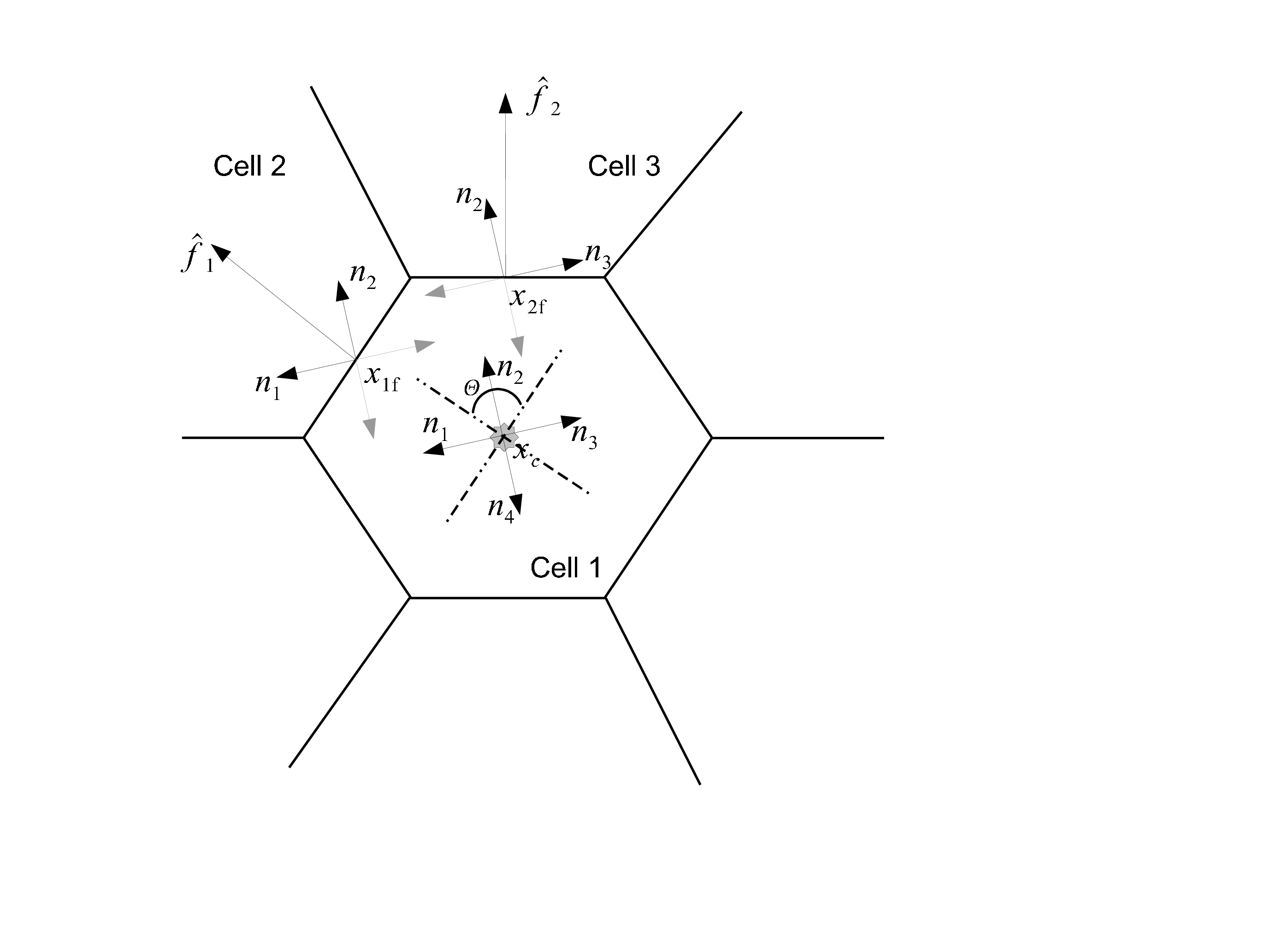}
\end{center}
\caption{This sketch illustrates the geometry and the vectors involved
  in our ``cone transport'', where the angular space is discretised into
  regions of equal angle (in 2D) or solid angle (in 3D).  In this
  example, only four cones in in 2D are used. The photon field is
  linearly decomposed into radiation fields corresponding to the four
  cones, which have symmetry axes $\vec{\hat n}_1,\, \vec{\hat n}_2,\,
  \vec{\hat n}_3,\, ... \, \vec{\hat n}_N$, where $N$ is the number of
  discrete cones or angles, i.e.~$N=4$ in the sketch. At each face of
  the mesh (here the normal vectors $\vec{\hat f}_1$ and $\vec{\hat
    f}_2$ are shown), photon fluxes for each of the partial fields are
  estimated. The photon propagation direction is taken to be parallel
  to the gradient of the total radiation intensity field, constrained
  to lie within the opening angle of the corresponding cone.
 \label{fig:Voronoi_vec}}
\efig

\bfig
\begin{center}
\includegraphics[width=0.41\textwidth,clip]{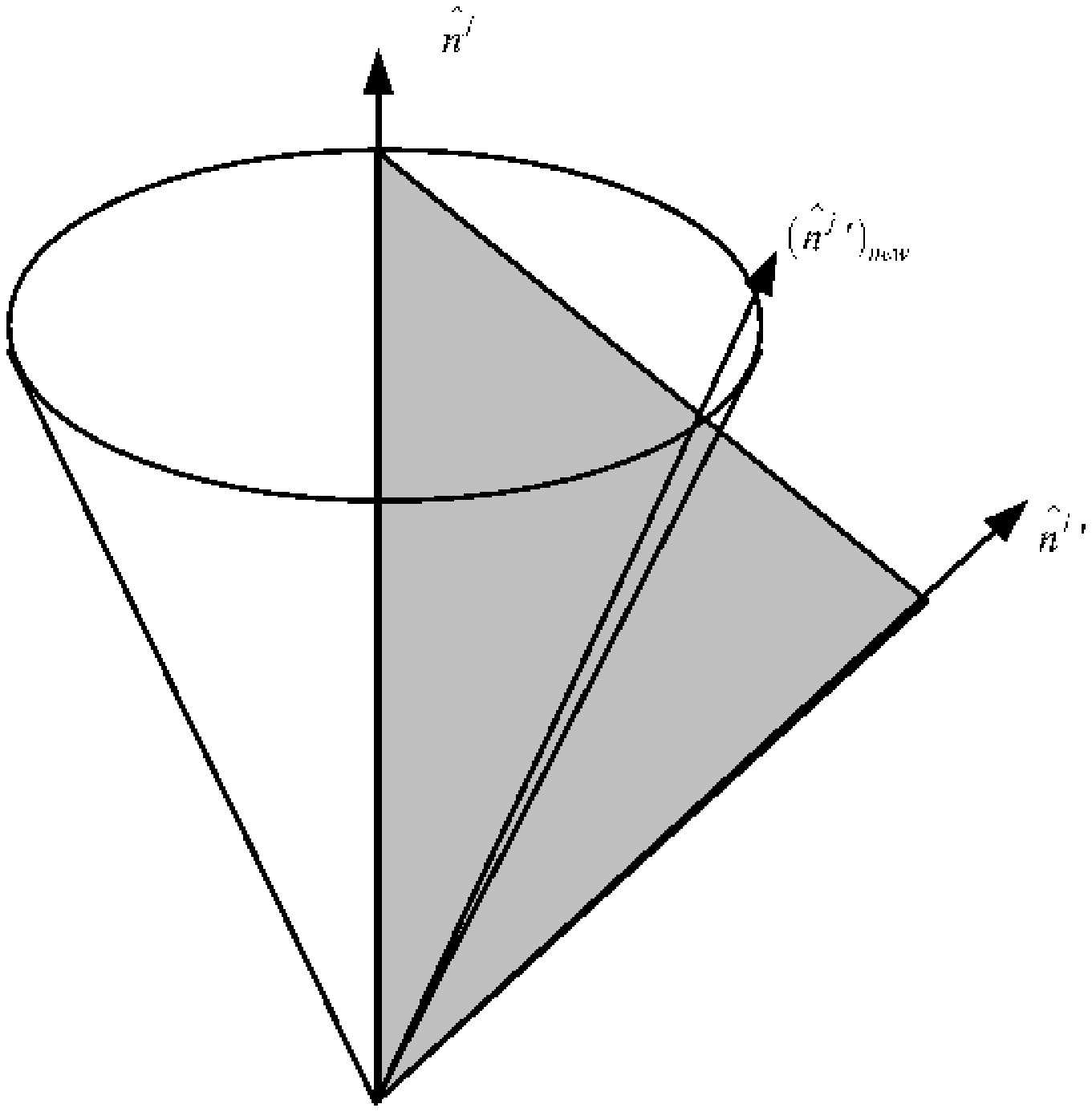}
\end{center}
\caption{A sketch illustrating the construction of the vector given by
  equation~(\ref{eqn:n_new}).  The symmetry axis of the solid-angle
  cone $j$ is given by $\vec{\hat n}_j$, while the gradient direction
  is $\vec{\hat n}_j'$. If the latter lies outside the cone, it is
  projected onto the cone to yield direction $(\vec{\hat n}_j')_{\rm
    new}$, which is then used in the local advection step for the
  cone's radiation field.
 \label{fig:cone_sketch}}
\efig

One may try to fix this problem by somehow randomising the direction
within the corresponding cone taken in single transport steps, or by
using higher-order quadratures in integrating the fluxes arising for a
given mesh geometry.  However, we have found that a simple trick can
be used to resolve this issue, and to obtain close to perfect results
even for unfavourable mesh geometries. To this end, we replace the
local advection direction $\vec{\hat n}_j$ appearing in equation
(\ref{eqn:advectionparallel}) with a modified direction $\vec{\hat
  n}_j'$, chosen along the gradient of the {\em total} radiation
density field, but constrained to lie within the cone
$j$. Specifically, we first adopt \be \vec{\hat n}_j' = - \frac{\nabla
  f_\gamma}{|\nabla f_\gamma|}, \label{eqn:direction} \ee and
calculate the angle between the gradient direction and the cone
direction as \be \phi = \arccos{(\vec{\hat n}_j' \cdot \vec{\hat
    n}_j)}. \ee If this angle is larger than the half opening angle of
the cone, \be \phi^{\rm max} = \sqrt{(4 \pi / N_{\rm pix}) / \pi} ,
\ee then we use the vector $(\vec{\hat n}_j')_{\rm new}$, which is
defined by the intersection of the plane spanned by $\vec{\hat n}_j'$
and $\vec{\hat n}_j$ with the cone of half-opening angle $\phi^{\rm
  max}$ (see Figure~\ref{fig:cone_sketch}). This vector is given by
\be (\vec{\hat n}_j')_{\rm new} = \sin{(\phi^{\rm max})} \vec{m} +
\cos{(\phi^{\rm max})} \vec{\hat n}_j, \label{eqn:n_new} \ee where \be
\vec{g} = \vec{\hat n}_j \times \vec{\hat n}_j', \ee \be \vec{m} =
\vec{g} \times \vec{\hat n}_j . \ee In other words, we transport the
radiation corresponding to a certain cone always in the direction of
the negative intensity gradient, constrained to lie within the solid
angle defined by the cone.

It is clear that this modification has the tendency to smooth out the
angular gradient of the radiation field within a cone, making it
uniform in the cone. For example, imagine that the transport has led
to some intensity excess along the principal direction of the
cone. This will then cause some of the transport steps to propagate
photons away from the symmetry axis of the cone, slightly more
sideways, until the cone is illuminated homogeneously again.  But
importantly, the constraint we imposed on the advection
direction means that all of the photons of any of the partial
radiation fields are always transported along a direction ``permitted''
by their corresponding angular cone. While the specific choice for
this direction may hence deviate slightly from the primary cone axis
$\vec{\hat n}_j$, this deviation is strictly bounded, and it will
automatically become smaller if a larger number of angular cones is
used. One may wonder why we base the initial calculation of the
transport direction in equation~(\ref{eqn:direction}) on the total
radiation intensity field, and not on the partial cone field
$f_\gamma^j$ alone. This is done to avoid possible boundary effects at
the edges of cones, for example when two neighbouring cones are both
homogeneously illuminated. Using the gradient of the total field will
in this case automatically work to eliminate any residuals from the
common boundary and to produce a seamless connection of the cones, a
feature that is not guaranteed when the gradient of the partial field
is used instead. In Section~\ref{sec:tests}, we will discuss a number
of test problems that illustrate that our simple approach works rather
well in practice.  

We note that the angular discretisation we outlined here is completely
independent of the total number of sources. Also, its angular
resolution is constant everywhere (at least in the present
implementation), even though the spatial resolution of the mesh can
vary as a function of position. Another interesting aspect of the
method is that it can work accurately both in the optically thin and
in the optically thick regime, as well as in the transition
region. Unlike in certain approximate treatments of RT, for example in
radiative diffusion, we have not made any approximation that changes
the fundamental character of the equations, apart from the use of a
spatial and an angular discretisation.  This suggests that the robustness
and the convergence of results obtained with this method can reliably
be tested by simply changing the grid and/or angular resolution, and
if convergence is achieved, then the method should converge to the
{\em correct} solution in the limit of high resolution.  The latter
property can not necessarily be expected for RT schemes that use more
drastic approximations.

\subsection{Source and sink terms}  \label{sec:sources}

We treat source and sink terms in the radiative transfer equation
through an operator splitting approach, where the evolution of the
homogeneous RT equation (which conserves photon
number) is alternated with an evolution of the source terms
alone. This greatly simplifies the calculation of the interaction of
the local radiation field with matter, and also allows accurate
balance equations that for example ensure that the number of photons
absorbed matches the number of atoms that are ionised. As an
illustrative example, we here detail our implementation of hydrogen
chemistry, which can be used in simple model calculations of cosmic
reionisation.

\subsubsection{Emission processes} 

Emission of ionising radiation in a cosmological simulation can be
based on a variety of source models, tied for example to star-forming
gaseous cells, star particles, or sink particles that represent
accreting supermassive black holes. Given the source luminosities and
their coordinates, we can simply find the cells in which the sources
fall, and inject the number of photons emitted by them over the timestep into
each of the corresponding host cells. We normally assume isotropic
sources where we distribute the total emissivity equally over all
angular cones, but in principle also beamed emission 
characteristics can be realised.

If our single/multiple point source approach is used instead, we
spread the source photons over a small region around the host cell
with a Gaussian-shaped kernel with a radius equal to a few effective
host-cell radii. This is done to avoid potential asymmetries in the
source's radiation field that otherwise can arise from the particular
geometry of the source cell.

\subsubsection{Absorption and hydrogen chemistry} \label{sec:absorption}

For simplicity, we here discuss a minimal chemical model that only
follows hydrogen and an ionising photon density field with a fixed
spectral shape.  Extensions to include helium and several ionising
frequencies to account for changes of the spectral shape can be
constructed in similar ways.

The neutral hydrogen fraction $\tilde n_{\rm HI}$ evolves due to
photo-ionisations, collisional ionisations and recombinations: \be
\frac{{\rm d} \tilde n_{{\rm HI}}}{{\rm d} t} = \alpha n_{{\rm H}} \,
\tilde n_{\rm e} \tilde n_{{\rm HII}} - \beta n_{{\rm H}} \, \tilde
n_{\rm e} \tilde n_{{\rm HI}} - c \sigma n_{{\rm H}} \, \tilde n_{\rm
  HI} \tilde n_\gamma , \label{eqn:chem}\ee where $\alpha(T)$ is the
recombination coefficient, $\beta(T)$ is the collisional ionisation
coefficient and $\sigma$ is the effective photo-ionisation
cross-section of neutral hydrogen for our adopted spectrum, defined as
\be \sigma = \left[\int \frac{4 \pi
    J_{\nu}(\nu)}{h\nu}\,\sigma_{\nu}(\nu)\,{\rm d}\nu \right]
\times \left[ \int \frac{4 \pi J_{\nu}(\nu)}{h\nu}\,{\rm d}\nu
  \right]^{-1}.  \ee Here $\sigma_{\nu}(\nu)$ is the frequency
dependent photo-ionisation cross-section of neutral hydrogen (with
$\sigma_{\nu}=0$ for frequencies $\nu < \nu_0$ below the
ionisation cut-off $\nu_0$).  The photon density on the other hand
evolves according to \be \frac{{\rm d} \tilde n_\gamma}{{\rm d} t} = -
c \sigma n_{{\rm H}} \, \tilde n_{\rm HI} \tilde
n_\gamma. \label{eqn:chem2} \ee Here the variables $\tilde n_{\rm
  HI}$, $\tilde n_{\rm HII}$, $\tilde n_{\rm e}$ and $\tilde n_\gamma$
express the corresponding abundance quantities in dimensionless form,
in units of the total hydrogen number density $n_{\rm H}$, for example
$\tilde n_\gamma \equiv {n_\gamma}/{n_{\rm H}}$.  If we consider only
hydrogen, we hence have the constraints $\tilde n_{\rm e} = \tilde
n_{\rm HII}$ and $\tilde n_{\rm HI } + \tilde n_{\rm HII} = 1$.

In order to robustly, efficiently and accurately integrate these stiff
differential equations, special care must be taken. This is especially
important if one wants to obtain the correct post-ionisation
temperatures, which requires an accurate treatment of the rapid
non-equilibrium effects during the transition from the neutral to the
ionised state \citep[e.g.][]{Bolton2005}. Also, one would like to
ensure that the number of photons consumed matches the number of
hydrogen photo-ionisations, and that the injected photo-heating energy
is strictly proportional to the number of photons absorbed.  We use
either an explicit, semi-implicit, or exact integration of equations
(\ref{eqn:chem}) and (\ref{eqn:chem2}) to achieve these goals,
depending on the current conditions encountered in each step.

Specifically, we start by first calculating an explicit estimate of
the photon abundance change over the next timestep, as \be \Delta
\tilde n_\gamma = \tilde n_\gamma^{i+1} - \tilde n_\gamma^{i} = - c
\sigma n_{\rm H}\, \tilde n_{\rm HI}^{i} \tilde n_\gamma^{i} \,\,
\Delta t ,\ee where $i$ enumerates the individual
timesteps. If the
implied relative photon density change is small, say $|\Delta \tilde
n_\gamma| < 0.05\, \tilde n_\gamma^{i}$, we are either in approximate
photo-ionisation equilibrium or the photon density is so large that it
does not change appreciably due to hydrogen ionisation losses during
the step. In this situation, we can calculate an estimate for the
neutral hydrogen density at the end of the step based on implicitly
solving \be \tilde n_{\rm HI}^{i+1} = \tilde n_{\rm HI}^{i} + [ \alpha
  (1 - \tilde n_{\rm HI}^{i+1} )^2 - \beta \tilde n_{\rm HI}^{i+1} (1-
  \tilde n_{\rm HI}^{i+1}) ] n_{\rm H} \Delta t + \Delta \tilde
n_\gamma
\label{eqn:impl}
 \ee 
for $\tilde n_{\rm
  HI}^{i+1}$.  If the implied relative change in $\tilde n_{\rm HI}^{i}$ is
again small, we keep the solution.

Otherwise, we first check whether the photon number is very much
smaller than the neutral hydrogen number, i.e.~whether we have $\tilde
n_\gamma < 0.01\, \tilde n_{\rm HI}$.  If this holds, the photons in
the cell cannot possibly ionise a significant fraction of the neutral
hydrogen atoms, but the photon abundance itself may still change
strongly over the step (for example because almost all of the photons
are absorbed).  We in this case first compute an estimate of the new
photon number at the end of the step, based on the implicit step \be
\tilde n_\gamma^{i+1} = \tilde n_\gamma^{i} - c \sigma n_H \tilde
n_{\rm HI}^i \tilde n_\gamma^{i+1}\, \Delta t
.  \label{eqn:n_gamma}\ee With the solution
for $\tilde n_\gamma^{i+1}$ in hand, we calculate again an implicit
solution for the new neutral hydrogen fraction at the end of the step,
using equation~(\ref{eqn:impl}).  If the predicted relative change in
the hydrogen ionisation state is small, we keep the solution,
otherwise we discard it.

Finally, if both of the two approaches to calculate new values for
$\tilde n_\gamma^{i+1}$ and $\tilde n_{\rm H}^{i+1}$ at the end of the
step have failed, we integrate the rate equations (\ref{eqn:chem}) and
(\ref{eqn:chem2}) essentially exactly over the timestep $\Delta t$,
using a 4-th order Runge-Kutta-Fehlberg integrator with adaptive step-size
control as implemented in the {\small GSL}
library\footnote{http://www.gnu.org/software/gsl}.  We note that this
sub-cycled integration is hence only done in timesteps where the
ionisation state changes rapidly in time and non-equilibrium effects
can become important, which is a very small fraction of all cells,
such that our updating scheme remains computationally very efficient.

\subsection{Photo-heating and radiative cooling}\label{sec:heating}

To calculate the evolution of the thermal energy, we can now inject
the photo-heating energy \be \Delta E_\gamma = (\tilde n_\gamma^i -
\tilde n_\gamma^{i+1}) n_{\rm H} V\, \epsilon_\gamma \ee into the
corresponding cell, where $V$ is the volume of the cell under
consideration, $(\tilde n_\gamma^i - \tilde n_\gamma^{i+1}) n_{\rm H}$
is the number density of photons consumed by ionising events over the
timestep, and $\epsilon_\gamma$ gives the average energy absorbed per
photo-ionisation event.  For our prescribed spectral shape, this
injection energy per ionisation event is given by the
frequency-averaged photon excess energy \citep{Spitzer1998} \be
\epsilon_\gamma = \left[ \int_{\nu_0}^\infty {\rm d\nu} \frac{4\pi J_\nu
   }{h\nu}\sigma_\nu(h\nu - h\nu_0) \right] \times \left[ \int_{\nu_0}^\infty
        {\rm d\nu} \frac{4\pi J_\nu} {h\nu}  \sigma_\nu\right]^{-1} \ee
        above the ionisation cut-off $\nu_0$.
For many of our test calculations,
we assume a black body spectrum with $T_{\rm 
  eff}=10^5\, \rm K$, which leads to $\epsilon_\gamma = 6.4\, \rm
eV$. 

The evolution of the thermal energy is then completed by a separate
cooling step that accounts for recombination cooling, collisional
ionisation, excitation cooling, and bremsstrahlung cooling
\citep[e.g.][]{Katz1996}. We implement these cooling rates with a
combination of an explicit and implicit timestep integrator, where an
explicit integration scheme is used as default, but if the temperature
change over the step becomes large, the cooling is instead calculated
with an unconditionally stable implicit solver.

\subsection{Time stepping and the reduced speed-of-light approximation}
 \label{sec:timestep}

As discussed above, we include the source terms into the time
integration of our RT solver by an operator splitting technique, where
the source and advection parts are treated separately. This technique
can be generalised also to the coupling of hydrodynamics and radiative
transfer, by alternatingly evolving the hydrodynamical density field
and the radiation field with its associated radiation chemistry. In
fact, this is the approach we follow in our radiative transfer
implementation in the hydrodynamical {\small AREPO} code.  As the
latter is a moving-mesh code, we however need to ensure that during
the hydrodynamical step the radiation field is left invariant. This
can be achieved by appropriate advection terms that compensate for the
mesh-motion during the hydrodynamical step.

\bfigs
\begin{center}
\includegraphics[width=0.3\textwidth,clip]{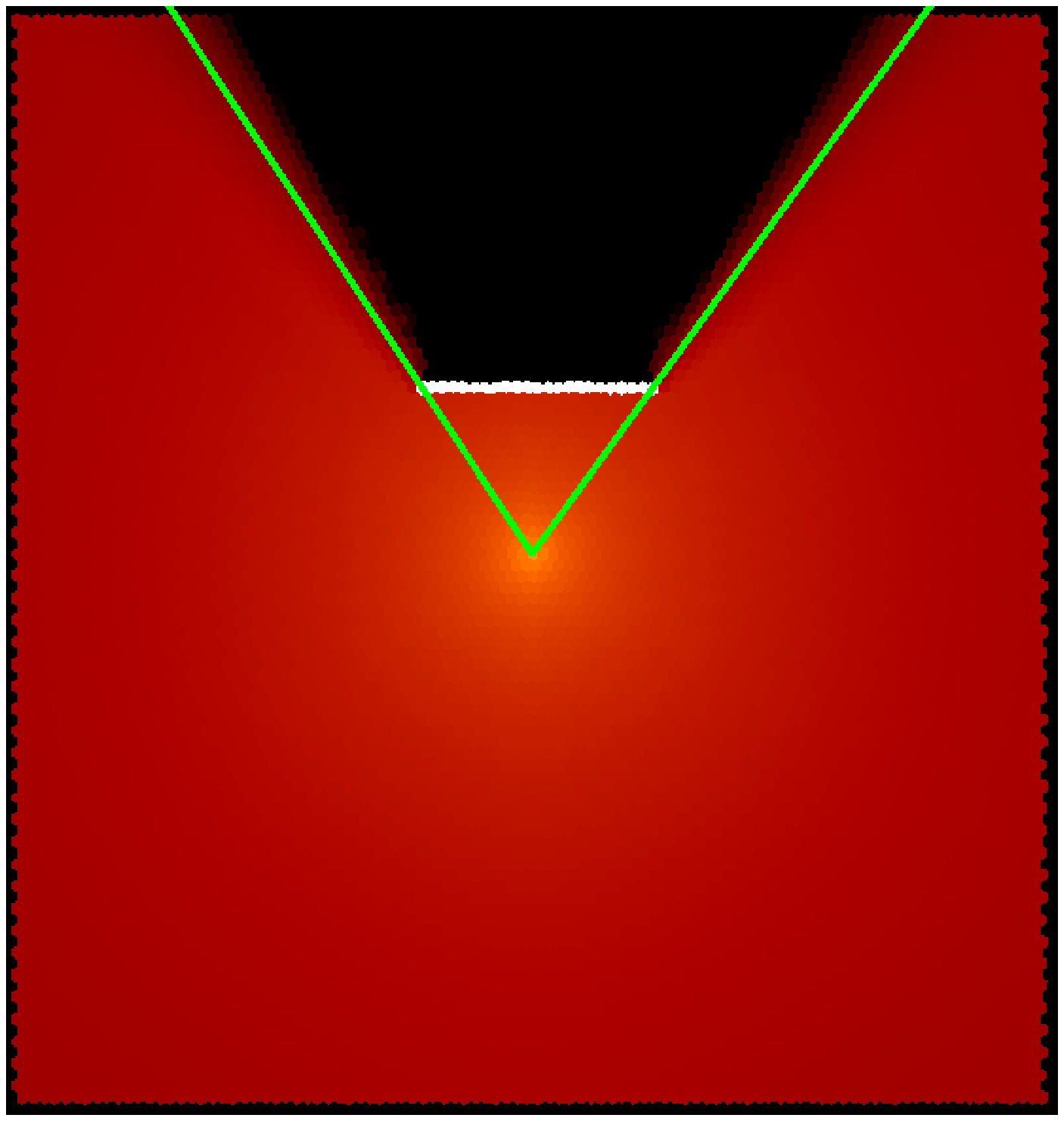}
\includegraphics[width=0.3\textwidth,clip]{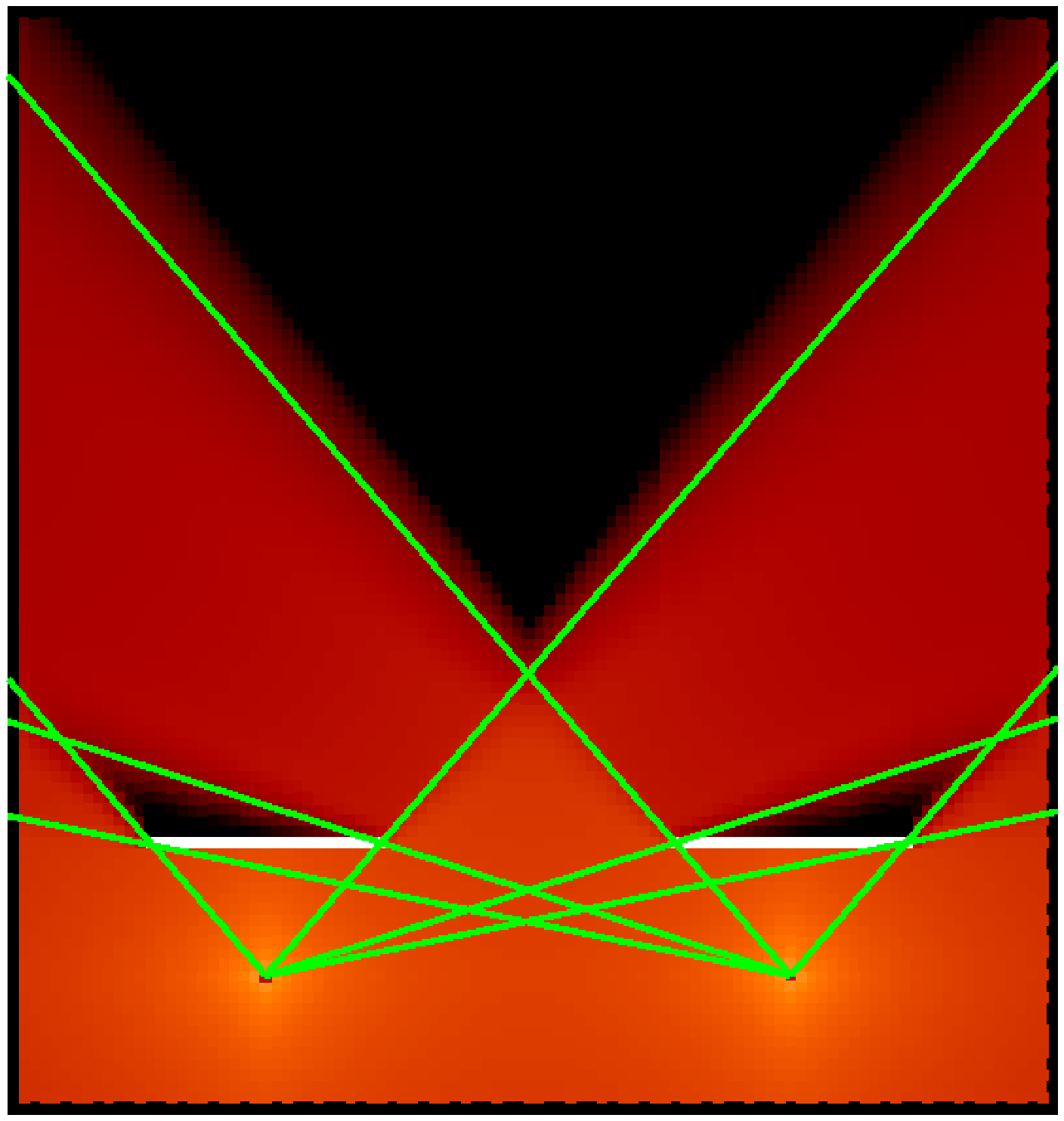}
\includegraphics[width=0.3\textwidth,clip]{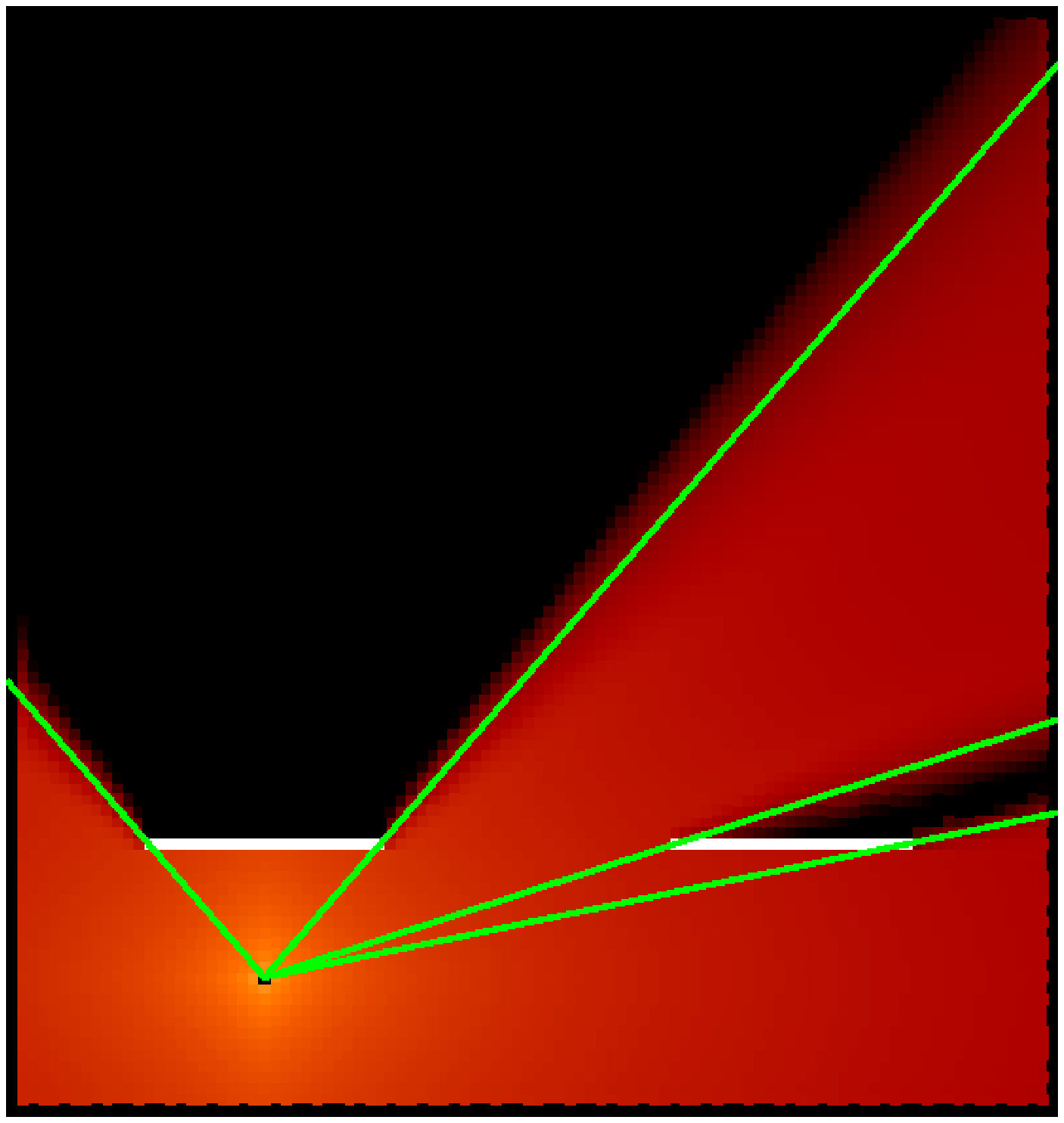}
\includegraphics[height=0.3\textwidth]{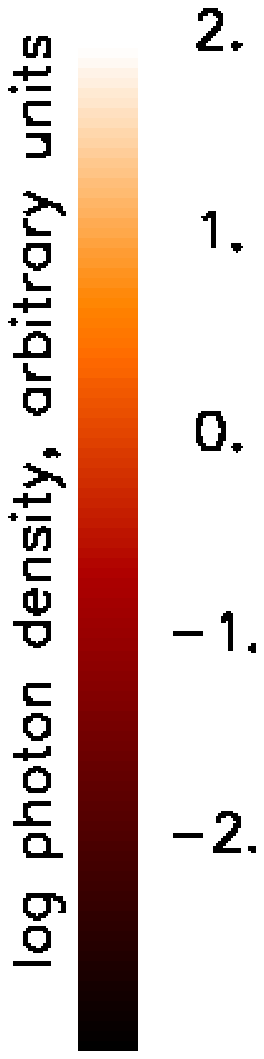}
\end{center}
\caption{Photon density maps of 2D shadowing tests in three different
  cases: single point source with a single obstacle (left panel), two
  point sources with two obstacles (middle panel), and a single source
  with two obstacles (right panel). The green lines indicate the
  geometric boundaries of the expected shadow regions, whereas the thick
  white lines mark the absorbing obstacles. \label{fig:shadow}} \efigs

For the time integration of the radiative source terms, we employ
implicit or semi-implicit methods, as described in
Section~\ref{sec:sources}, that are stable even for very large time
steps, and in selected situations, adaptive numerical integration of
the stiff ordinary differential equations that describe the chemical
networks. The latter is essential to accurately account for
non-equilibrium effects.  As these processes are completely local,
this does usually not incur a very significant computational cost,
provided the exact integration is only done where really needed.  In contrast, the
timestepping of the radiation advection step poses more severe
computational requirements. This is because this step is based on an
explicit time integration scheme, whose timestep needs to obey a
Courant criterion of the form \be \Delta t _{\rm advect} < C_k
\frac{\Delta x \, a}{c}, \label{eqn:stability}\ee where $0 < C_k < 1$
is the Courant factor, and $\Delta x$ is taken to the smallest comoving size of
a cell in the simulated volume.

In ordinary hydrodynamics, a similar time step constraint is
encountered, except that the speed of light is replaced with the speed
of sound. Since we are primarily interested in non-relativistic gas
dynamics in cosmological structure formation, the speed of light will
typically be a factor $\sim 10^2$ to $10^4$ larger than the
hydrodynamical sound speed. The resulting reduction in the allowed
time step size can hence make a simulation prohibitively expensive
when the RT is coupled to the hydrodynamics over significant fractions
of the Hubble time.  However, in many applications of interest this
problem can be greatly alleviated by resorting to an artificially
reduced speed of light $c'$, which is introduced instead of the
physical speed of light both in the transport equation and the
ionisation equation. As \citet{GA2001} and \citet{AT2007} discuss in
detail, this reduced speed-of-light approximation is especially
attractive for cosmic reionisation problems because it here does not
modify the propagation speed of I-fronts, except perhaps in
the very near field region around a source directly after it turns on,
but this introduces a negligible timing error. In general, the reduced
speed-of-light approximation can be expected to yield reasonable
accuracy in many radiation hydrodynamic problems as long as $c'$ remains
significant larger than the maximum sound speed occurring in the
simulation.

\subsection{Implementation aspects in the moving-mesh code {\small AREPO}}
 \label{sec:imple}

We have implemented the different variants of our radiation advection
solver in the moving-mesh code {\small AREPO} \citep{Arepo}. This code
treats hydrodynamics with an ordinary finite-volume approach and a
second order accurate Godunov scheme, similar to many Eulerian grid
codes.  However, {\small AREPO} works on an unstructured mesh created
with a tessellation technique.  The particular mesh used is the
Voronoi tessellation created by a set of mesh-generating points. Using
such a mesh offers a number of advantages compared to traditional grid
codes in that its mesh can flow along with the gas.  As a result of
the induced dynamic mesh motion, {\small AREPO} exhibits considerably
lower advection errors than ordinary mesh codes, and also avoids the
introduction of preferred spatial directions.  Also, the cell size
automatically and continuously adjusts to the density in a Lagrangian
sense, and is hence decreased in regions where typically more
resolution is required even without doing adaptive mesh refinement.

For implementing our RT transfer scheme as described above, we can
readily employ the infrastructure and communication algorithms
provided by the fully parallelised {\small AREPO} code, making it an
ideal base for a first demonstration of the method. This in particular
applies to the gradient estimation, the spatial reconstruction of the
photon intensity fields, and the parallelisation for distributed
memory computers. A full description of these aspects of our code can
hence be found in \citet{Arepo}.  We carry out a RT step on every
top-level synchronisation point of the {\small AREPO} code, which
means on the longest time step $\Delta t_{\rm max}$ allowed by the
gravitational and hydrodynamical interactions followed by the code.
If $\Delta t_{\rm advect}$ is smaller than the top-level simulation
time-step $\Delta t_{\rm max}$, the radiation transfer step is
calculated in several sub-cycling steps equal to or smaller than
$\Delta t_{\rm advect}$, as needed.

Note that these sub-cycling steps do not require a new construction of
the Voronoi mesh, or a new gravity calculation, hence they are in
principle quite fast compared to a full step of the hydrodynamic
code. However, this advantage can be quickly  (over)compensated by the
need to carry out multiple flux calculations for each of the angular
components of the radiation field, and the additional need to do
subcycling in time to ensure stability of the explicit time integration
used in the advection steps. Furthermore, if a multiple frequency
treatment is desired, the cost of the radiative transfer calculations
will scale linearly with the number of frequency bins
employed, simply because the dominating advection part of the
radiative transfer problem needs to be carried out for each frequency
independently. The additional storage requirements for a multiple
frequency treatment should also not be overlooked, which again scale
linearly with the number of frequency bins, likewise with the number
of solid-angle bins used in the angular discretisation. It is hence
clear that multi-frequency radiative transfer at high angular
resolution clearly remains expensive with the discretisation scheme
proposed here. However, the relative cost increase compared to
hydrodynamics alone is a constant (and at least for sufficiently
interesting problems still affordable) factor that is nearly
independent of spatial resolution. This, together with the ability of
our scheme to cope with essentially arbitrary source functions, makes
it an interesting new technique for cosmological hydrodynamics.

\section{Basic test problems}\label{sec:tests}

\subsection{Shadows around isolated and multiple point sources}
\label{sec:shadow}

We begin our investigation of the accuracy of our proposed radiative
transfer algorithms with isolated point sources in an optically thin
medium that includes some regions with absorbing obstacles. This
serves both as a verification that an isolated point source produces a
radiation field $n_\gamma \propto 1/r^2$ (in 3D, and $n_\gamma \propto
1/r$ in 2D), with sufficiently spherical isophots, and as a test
whether the method can cast sharp shadows behind obstacles. The latter
is often difficult for RT transfer schemes, especially the ones that
are diffusive in character such as the OTVET scheme
\citep[e.g.][]{GA2001,Petkova2009}.

In Figure~\ref{fig:shadow}, we show such shadowing tests for three
different cases, which for visualisation purposes have been done in 2D
space. In the left hand panel, we consider the shadow that is produced
by an obstacle when it is illuminated by a single source in the middle
of the panel. The green lines show the geometric boundaries of the 
theoretically expected position of the shadow. We see that the
obstacle produces a rather sharply defined shadow with only a small
radiation leak into the shadowed region due to numerical advection and
discretisation errors along the shadow boundaries. In the unshadowed
regions, the radiation intensity falls of as $\propto 1/r$, as
expected.

\bfig
\begin{center}
\includegraphics[width=0.3\textwidth,clip]{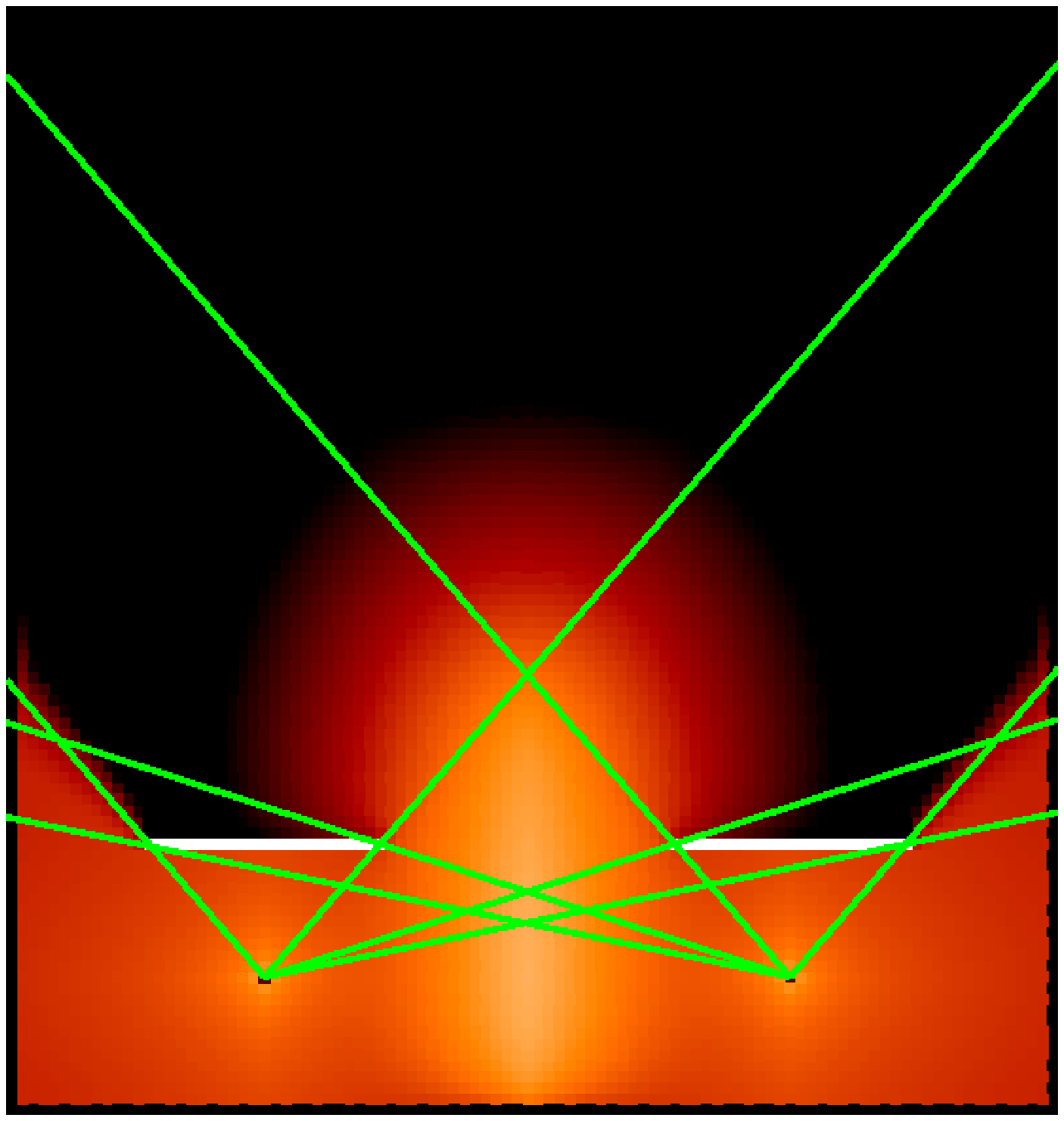}
\includegraphics[height=0.3\textwidth]{images/phot_legend.eps}
\end{center}
\caption{Radiation field around two point sources and two absorbing
  obstacles, for our hybrid treatment of point-sources and radiative
  diffusion. In this example, only the brightest source seen from a
  given cell was treated explicitly as a point source, while the other
  radiation was dumped into a background field transported with
  radiative diffusion.
\label{fig:shadow_back}}
\efig

\bfig
\begin{center}
\includegraphics[width=0.3\textwidth,clip]{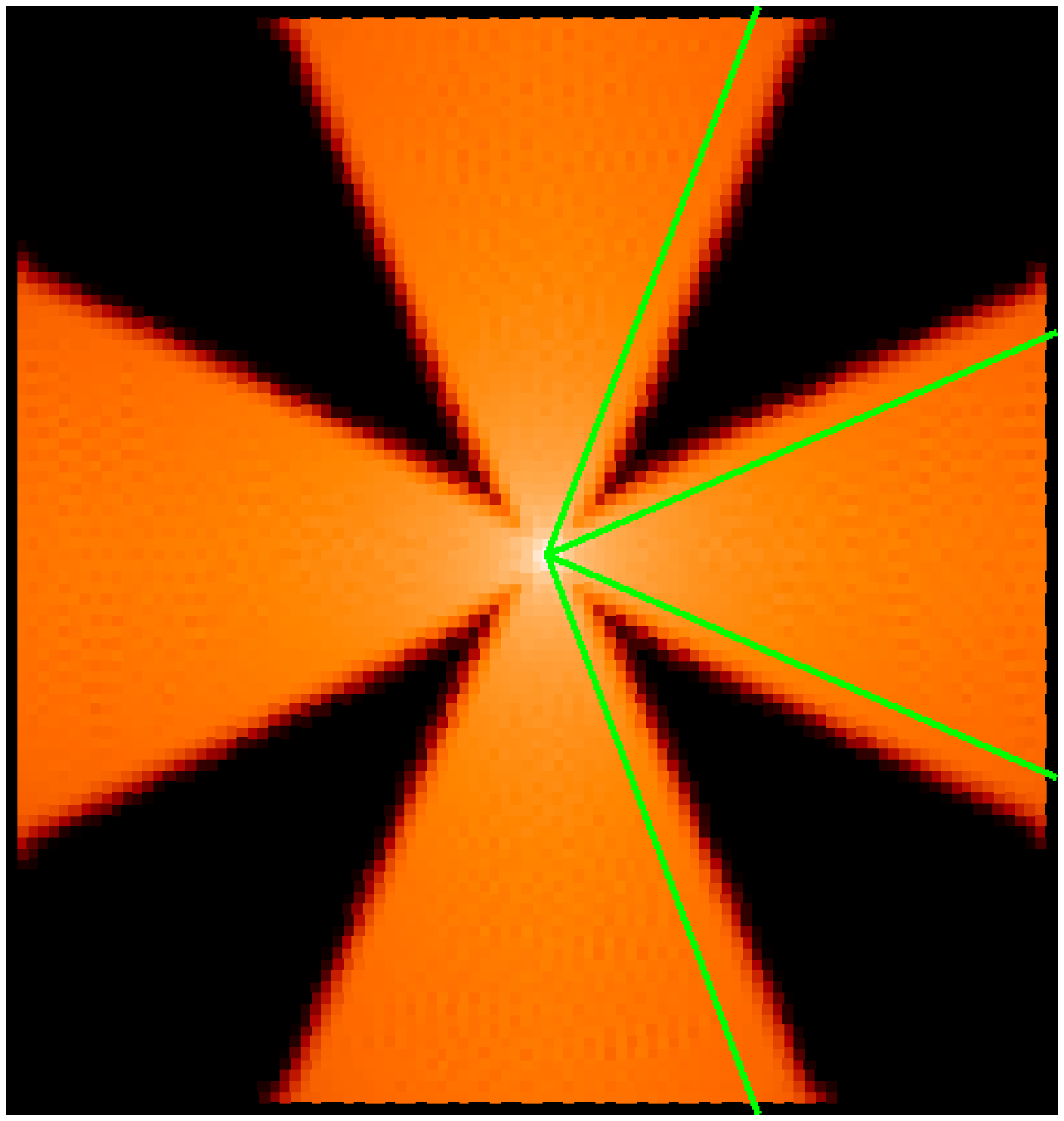}
\includegraphics[height=0.3\textwidth]{images/phot_legend.eps}
\end{center}
\caption{Illustration of our cone transport scheme, and the accuracy
  with which it can represent homogeneously illuminated radiation
  cones.  The panel shows a map of the 2D photon density around a
  single source positioned at the center of the field. Radiation
  transport was calculated by partitioning the full $2 \pi$ angle into
  eight regions of size $\pi/4$ each, with each of the fields
  transported individually. To show that the cones produce a
  homogeneous field, the source luminosity was only injected into
  every second cone and some of the cone boundaries were marked with
  green lines. \label{fig:cone}} \efig

Equally good results are also obtained when multiple sources are
considered in our ``linear sum'' approach to the total radiation field,
where the total photon density is computed as a linear sum of the photon
fields from each source, and the transport of each partial field is
treated independently.  Examples for this are shown in the middle and
right panels of Figure~\ref{fig:shadow}, where two obstacles and one or
two sources are used in different configurations.  Again, the shadows
agree very well with the expected boundaries shown with green lines,
with only a small amount of residual diffusion into the shadowed regions.
If the spatial mesh resolution is improved, the shadows become
progressively sharper still. 

We note that the above success essentially holds in this approach for
an arbitrary set of absorbing regions, and an arbitrary combination of
point sources. It hence provides a general and highly accurate
solution to the radiative transfer problem, even though it can
certainly get expensive to obtain it, especially for a large number of
source. It is important to note however that the radiation fields
produced by our scheme are essentially noise-free, which is a drastic
improvement compared to results obtained from schemes that rely on
Monte-Carlo methods \citep[e.g.][]{MFC2003}, or on randomised cone transport
\citep[]{Pawlik2008}.

As we discussed earlier, for many problems in astrophysics the number
of sources is too large to make the linear sum approach a viable
solution technique. In our first approach to work around this
limitation, we only treat the photons from the brightest sources at a
given cells as independent point sources in the transport scheme,
while all other incoming photons form fainter sources are added to a
background radiation field, which is then diffused from cell to
cell. In Figure~\ref{fig:shadow_back}, we show a (somewhat extreme)
example for how this can change the results. We repeat the test shown
in the middle panel of Fig.~\ref{fig:shadow}, which has two sources
and two obstacles, but this time we only allow the code to treat the
locally brightest $N_{\rm br}=1$ sources as explicit point sources,
while the rest of the radiation needs to treated with radiative
diffusion. As we can see from Figure~\ref{fig:shadow_back}, the
radiation field near to the two sources is unchanged, as expected, but
at the mid-plane, where the sources have equal intensity, half of the
flux is dumped into a diffusive reservoir. The diffusion approximation
then lets the radiation spread from the mid-plane more slowly, causing
an incorrect increase of the radiation intensity there. A second
effect is that the shadows behind the obstacles are not sustained as
nicely any more, instead they are partially illuminated by the
radiative diffusion. It is important to be aware that unlike in the
pure transport scheme considered earlier, these errors will not become
smaller for an improved mesh resolution, rather, one would simply
converge to a wrong solution in this case.

The example studied in Fig.~\ref{fig:shadow_back} is deliberately
extreme in the sense that $N_{\rm br}$ was kept very low. Much better
results can be expected for a sizable value of $N_{\rm br}$, say
$5-10$, because then the flux that needs to be treated with the
diffusion approximation should become locally sub-dominant
everywhere. Nevertheless, for general radiation fields and smoothly
distributed source functions, we prefer our ``cone transport'' scheme,
which we now begin to evaluate in the context of shadowing.

\bfigs
\begin{center}
\includegraphics[width=0.3\textwidth,clip]{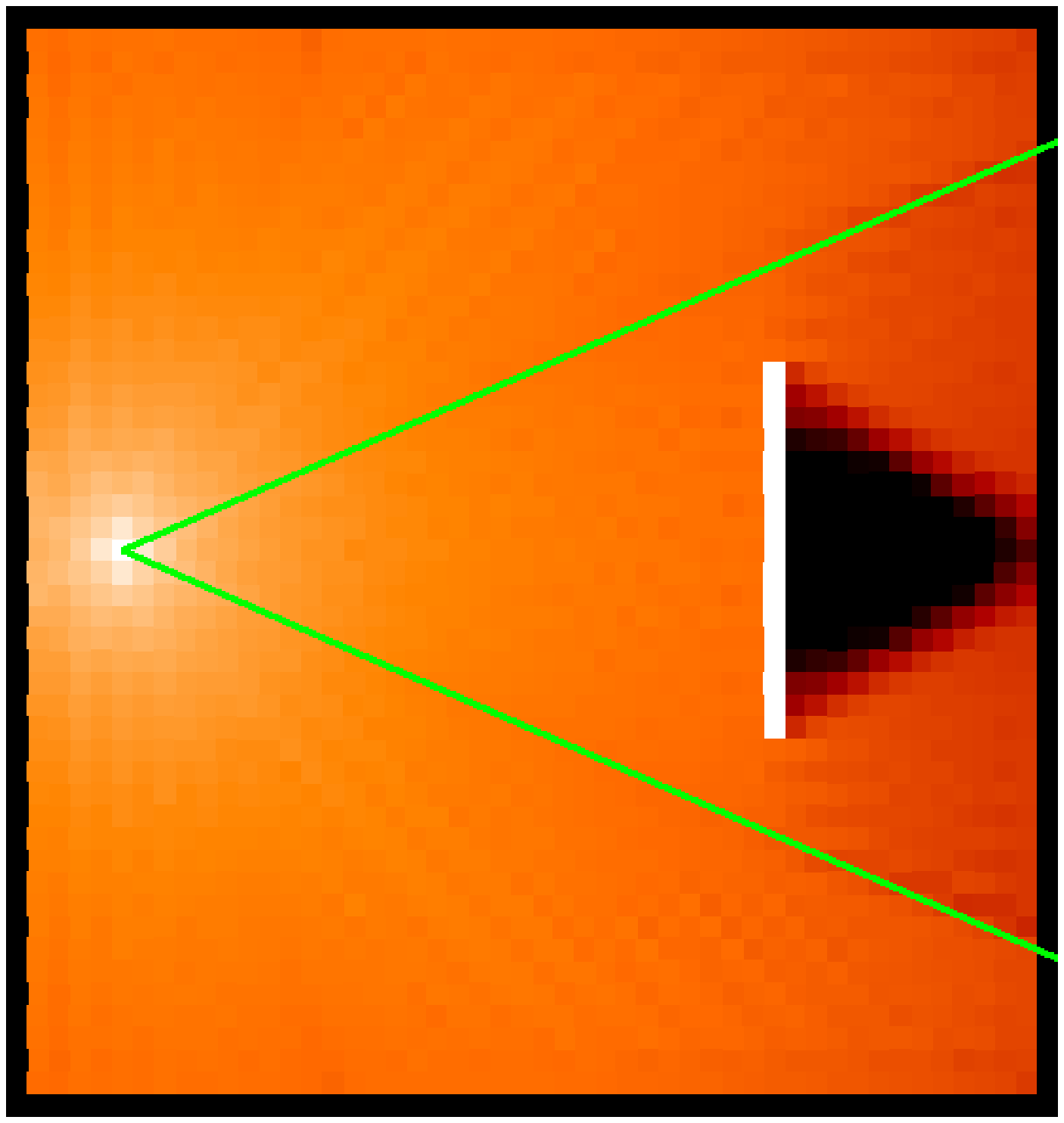}
\includegraphics[width=0.3\textwidth,clip]{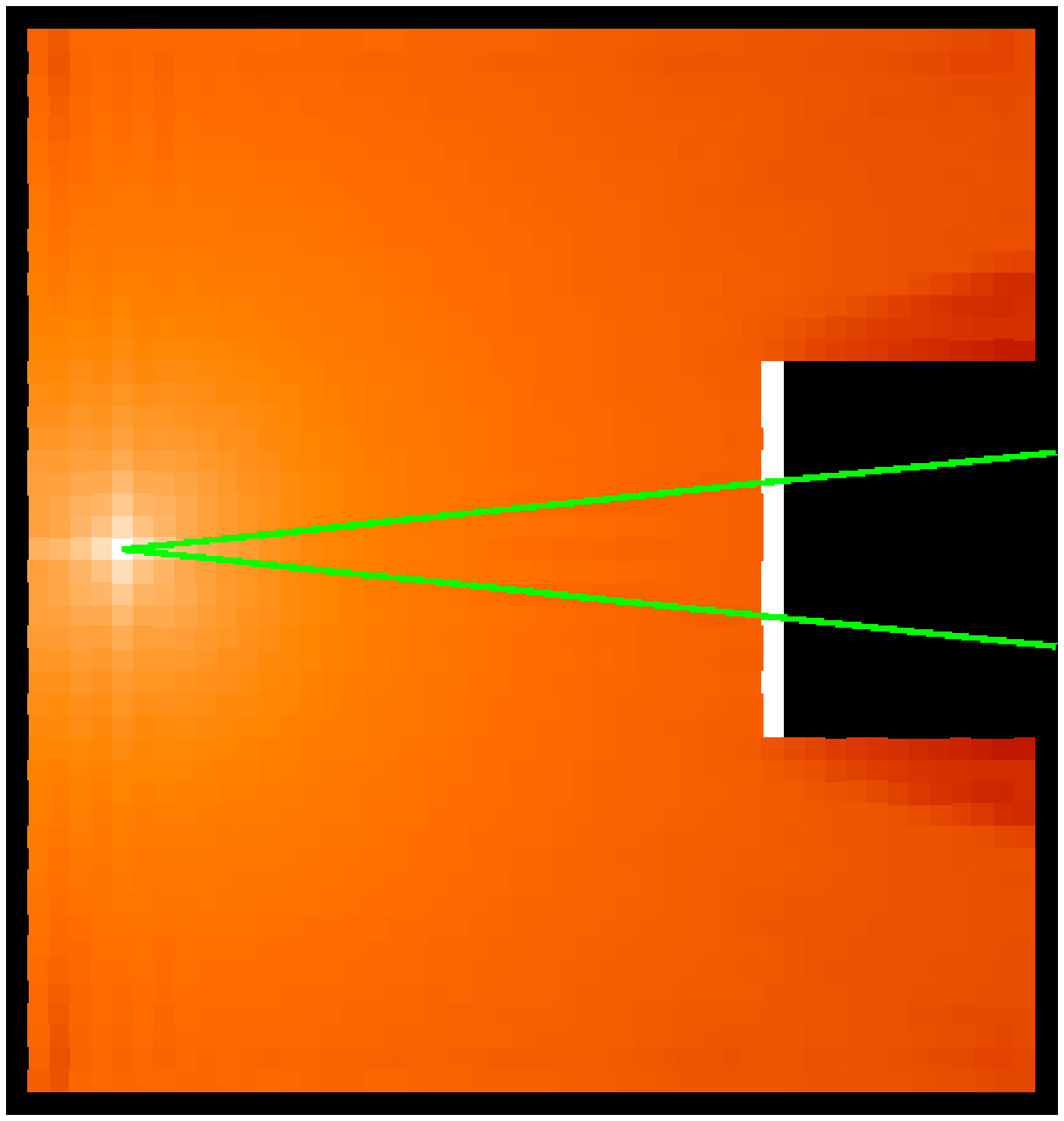}
\includegraphics[width=0.3\textwidth,clip]{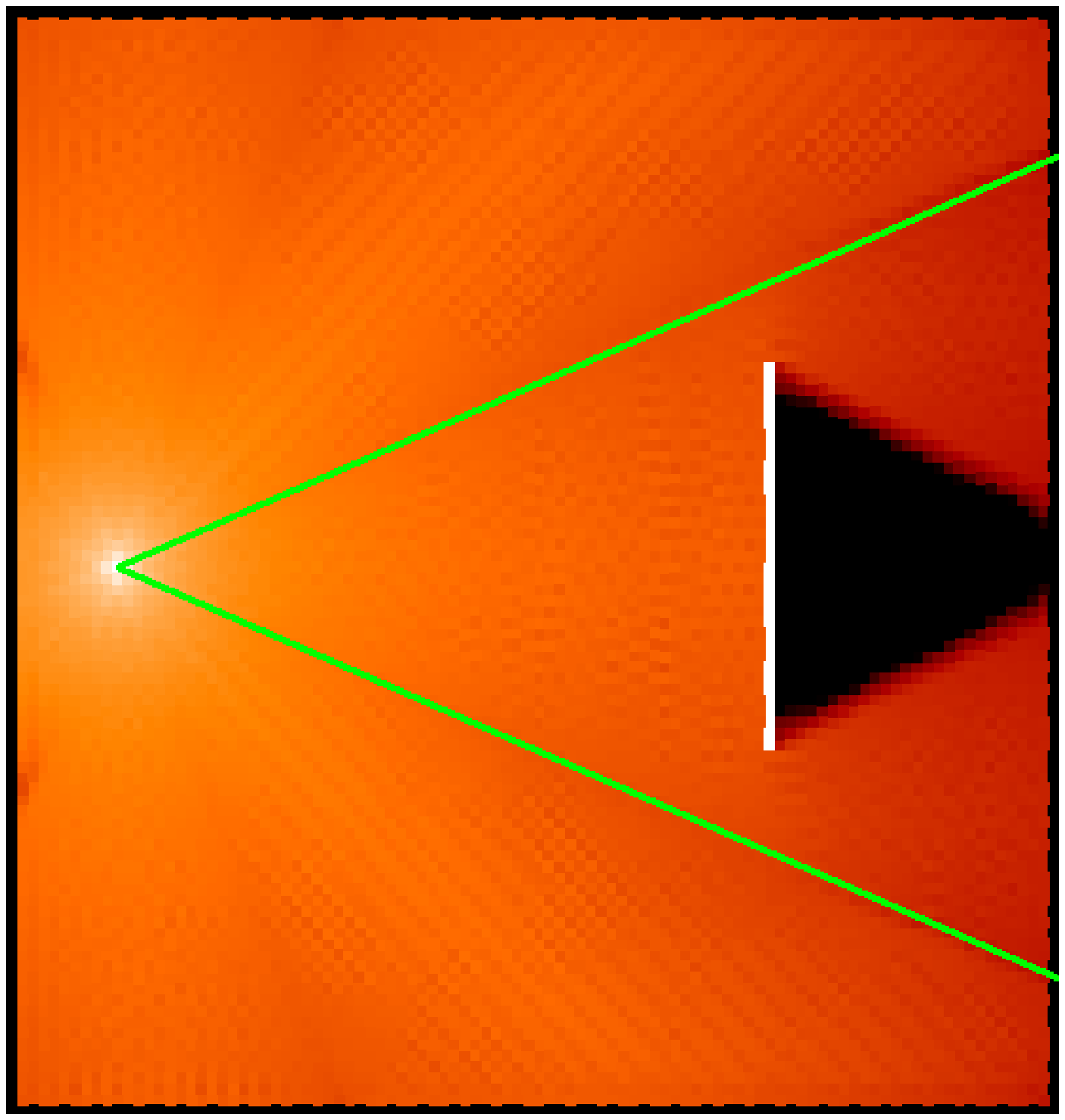}
\includegraphics[height=0.3\textwidth]{images/phot_legend.eps}
\end{center}
\caption{Maps of the photon density field obtained with our ``cone
  transport'' scheme for a point source at coordinates $(0.1,0.5)$ and
  an obstacle (shown in white) centered at coordinates
  $(0.72,0.5)$. The three panels differ in the mesh resolution used
  and the angular resolution employed for the radiation transfer.  In
  all panels, the angular size of the cones employed in the angular
  discretisation are shown with green lines. In the left panel, where
  eight cones are used, the obstacle's opening angle as seen from the
  source is smaller than the fundamental cone, and therefore no
  complete shadow is formed.  In the middle panel, 32 cones are used
  instead, such that the obstacle's opening angle is now larger than
  the angular resolution, allowing a full shadow to be formed.
  Finally, in the right hand panel, the spatial mesh resolution has
  been doubled in each dimension compared to the panel on the left,
  while the number of angular resolution elements has been kept at
  eight. Again there is no complete shadow formed, as expected, but
  the boundary of the shadow region behind the obstacle is now more
  sharply defined. \label{fig:mesh}} \efigs

\bfigs
\begin{center}
\includegraphics[width=0.24\textwidth,clip]{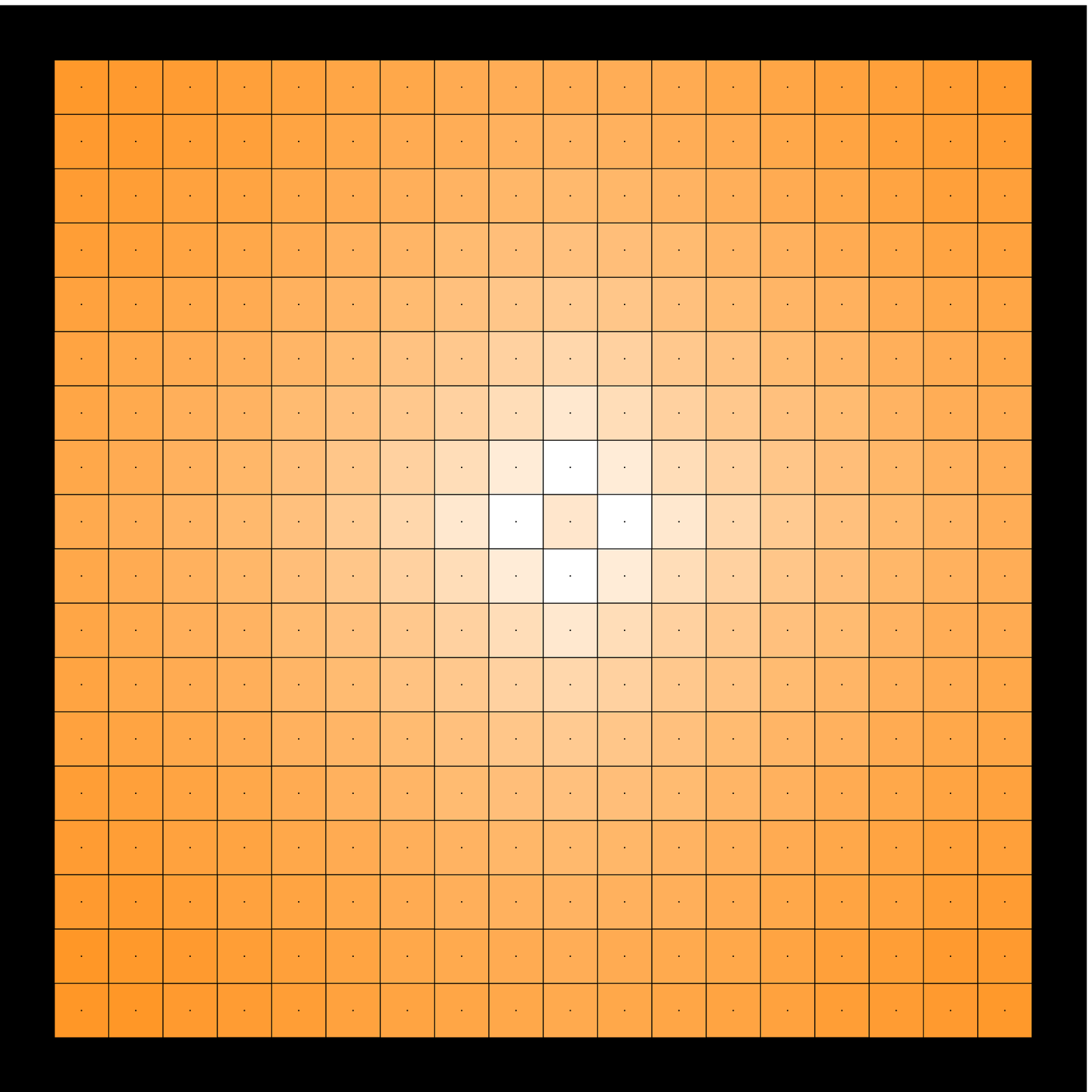}
\includegraphics[width=0.24\textwidth,clip]{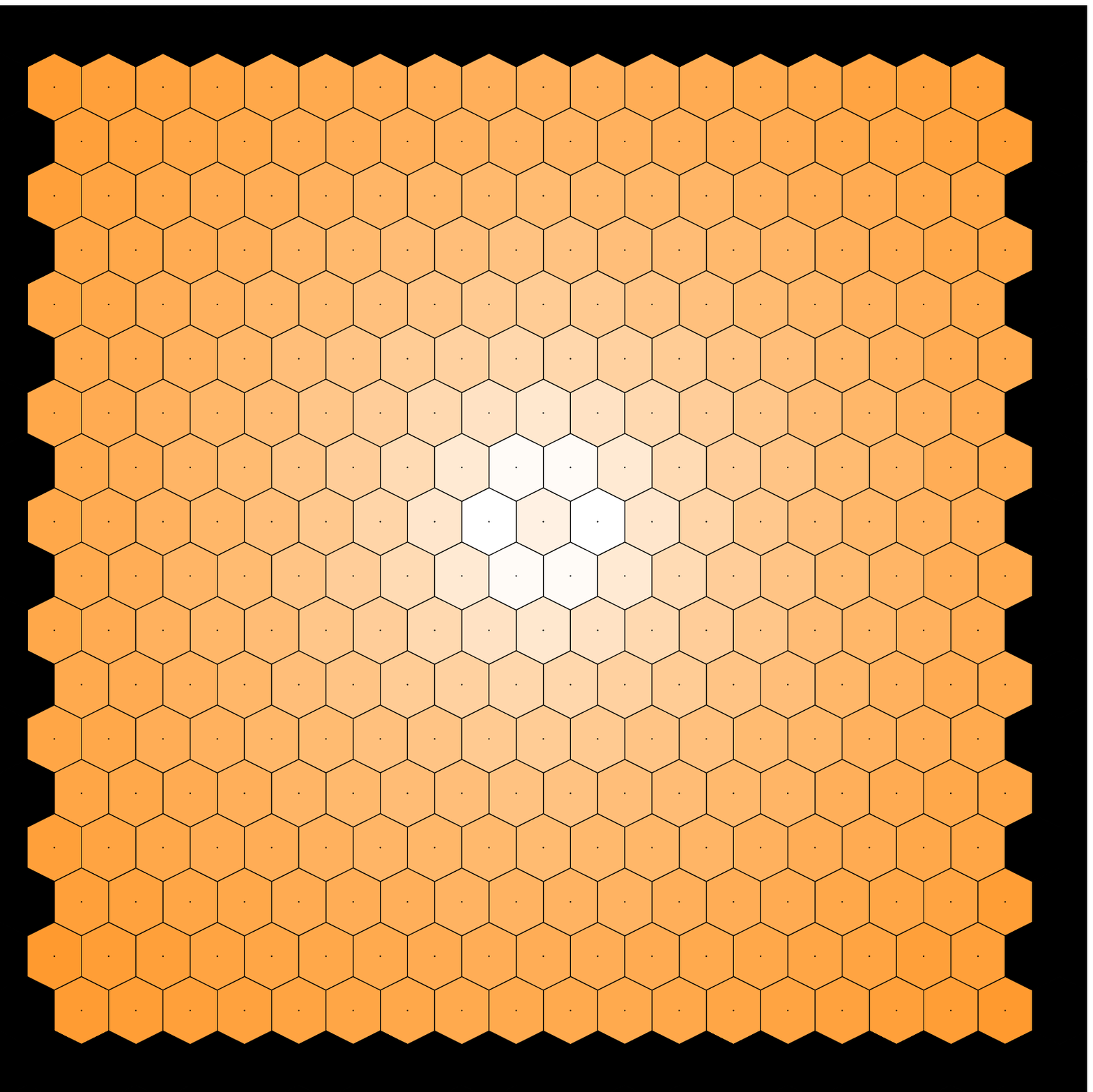}
\includegraphics[width=0.24\textwidth,clip]{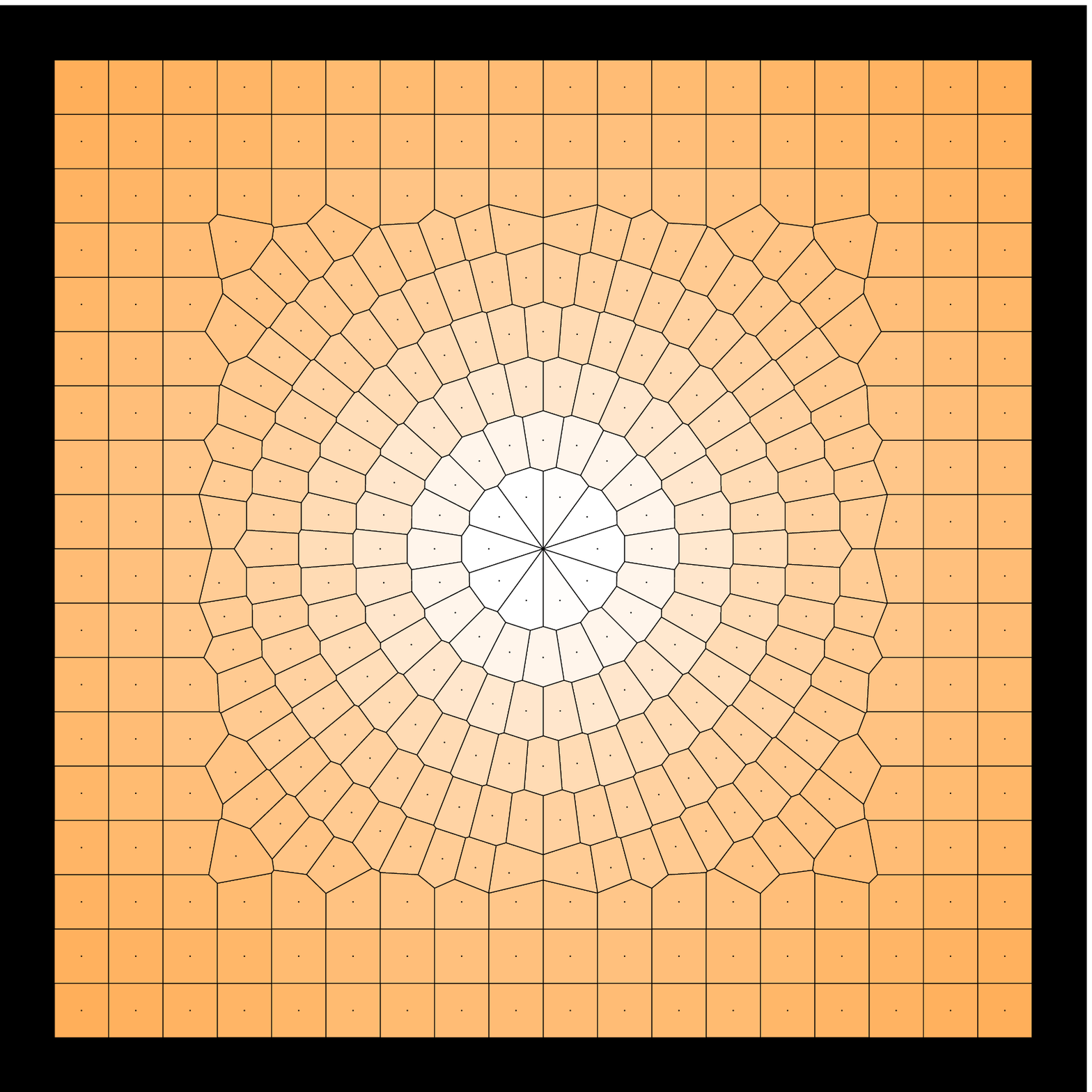}
\includegraphics[width=0.24\textwidth]{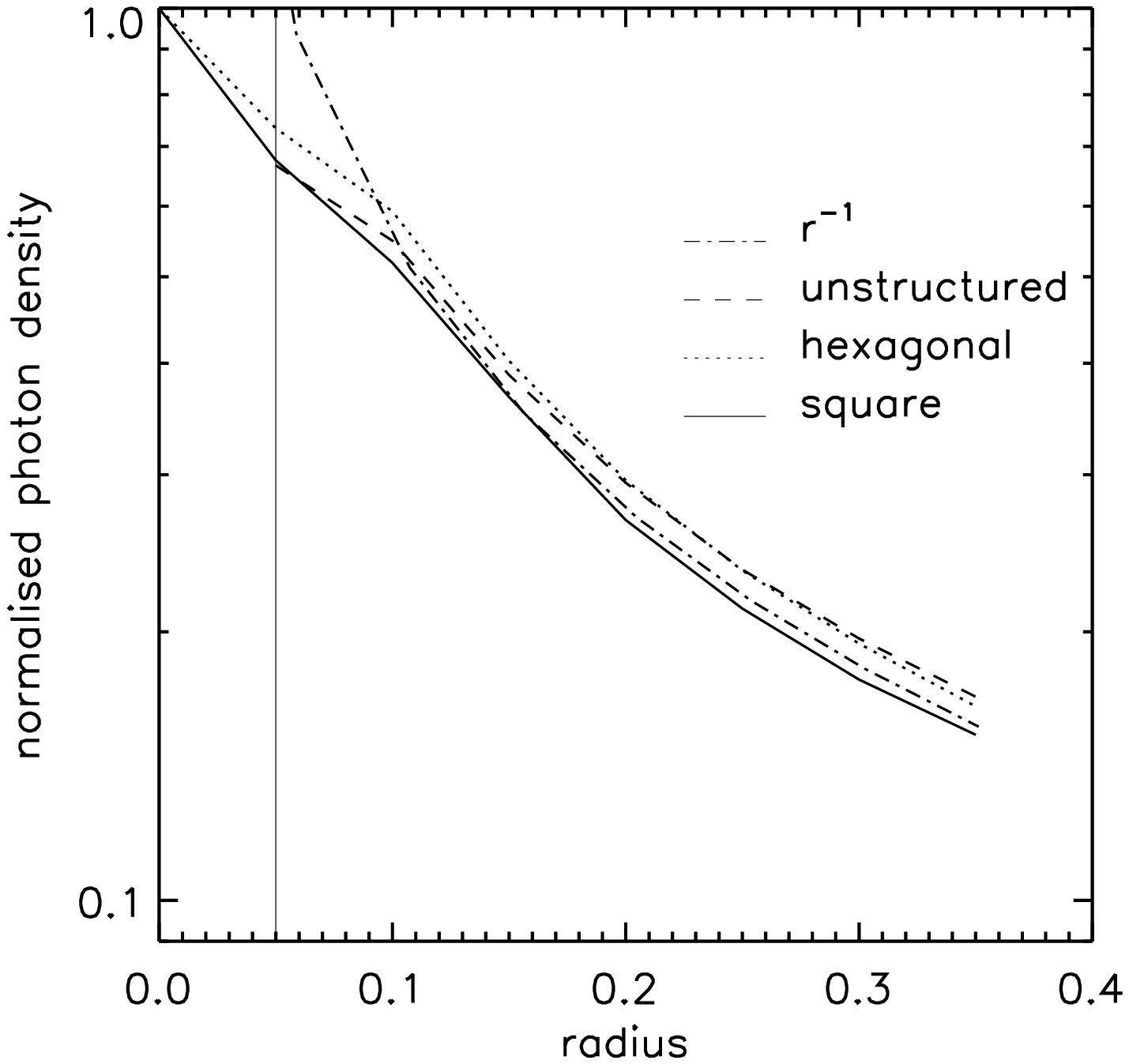}
\end{center}
\caption{Maps of the photon distribution for 2D simulations with three
  different cell shapes: Cartesian square, hexagonal and
  azimuthal/unstructured. The line plot on the right shows the photon
  intensity profile, overplotted with the expected $r^{-1}$ law. The
  vertical line indicates the average cell size. Results from all mesh
  shapes agree well with the analytical
  prediction (dot-dashed line). \label{fig:mesh_shape}} \efigs

In Figure~\ref{fig:cone}, we illustrate the ability of this transport
scheme to produce homogeneously illuminated radiation cones with an
opening angle equal to the angular resolution adopted for the scheme. In
this example, a single source was placed in the center of a 2D unit
square, and angular space has been divided into eight equal sized
regions, with the source radiation only injected into four of them,
alternating between an ``empty'' and a ``full'' cone. The green lines in
the plot show some of the geometric boundaries of the angular
discretisation as seen from the source. We see that the cone transport
succeeds in producing a flat intensity profile as a function of angle
within every illuminated cone, while at the same time the leaking of
radiation into cones that should remain dark as seen from the source is
very small. We note that if we let the source inject radiation into two
adjacent cones with equal luminosity, the radiation field shows no
trace of the angular boundary between the cones, thanks to our use of
the total intensity field in calculating the local advection direction
for the radiation of each partial field.

An interesting question now arises how this transport scheme deals
with obstacles and the problem of shadowing. We illustrate the salient
points with a few tests in Figure~\ref{fig:mesh}. Here, we illuminate
an obstacle (shown in white) by a single source in the left part of
the simulated 2D space. We vary both the angular and the spatial mesh
resolution in order to study how the shadowing performs in the cone
transport scheme. In the left panel, we have used $50^2$ cell and
eight angular regions. The fundamental cone size is shown by the green
lines. In this setup, the opening angle of the obstacle as seen from
the source is hence smaller than the angular resolution of the RT,
making the obstacle ``unresolved''.  In this case, the obstacle
absorbs the correct amount of radiation expected for its size, but it
will not form a correct shadow behind it. Instead, the ``downstream
region'' behind the obstacle will get refilled with photons. As this
can happen only by photons transported within the same geometric cone,
a partial shadow is formed behind the object, with boundaries that are
in principle parallel to the cone boundaries.  In the middle panel, we
repeat the test with the same spatial mesh resolution, but we have
increased the number of cones to 32. In this way the angular size of
the obstacle as seen from the source becomes larger than the angular
resolution, allowing it to be resolved. As a result, a complete shadow
is being formed, but this shadow is in general a bit smaller than the
correct geometric shadow, with the difference being filled by a
partial shadow, created in the cones that are only partially obscured
by the obstacle.  Finally, in the right hand panel, we have repeated
the test on the left a second time, but now doubling the spatial
resolution to $100^2$ cells while the angular resolution was kept
unchanged. The primary difference this makes is that the borders of
the partial shadow that is formed are now more sharply defined
compared to the case with lower spatial resolution, as expected.

Finally, we examine how well our transport scheme can cope with
different mesh geometries, which naturally arise in simulations with
the {\small AREPO} code. In Figure~\ref{fig:mesh_shape} we show the
radiation fields developing around a point source embedded in
different mesh geometries: a Cartesian mesh, a hexagonal mesh (which
is akin to the mesh geometry developing in {\small AREPO} in regions
of constant resolution), and an azimuthal/unstructured mesh. For all
four cases, we compare the created radiation fields to the expected
profile in 2D, obtaining good agreement. This confirms the ability of
our approach to work well with the unstructured Voronoi meshes
produced by the {\small AREPO} code.

\subsection{Isothermal ionised sphere expansion} \label{sec:isphere}

We now turn to a test of our basic radiation advection scheme that
involves both sources and sinks.  To this end, we perform an ionised
sphere expansion test in three dimensions, which is arguably the most
fundamental and important test relevant for cosmic reionisation
codes. 

The expansion of an ionisation front in a static, homogeneous and
isothermal gas is the only problem in radiation hydrodynamics that has a
known analytical solution and is therefore indeed the most widely used
test for RT codes. We adopt a monochromatic source that 
steadily emits 
$\dot{N}_\gamma$ photons with energy $h\nu=13.6\, \rm eV$ per second
into an initially neutral medium with constant gas density $n_{\rm
  H}$. Then the Str\"omgren radius at which the ionised sphere around
the source  reaches its maximum radius is defined as \be r_{\rm S,0} =
\left(\frac{3\dot{N}_\gamma}{4\pi\alpha_{\rm B} n_{\rm
    H}^2}\right)^{1/3} ,
\label{rs1}
\ee where $\alpha_{\rm B}$ is the recombination coefficient.
If we approximate  the I-front is infinitely thin, i.e.~features
a discontinuity in the ionisation fraction, then the temporal expansion of the
Str\"omgren radius can be solved analytically in closed form,
with  the I-front radius
$r_{\rm I,0}$  given by \be r_{\rm I,0} = r_{\rm S,0}[1-\exp(-t/t_{\rm
    rec})]^{1/3},
\label{rI}
\ee
where\be
t_{\rm rec} = \frac{1}{n_{\rm H}\alpha_{\rm B}}
\ee
is the recombination time and $\alpha_{\rm B}$ is the recombination
coefficient.

More accurately, the neutral and ionised fraction 
as a function of radius of the Str\"omgren sphere can be calculated analytically
\citep[e.g.][]{OF2006} 
from the equation
\be
\frac{\tilde n_{\rm HI}(r)}{4\pi r^2}\int {\rm
  d}\nu\,\dot{N_\gamma}(\nu)\,
e^{-\tau_\nu(r)} \, \sigma_\nu\,=\,\tilde n_{\rm HII}^2(r) \, n_{\rm
  H}
\, \alpha_{\rm B},
\label{eqn:fraction}
\ee
where $\tilde n_{\rm HI}$ is the neutral fraction, $\tilde n_{\rm
HII}$ is the ionised fraction and
\be
\tau_\nu(r)\,=\, n_{\rm H} \, \sigma_\nu \, \int_0^r  {\rm d}r' \,
\tilde
n_{\rm HI}(r').\ee
Moreover, 
we can
 analytically solve for the radial profile of the photon density 
$n_\gamma(r)$,
yielding
\be
n_\gamma(r) = \frac{1}{c}\frac{\dot N_\gamma}{4\pi r^2}\,{\rm exp}
\left\{
-\int_0^r\kappa(r')\, {\rm d}r'  \right\}.
\label{exact}
\ee
From this we can also obtain the profile of the ionised fraction 
$\tilde n_{\rm HII}(r)$ as a function  of time.
We note that the Str\"omgren radius obtained by direct integration of
equation (\ref{eqn:fraction}) differs from the approximate
expression~(\ref{rs1}) because it does not approximate the ionised
region as a top-hat sphere with constant ionised fraction.  

For definiteness, we follow in our tests the expansion of the ionised
region around a source that emits $\dot N_\gamma = 5 \times 10^{48}
\,{\rm s}^{-1}$ photons. The surrounding hydrogen number density is set to
$n_{\rm H} = 10^{-3} \, \rm cm^{-3}$ at a temperature of $T = 10^4 \,
\rm K$. At this adopted temperature, the case B recombination coefficient is
$\alpha_{\rm B} = 2.59 \times 10^{-13} \, \rm cm^3 \, s^{-1}$. Given
these parameters, the recombination time is $t_{\rm rec} = 125.127 \,
\rm Myr$ and the expected Str\"omgren radius is $r_{\rm S,0} = 5.38 \, \rm
kpc$.

\bfig
\begin{center}
\includegraphics[width=0.45\textwidth]{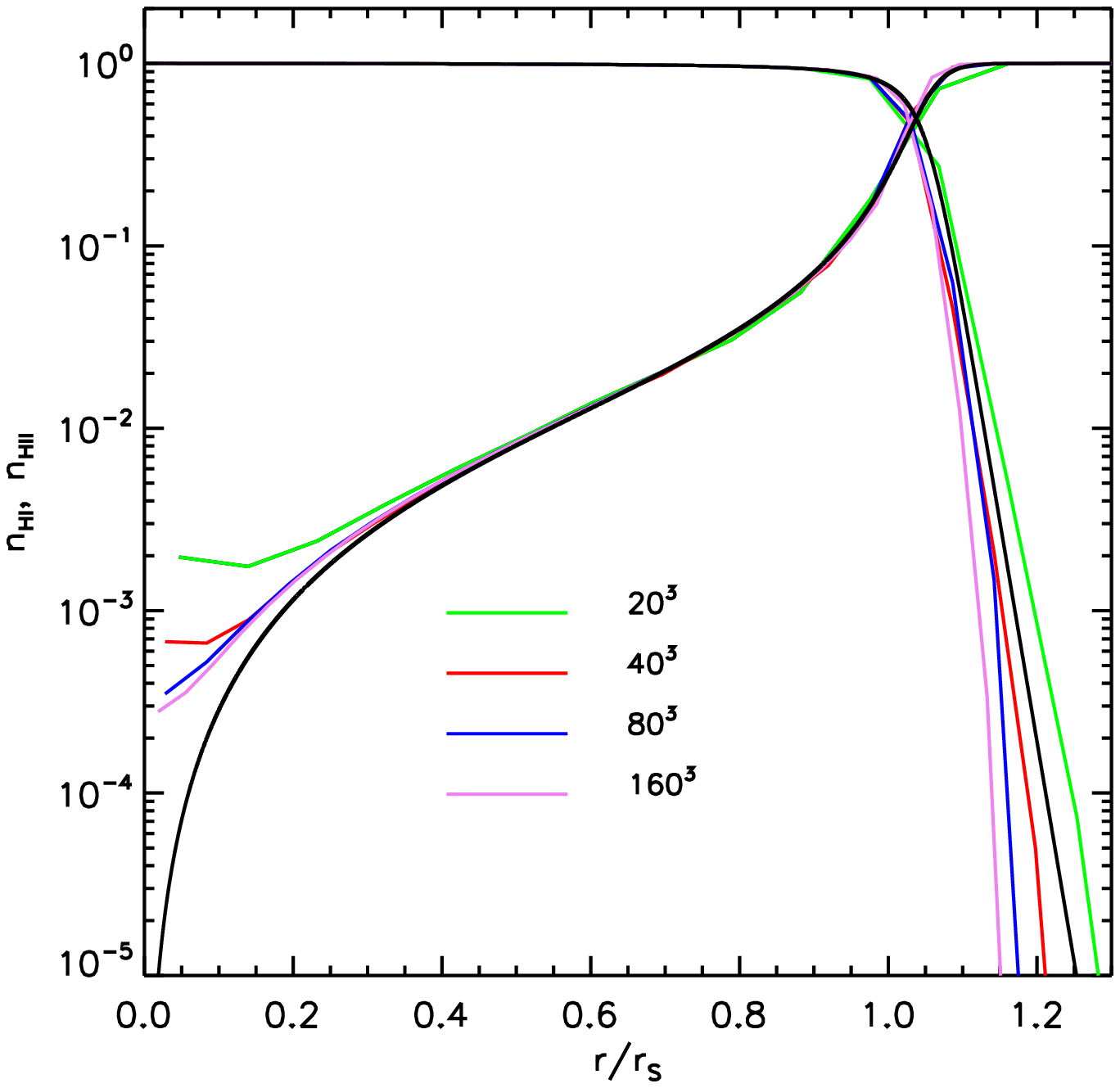}
\end{center}
\caption{Profiles of the neutral and ionised fraction at the end of the
  ionised bubble expansion test, when the Str\"omgren radius $r_{\rm S,0}$
  has been reached. The black line is the analytic solution based on
  equation~(\ref{eqn:fraction}), while the coloured lines are the
  numerical results for mesh resolutions of $20^3$, $40^3$, $80^3$, and
  $160^3$ cells, as labelled.\label{fig:multi}} \efig

\bfig
\begin{center}
\includegraphics[width=0.45\textwidth]{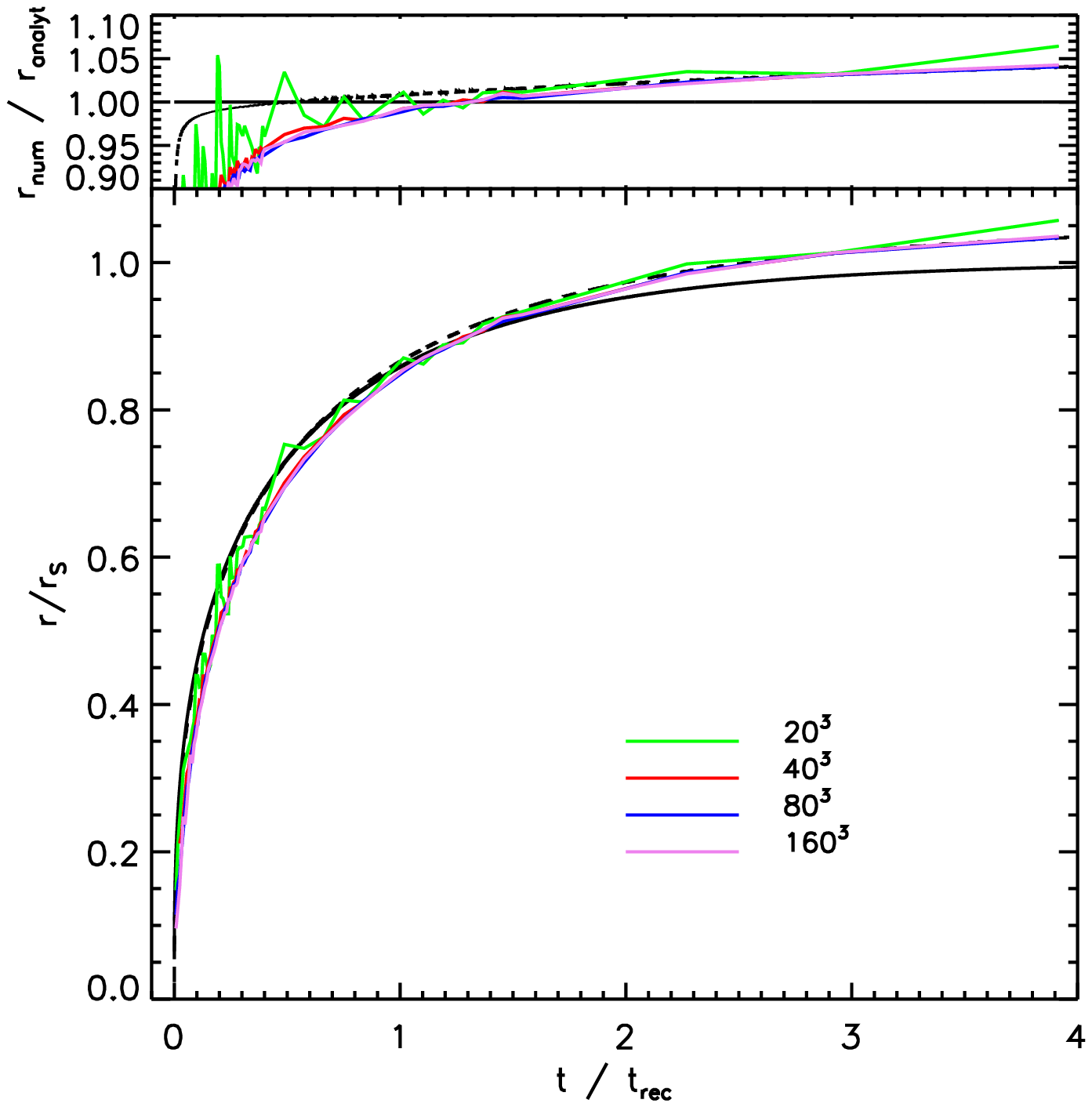}
\end{center}
\caption{Radius of the ionised region as a function of time, in units of
  the recombination time $t_{\rm rec}$. The solid black line is the
  analytic solution from equation~(\ref{eqn:fraction}), while the dashed
  black line gives the simple approximation of equation~(\ref{rI}).  The
  coloured lines give our numerical results for different mesh
  resolutions equal to $20^3$, $40^3$, $80^3$, and $160^3$ cells. We see
  that the ionising front is slower than the theoretical prediction in
  the beginning of the expansion, as a result of the artificially
  reduced speed of light adopted here. At late times, the numerical
  result agrees however very well with the analytical
  solution.\label{fig:Ifront}} \efig

\bfig
\begin{center}
\includegraphics[width=0.3\textwidth]{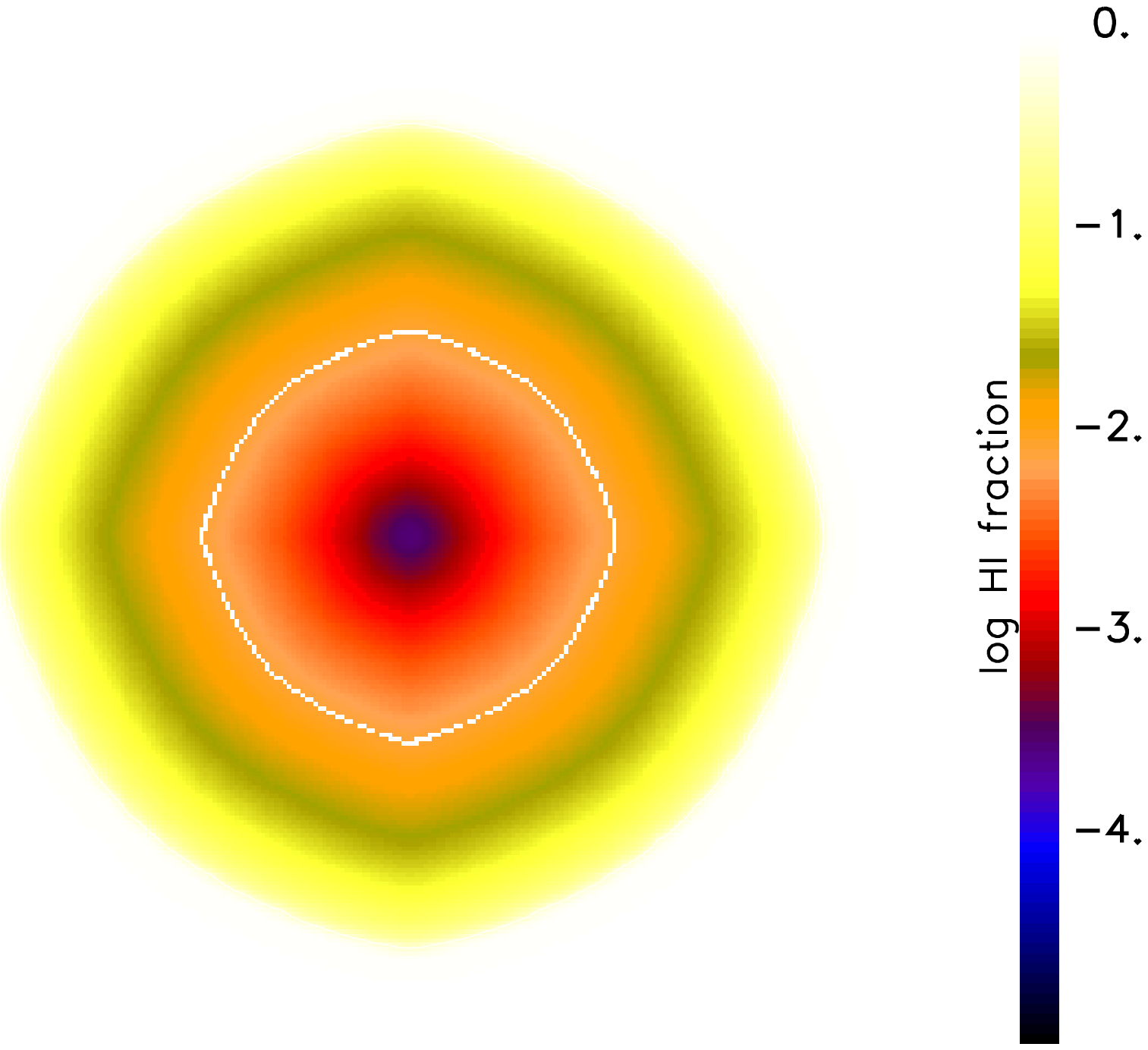}
\end{center}
\caption{Map of the neutral fraction in a slice through the center of
  our Str\"omgren sphere test (based on our point-source advection
  scheme), for our highest resolution simulation with $160^3$ mesh
  cells. The white line shows the contour at a neutral fraction of
  0.5. \label{fig:ion_slice_SST}} \efig

\bfig
\begin{center}
\includegraphics[height=0.4\textwidth]{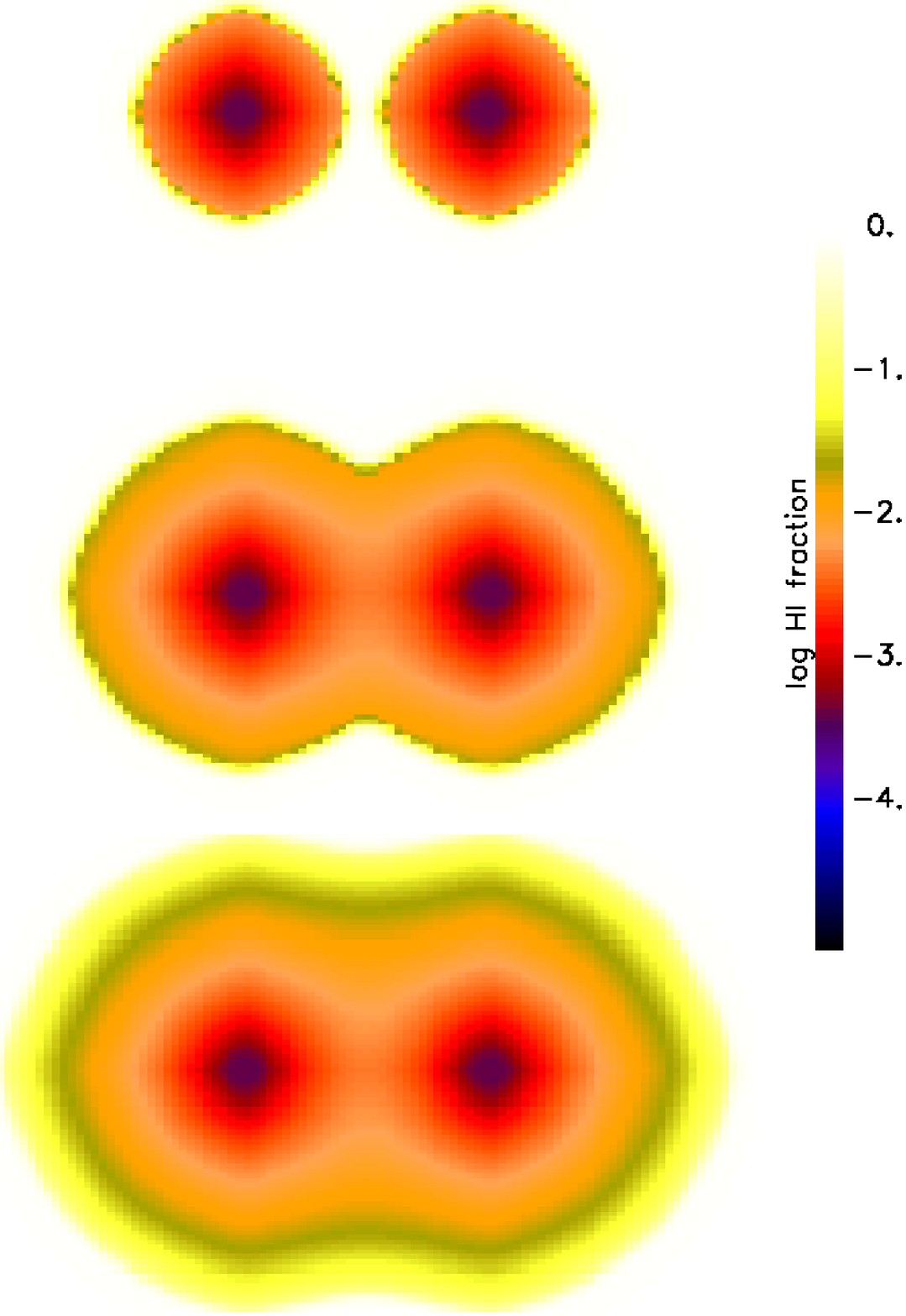}
\end{center}
\caption{Neutral fraction in a slice through the center of two nearby
  sources of equal luminosity that ionise neutral gas.  The three panels
  from top to bottom show different evolutionary stages. The top panel
  shows a stage before the ionised spheres overlap ($t=25\, \rm
  Myr$). Here they have exactly the same shape and do not influence each
  other yet. In the middle panel, the two have begun to overlap
  ($t=100\, \rm Myr$), while in the bottom panel the final state is
  shown, where the ionised region becomes time invariant ($t=500 \, \rm
  Myr$). \label{fig:ion_slice_2}} \efig

\bfig
\begin{center}
\includegraphics[width=0.45\textwidth]{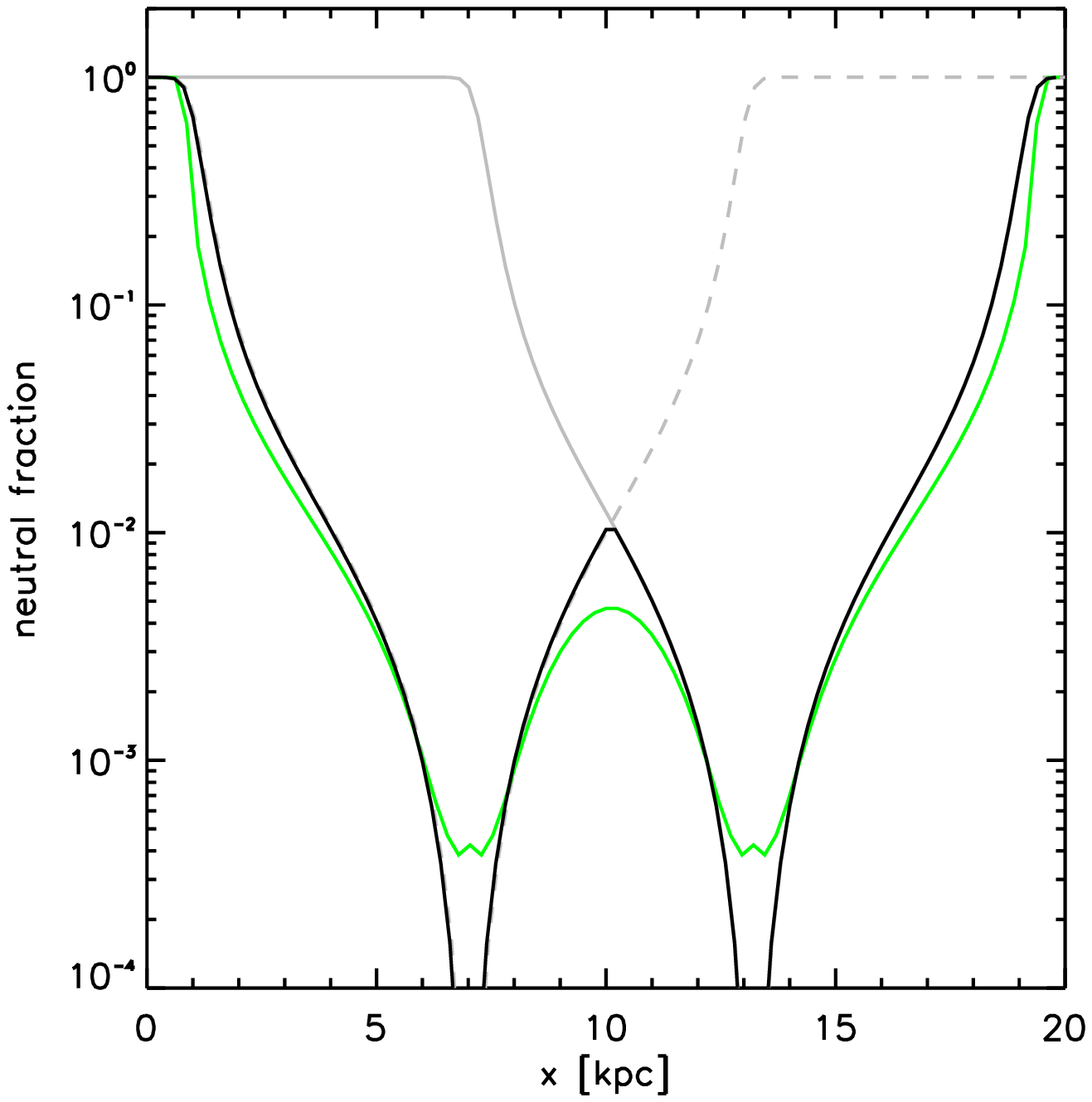}
\end{center}
\caption{Neutral fraction along a line passing through the centres of
  two nearby sources that ionise the background gas. The green line
  shows the numerical result, whereas the black lines are a simple
  composite model for the expected structure of the solution based on
  superposing the analytic solution for each of the sources (gray
  dashed line for the left source and gray solid line for the right
  source). This superposition of the individual sources describes the
  numerical solution reasonably accurately, but we note that it is not
  the correct solution; the latter can only be obtained numerically
  for this problem. \label{fig:profile_2}} \efig

\bfig
\begin{center}
\includegraphics[width=0.3\textwidth]{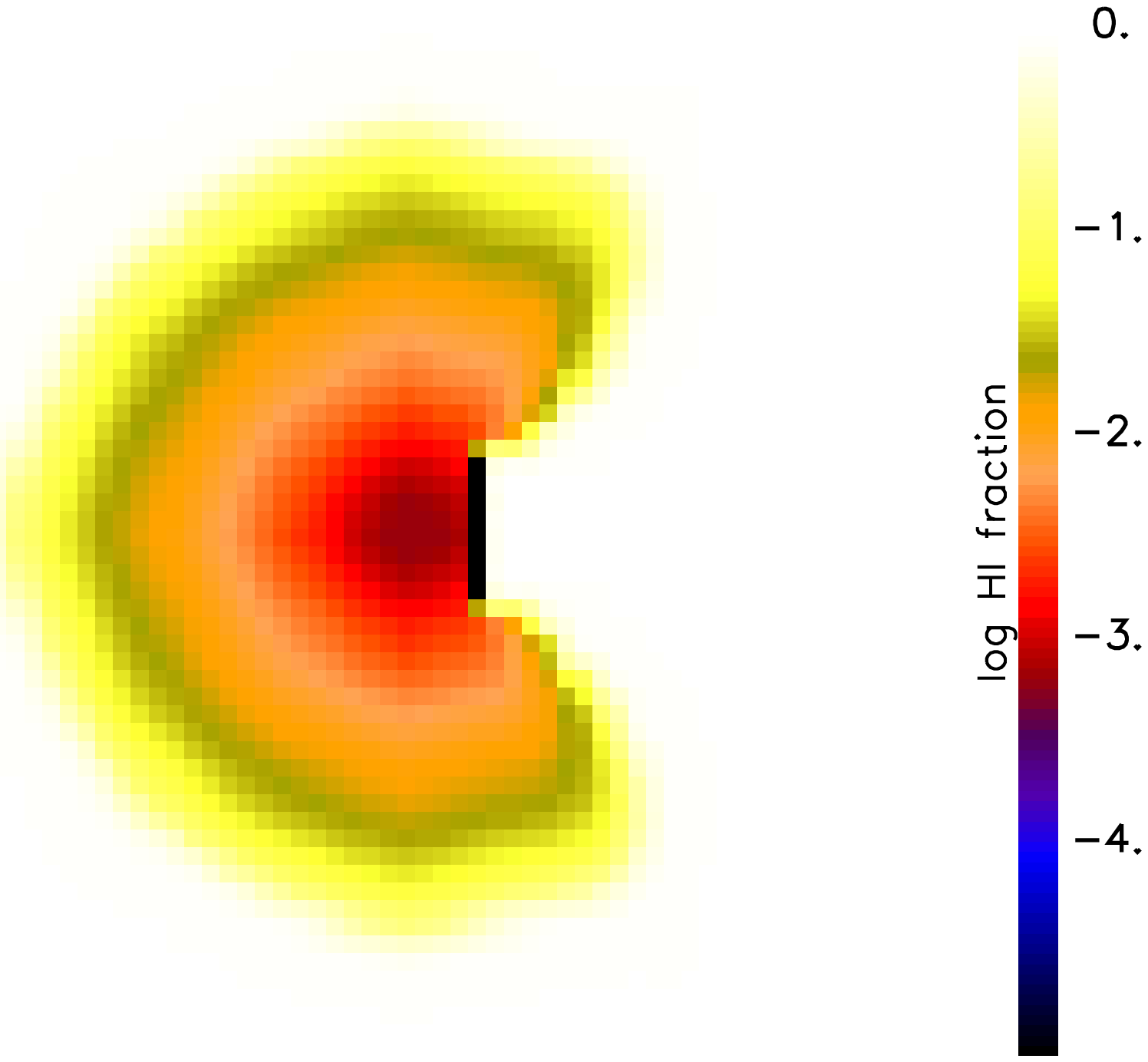}
\end{center}
\caption{Map of the neutral fraction in a slice through the center of a
  Str\"omgren sphere. Unlike in our previous tests, an absorbing obstacle
  in the form of an optically thick disk was included as well (shown as
  a black line). We find that a nice shadow is produced behind the
  disk, with the inobscured directions developing as in the Str\"omgren sphere
  without an obstacle. \label{fig:ion_slice_shadow}} \efig

In Figure~\ref{fig:multi}, we show the profiles of ionised and neutral
fraction at the end of the ionised sphere expansion, when the
Str\"omgren radius has been reached. We present results for simulations
with four different spatial resolutions, using grids with $20^3$,
$40^3$, $80^3$ and $160^3$ cells, respectively, using our point-source
advection scheme. The results for all resolutions agree well with the
analytical solution. The largest errors occur close to the central point
source, but with better spatial resolution they become progressively
smaller. In Figure~\ref{fig:Ifront}, we show the time evolution of the
ionising front, for the same simulations. The position of the front is
determined as the distance from the source at which the ionised
fraction equals $0.5$. The agreement with the
analytical solution is generally good and improves with better
resolution. However, in the beginning, the ionisation front moves
noticeably slower than expected, which is due to our use of the reduced
speed of light approximation with $c' = c/1000$. At later times, this
initial error becomes unimportant, however, and the numerical solution
matches the analytic expectation well. Making the start-up error
vanishingly small would be possible, if desired, but requires using $c' =
c$.

In Figure~\ref{fig:ion_slice_SST}, we show a map of the neutral
fraction in a slice through the source plane for the resolution
$160^3$. We notice that the isophotal shapes exhibit small departures
from a perfectly spherical shape, which originate in spatial
discretisation errors close to the source. In fact, these deviations
depend on the geometry of the source cell itself. For a hexagonal mesh
structure as it occurs for a regularised Voronoi mesh in 2D
dimensions, the errors are noticeably smaller than for the Cartesian
mesh employed here. Higher spatial resolution alone will normally not
be able to decrease the deviations to arbitrarily small levels, but
spreading the point source over multiple cells (effectively resolving
the source geometry) can make the isophots perfectly round if desired.
We note that our cone transport scheme also does a good job in
producing round isophots, even when a single cell is used as source.

As a simple variant of the isolated source case, we have also
considered the evolution of the ionised regions around two sources
that are $4\,\rm kpc$ apart, using our multiple point-source scheme.
The density of the gas and the luminosity of each source are the same
as in the previous test. In Figure~\ref{fig:ion_slice_2}, we show maps
of the neutral fraction in a slice through the source at three
different times: $t=25\,\rm Myr$ (top), $t=100\,\rm Myr$ (middle), and
$t=500\,\rm Myr$ (bottom). An important point of this test is that the
proximity of the sources does not affect the shape of the ionised
regions at all until they begin to overlap.  This is very different in
the OTVET scheme, for example, where the early expansion is distorted
because the Eddington tensor estimates already ``feel'' nearby sources
even though they may still be completely hidden in their own
ionisation bubbles.  In Figure~\ref{fig:profile_2}, we show the
neutral fraction along a line passing through both sources at the
final time. A simple model for the expected neutral fraction based on
the superposition of the analytic single source solution is shown in
black, while the numerical solution is shown in green. While the
superposition model does a reasonably good job in describing the
numerical solution, we note that the latter is showing important
differences, for example for the radiation intensity between the
sources. Our method allows an accurate calculation of this quantity,
and similarly for more complicated setups.

In Figure~\ref{fig:ion_slice_shadow} we show a further map of the
neutral fraction in a slice through the source plane in a simple
single-source Str\"omgren test. However, in this test we included an
obstacle in the form of an optically-thick three dimensional plate,
located $2\,\rm kpc$ from the source (shown in black in the
figure). The setup is meant to test shadowing in 3D for a problem with
non-trivial source function, and is designed to match the parameters
of an equivalent test in \citet{Pawlik2008}. We can see that our
obstacle produces a clear shadow that remains fully neutral, as
expected. Comparing our result to those of \citet[][see their
  Fig.~10]{Pawlik2008}, we find good qualitative agreement but much
reduced numerical noise.

Finally, we check whether using the cone transport scheme described in
Section~\ref{sec:cones} is equally well capable of accurately solving
the Str\"omgren sphere problem. To this end we have repeated our
standard setup for the ionised sphere expansion of a single source,
but this time employing direct discretisation of angular space using
12 cones for the full $4\pi$ solid angle, and a spatial mesh
resolution of $40^3$. In the top and bottom panels of
Figure~\ref{fig:ifrontall_cone} we show the profiles of ionised and
neutral fraction at the end of the ionised sphere expansion, and the
temporal evolution of the ionising front, respectively. The numerical
results agree well with the analytical solutions, with an overall
accuracy that is comparable to that of our point source treatment.

\bfig
\begin{center}
\includegraphics[width=0.45\textwidth]{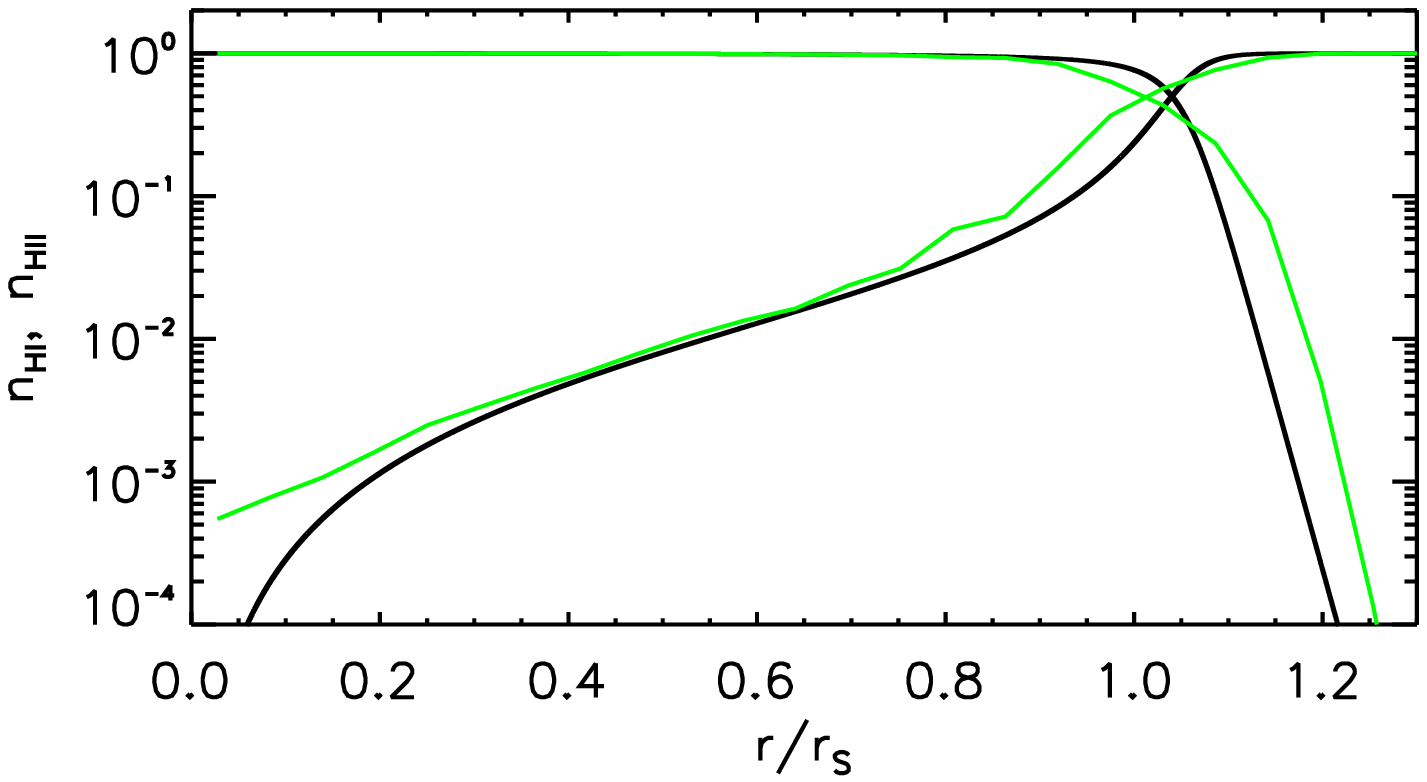}
\includegraphics[width=0.45\textwidth]{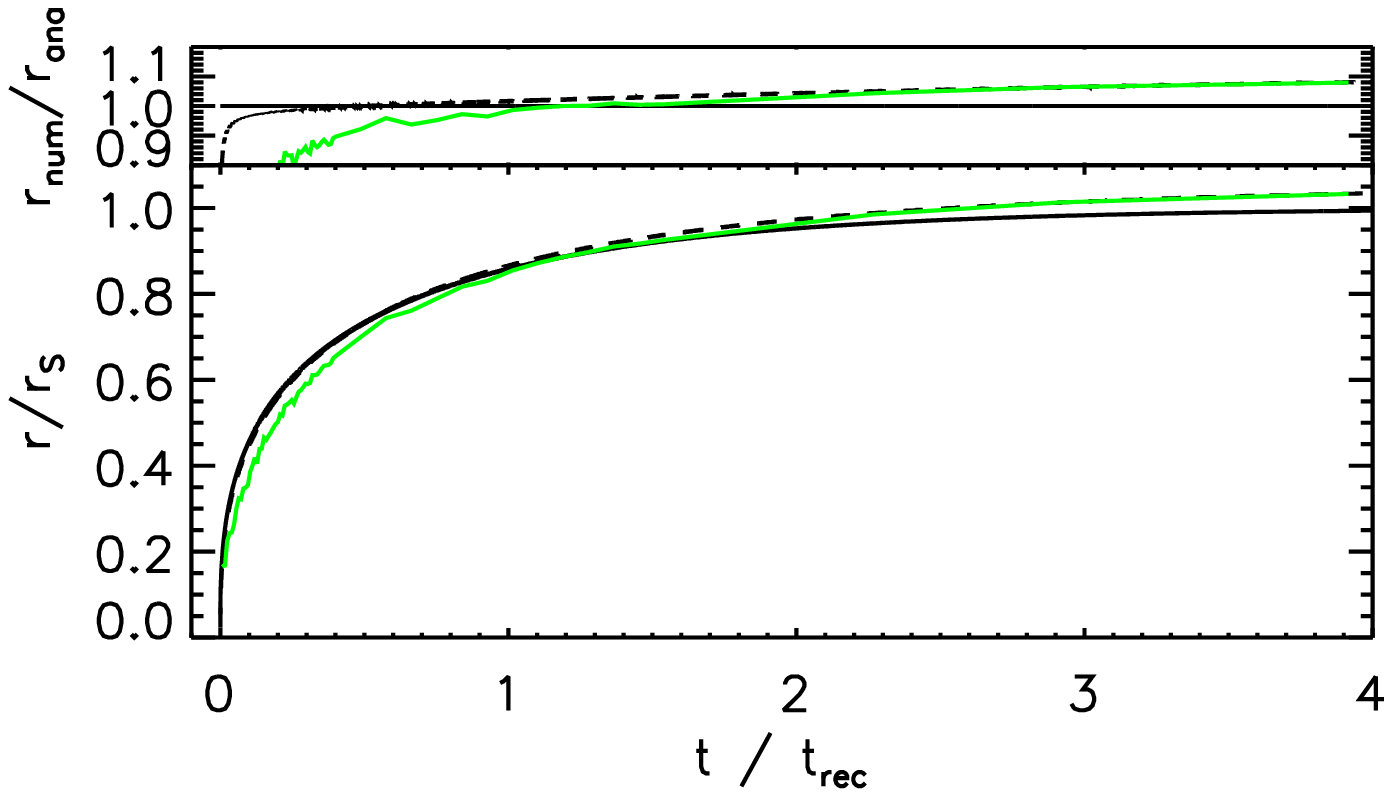}
\end{center}
\caption{Str\"omgren sphere test for our `cone transport' scheme where
  the angular space is decomposed into cones of equal solid angle.  The
  panel on top shows the profiles of ionised and neutral fractions
  (green lines) versus distance from the source, in units of the
  Str\"omgren radius $r_{\rm S,0}$ at the end of the expansion at $t=500\,
  \rm Myr$. The analytical solution is given by the solid black line.
  The bottom panel shows the ionisation front radius as a function of
  time, relative to the Str\"omgren radius $r_{\rm S,0}$ as a function of
  time in units of the recombination time $t_{\rm rec}$. The numerical
  solution is shown in green and the analytical one in black.
\label{fig:ifrontall_cone}}
\efig

\subsection{Ionising front trapping in a dense clump}  \label{sec:trap}

In our next test, we study the behavior of the code in a more challenging
setting taken from the RT code comparison study of
\citet{Iliev2006comparison}. A plane-parallel front of ionising
radiation  is incident on a dense, cold clump. The I-front
penetrates the clump, ionises it and heats it up. Eventually, the
I-front gets trapped half-way through the clump, and as the it is
stopped inside the obstacle, a shadow is produced behind the clump.

\bfig
\begin{center}
\includegraphics[width=0.2\textwidth]{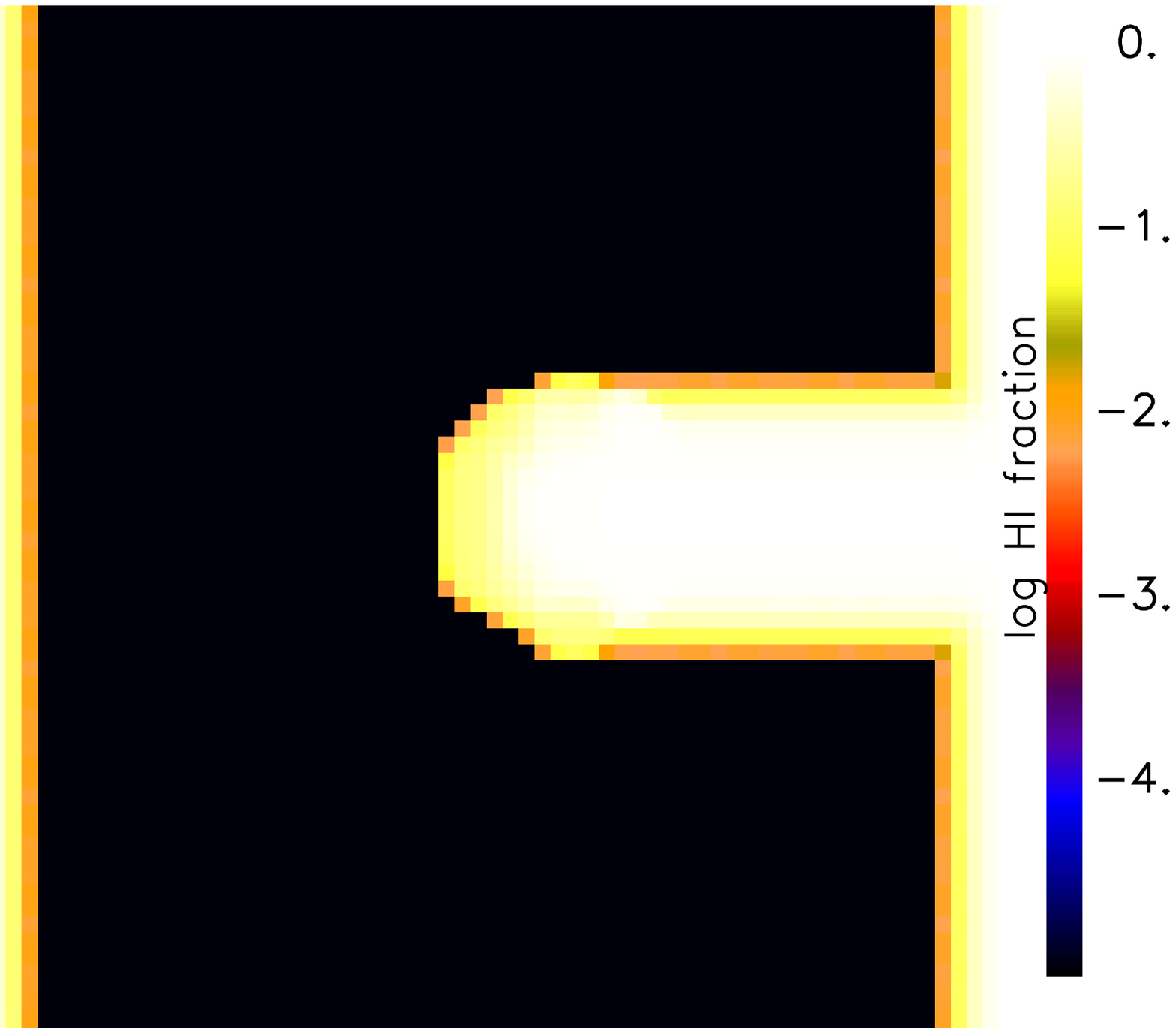}
\includegraphics[width=0.2\textwidth]{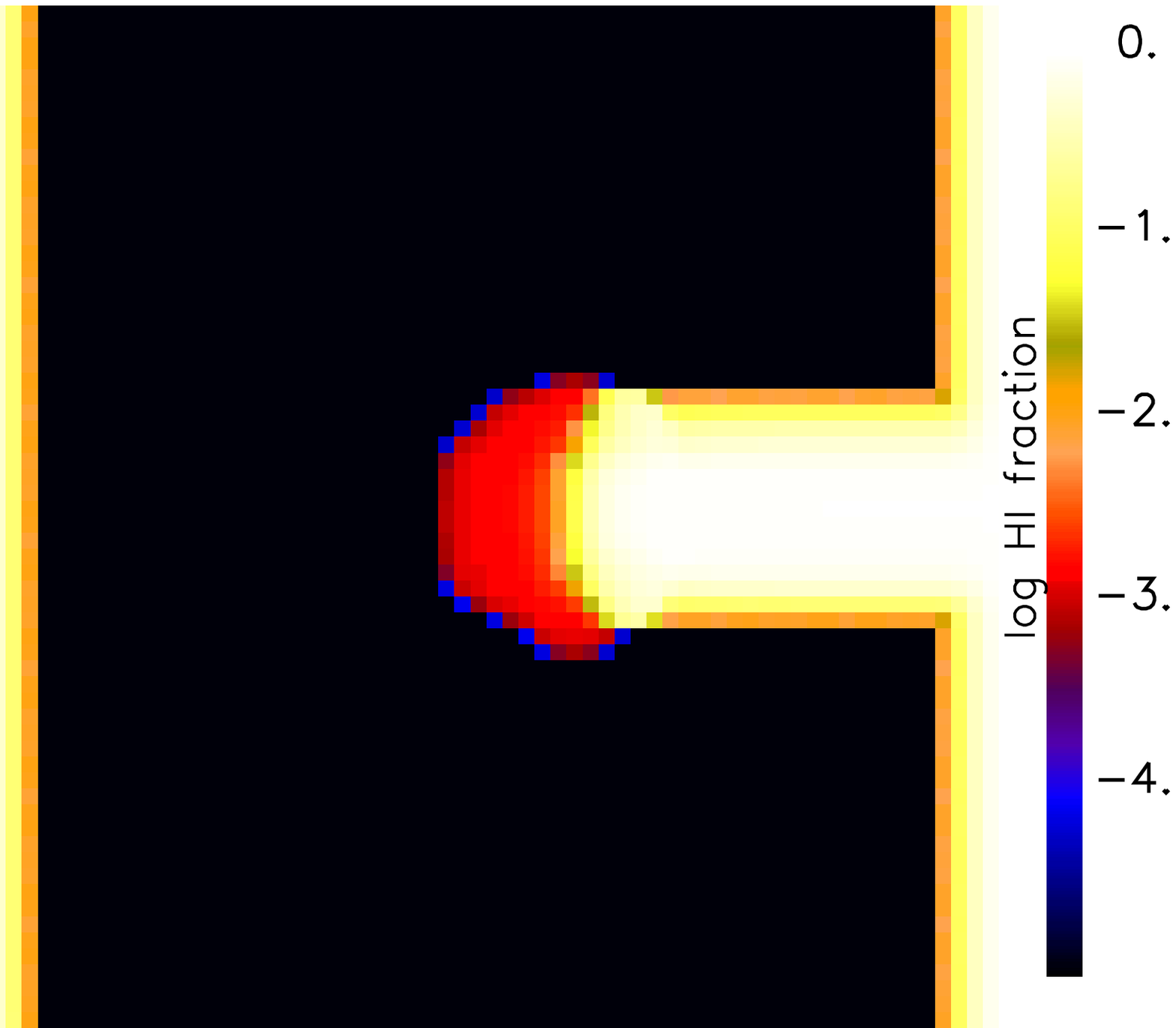}\\
\includegraphics[width=0.2\textwidth]{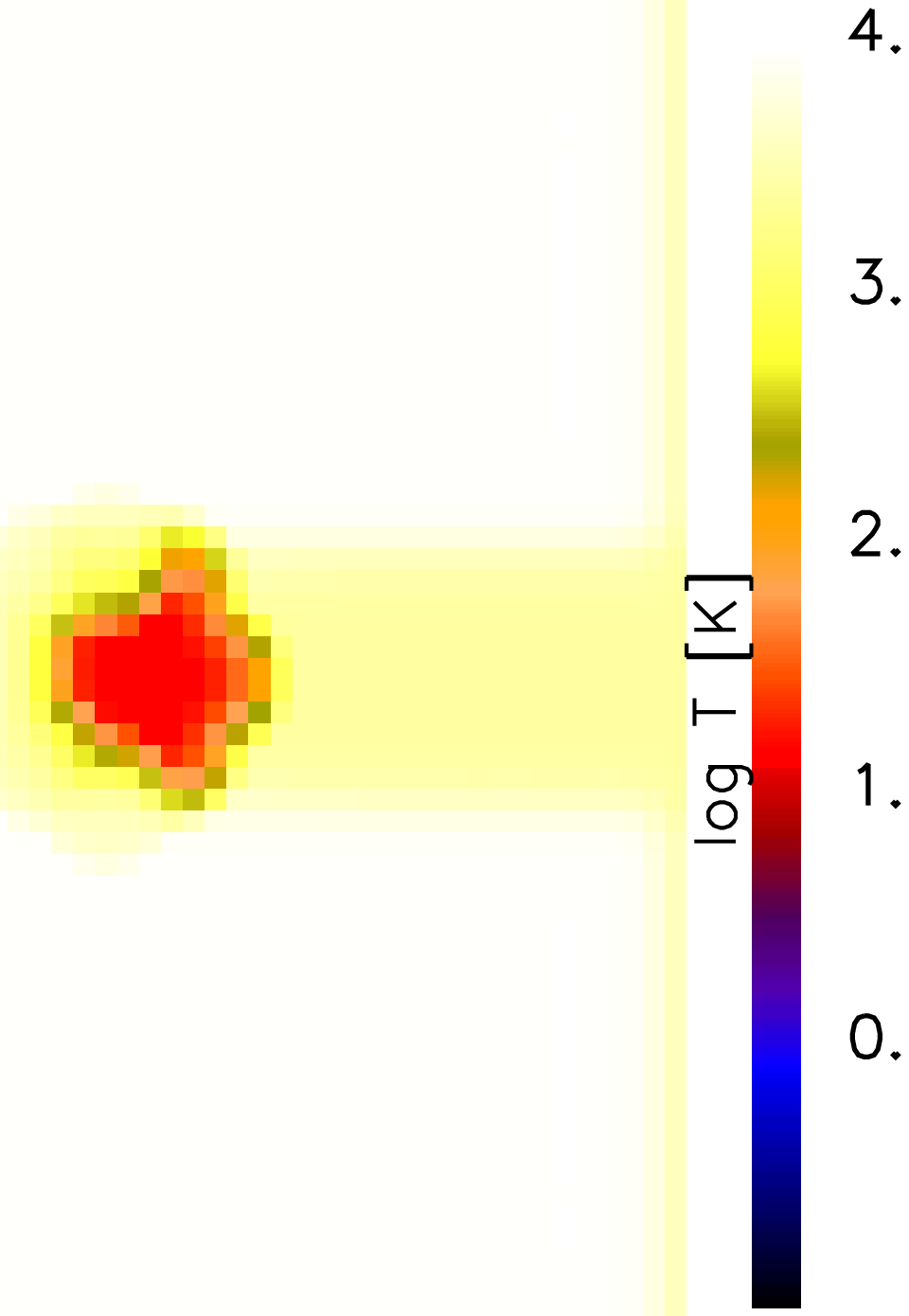}
\includegraphics[width=0.2\textwidth]{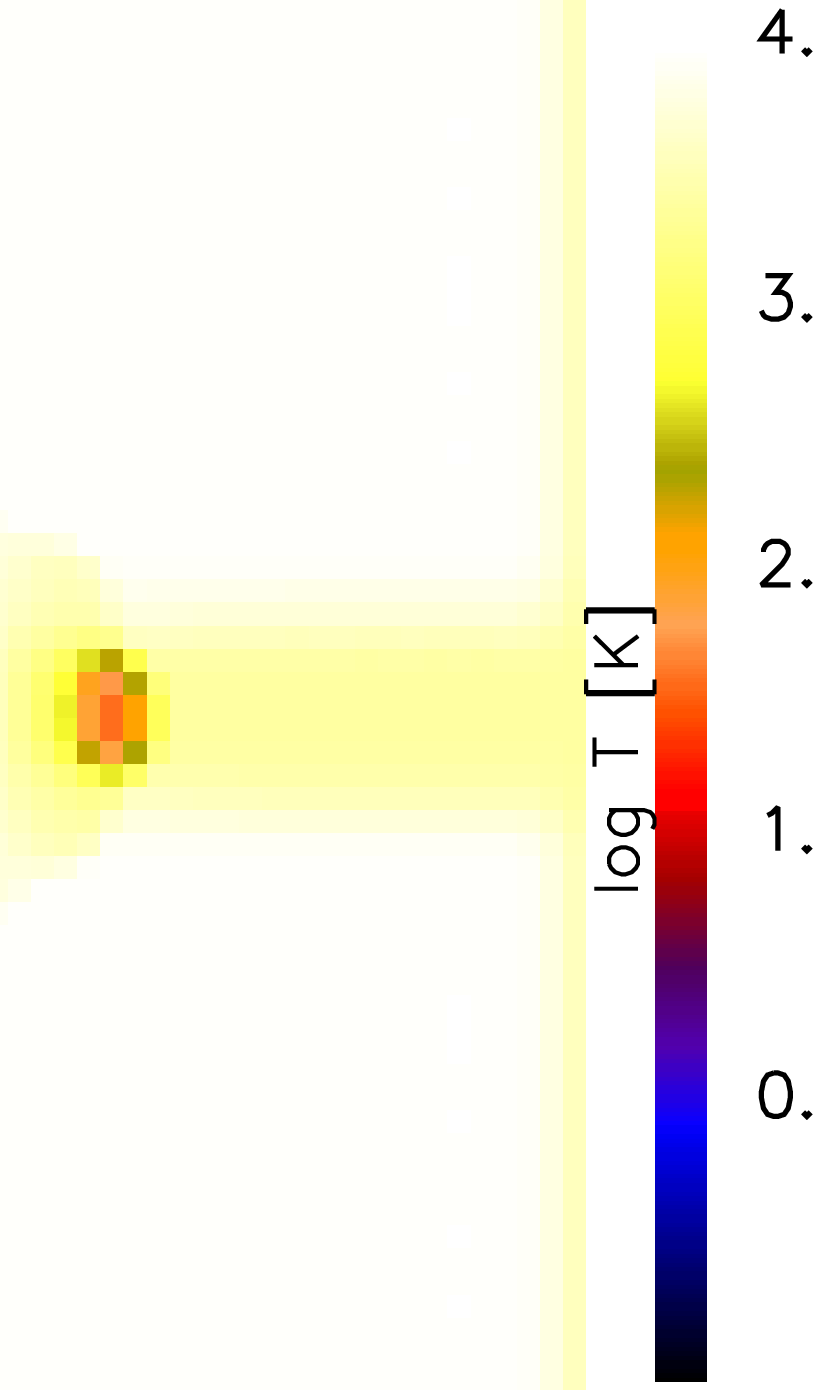}
\end{center}
\caption{Maps of neutral fraction (top row) and temperature (bottom
  row), in a simulation of the interaction of a plane-parallel
  ionisation front with a dense clump. The two columns show our
  simulation results at two different times, $t = 1\, \rm Myr$ (left)
  and $t = 15\,{\rm Myr}$ (right). We note that already at the earlier
  time the background gas has been fully ionised. The I-front gets
  however stuck in the clump, producing a shadow behind
  it. \label{fig:trap}} \efig

\bfig
\begin{center}
\includegraphics[width=0.45\textwidth]{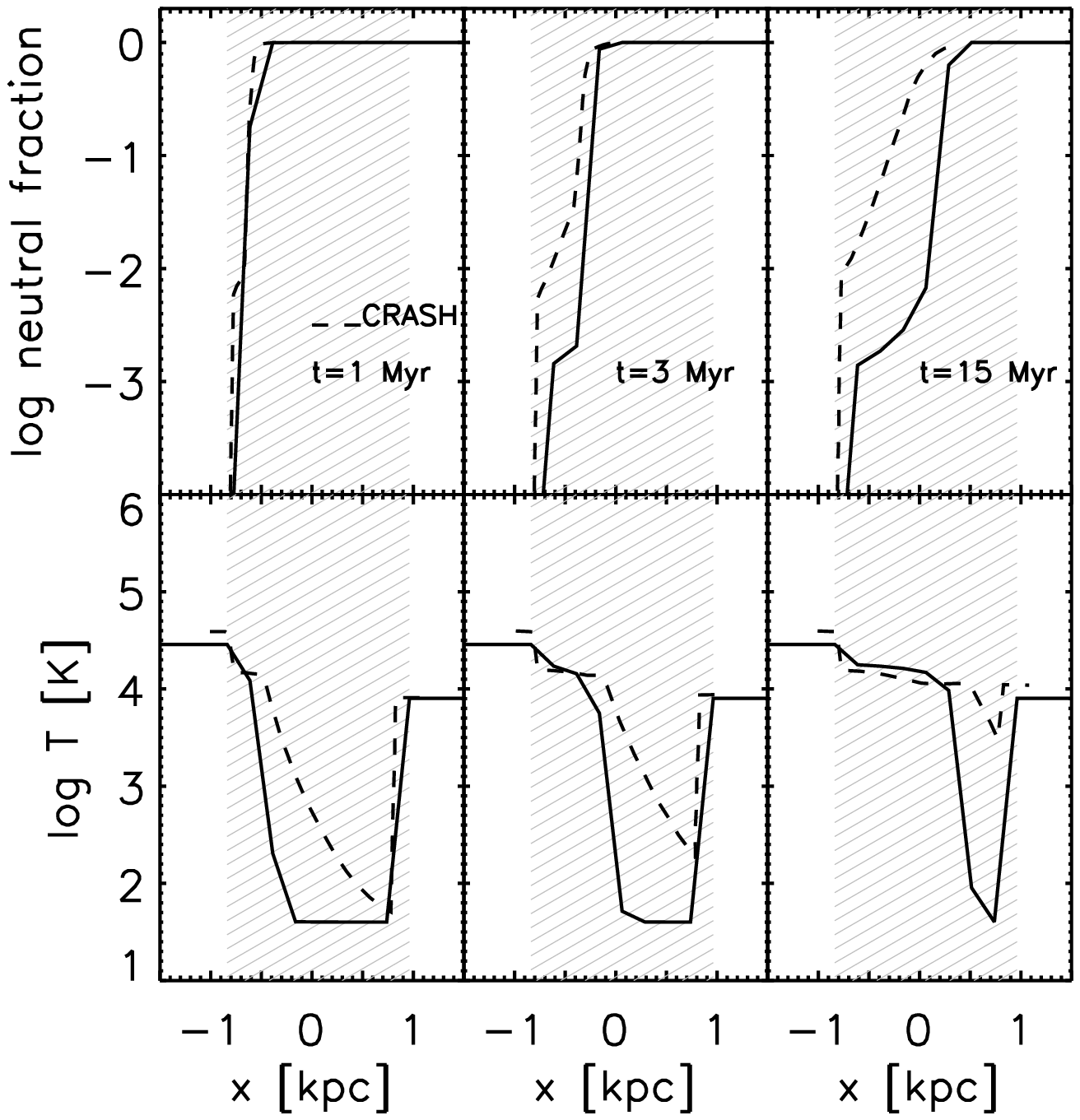}
\end{center}
\caption{Neutral fraction and temperature as a function of distance from
  the center of the dense clump, at three different times: $t = 1\, \rm
  Myr$ (left column), $3\,{\rm Myr}$ (middle) and $15\, {\rm Myr}$
  (right). The shaded area shows the geometric extension of the
  clump. Results obtained with the code {\small CRASH} in the RT code
  comparison project are also included and shown as dashed
  lines. \label{fig:clump}} \efig

\bfig
\begin{center}
\includegraphics[width=0.45\textwidth]{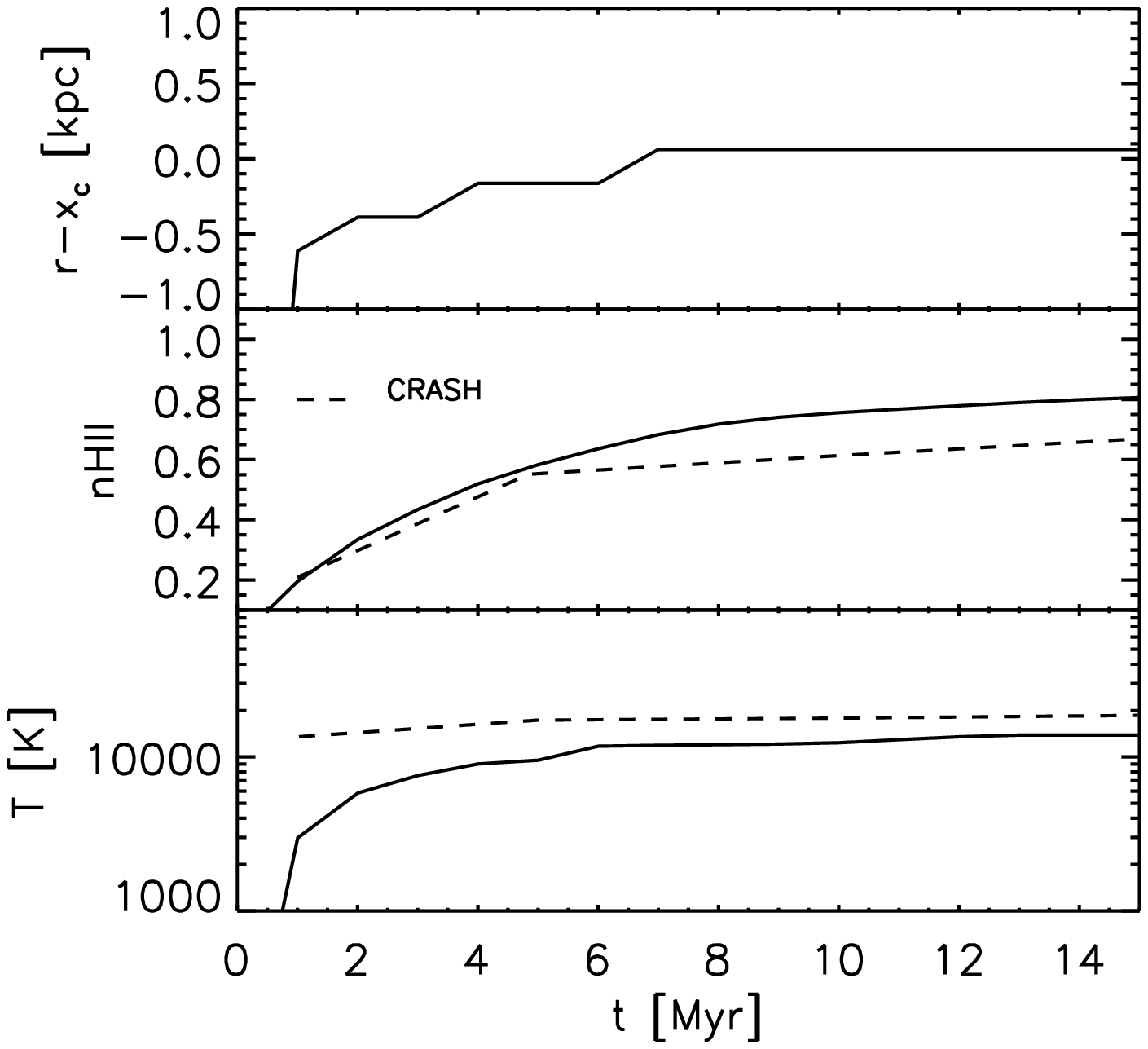}
\end{center}
\caption{Time evolution of the temperature, neutral fraction, and ionised
  front (solid lines) in the dense clump that is ionised by an impinging
  I-front. We also include results obtained with the code {\small CRASH}
  in the RT code comparison project, which are shown as dashed lines for
  comparison.\label{fig:clump_time}} \efig

Our set-up of this test problem is as follows. We simulate a
plane-parallel I-front with flux $F_\gamma = 10^6 \, \rm photons
\, s^{-1} \, cm^{-2}$ that is incident on a dense clump, located $5\,\rm
kpc$ away from the edge of the simulation domain. The ambient
background gas has density $n_{\rm H} = 2 \times 10^{-4} \, \rm cm^{-3}$
and temperature $T=8000\, {\rm K}$.  The radius of the clump is $r_{\rm
  clump} = 0.8 \, \rm kpc$, with a density of $n^{\rm clump}_{\rm H} =
200\, n_{\rm H}$, and a temperature $T_{\rm clump} = 40\,{\rm K}$.  We
note that in this test, following \citet{Iliev2006comparison}, the gas
is not allowed to move due to pressure or gravitational forces, hence
only radiative transfer is tested.  The system is evolved for a period of
$15 \, \rm Myr$ with a resolution of $40^3$ cells.
In Figure~\ref{fig:trap}, we show the neutral fraction and the temperature
of the system in slices through the centre of the clump at times $t =
1\, {\rm Myr}$, and $15\, {\rm Myr}$. The I-front approaches from
the left, moving to the right. At time $t=1\, \rm Myr$, already the
whole box has been swept up by the ionising photons, with the clump
producing a clear shadow in the downstream direction on the right hand
side of the clump. As time advances further, the clump becomes more
ionised and continues to heat up, but the shadow is preserved
throughout the simulated time span without being filled in by
diffusion.

\bfigs
\begin{center}
\includegraphics[width=0.33\textwidth]{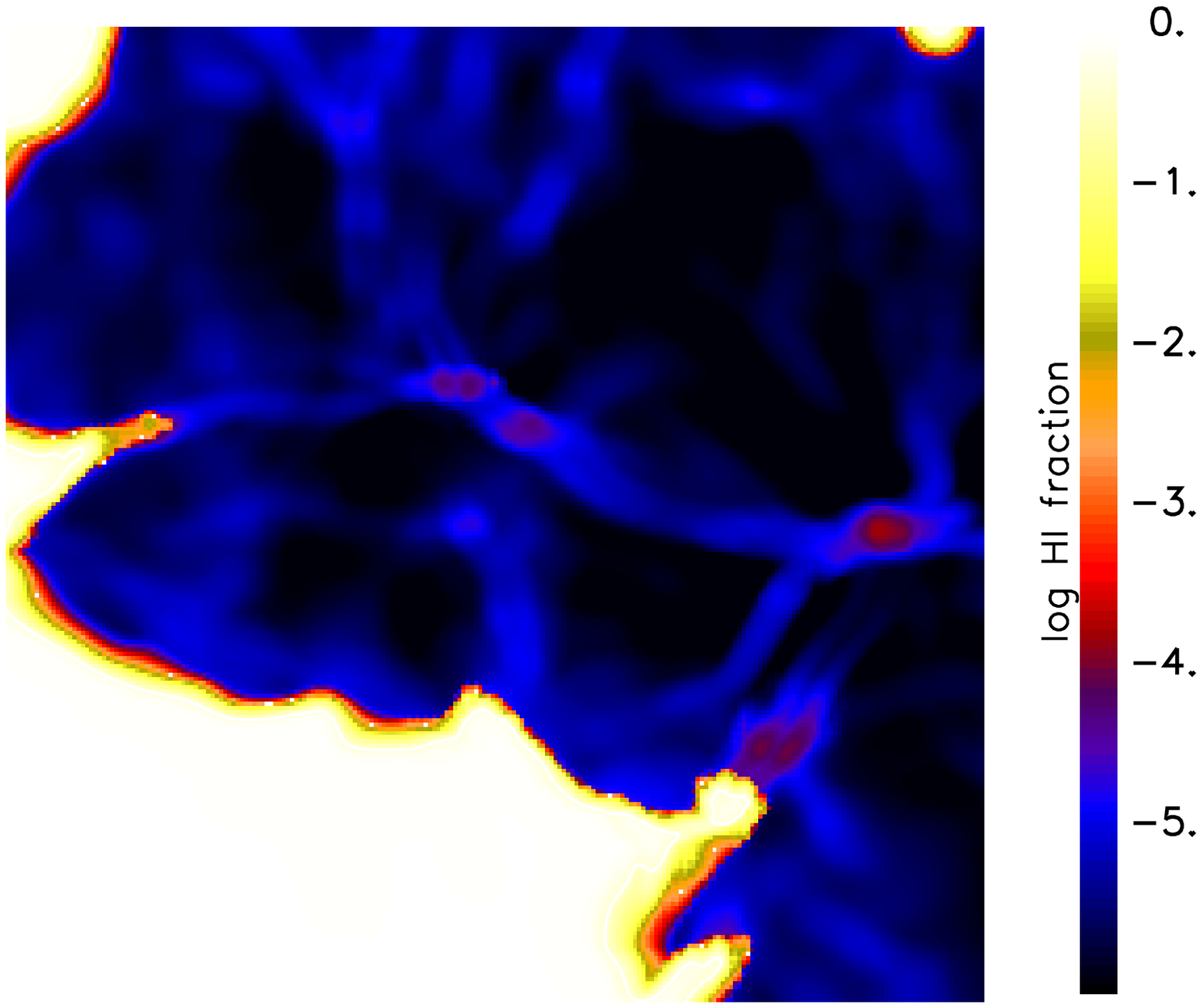}
\includegraphics[width=0.33\textwidth]{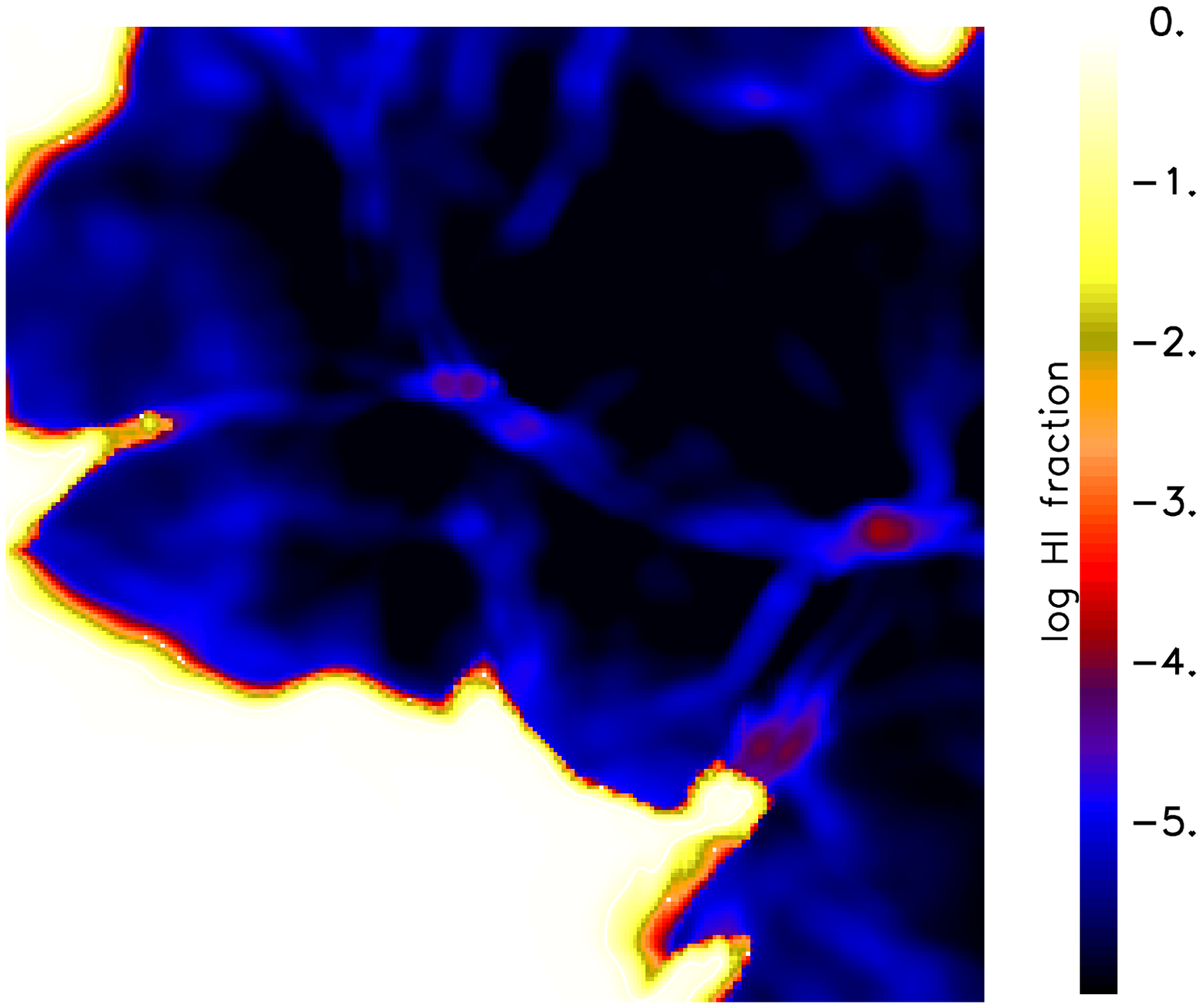}
\includegraphics[width=0.33\textwidth]{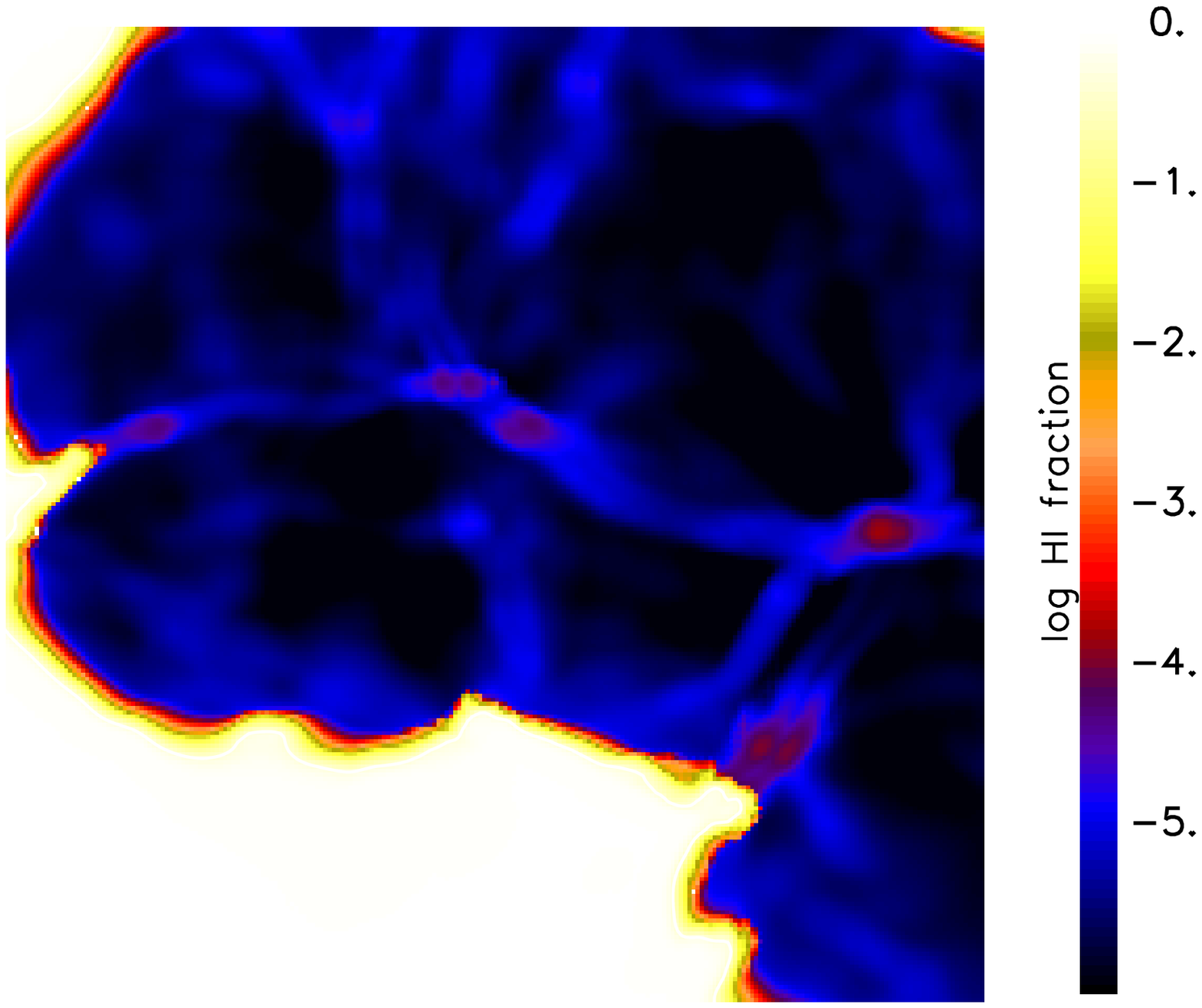}\\
\includegraphics[width=0.33\textwidth]{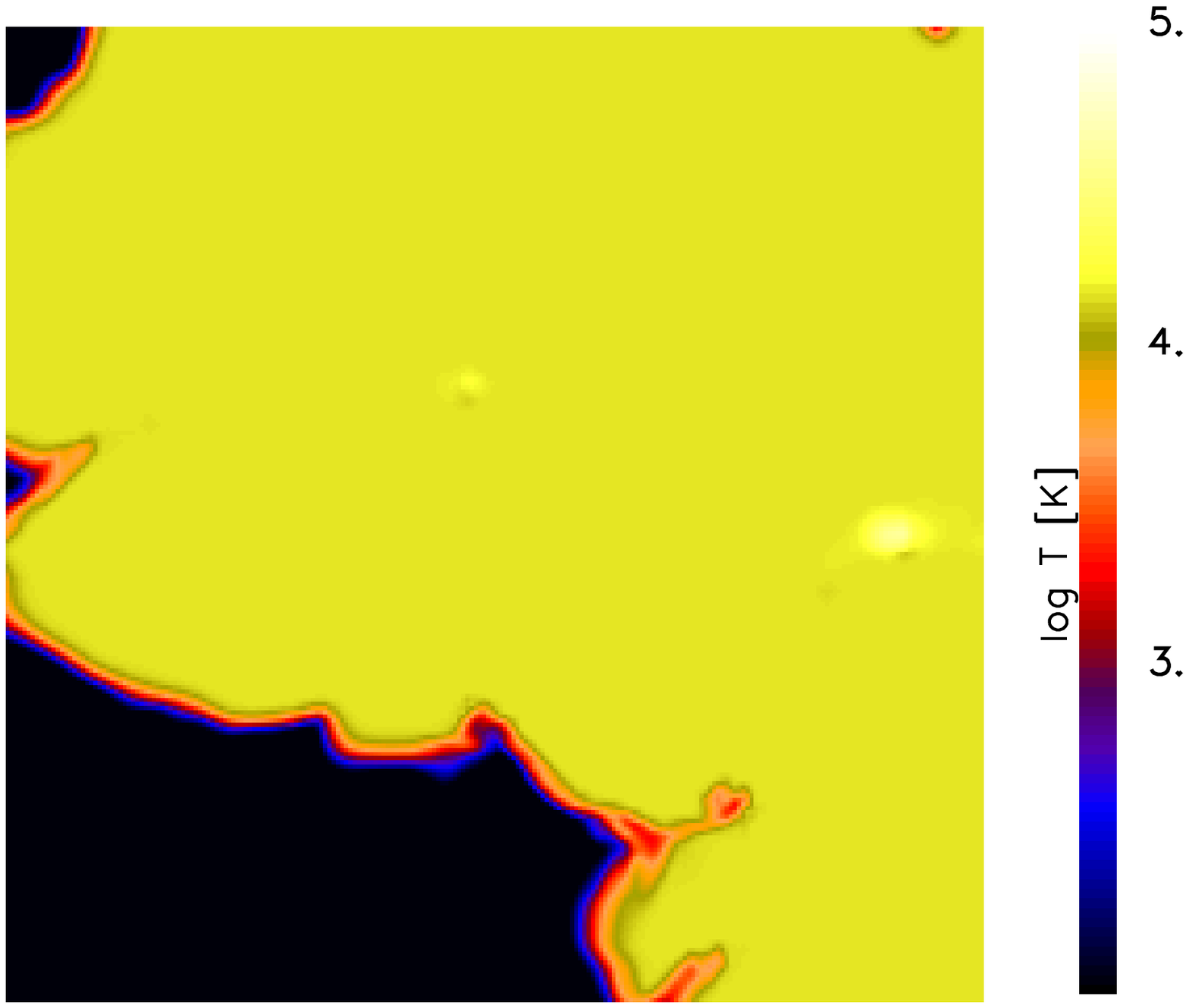}
\includegraphics[width=0.33\textwidth]{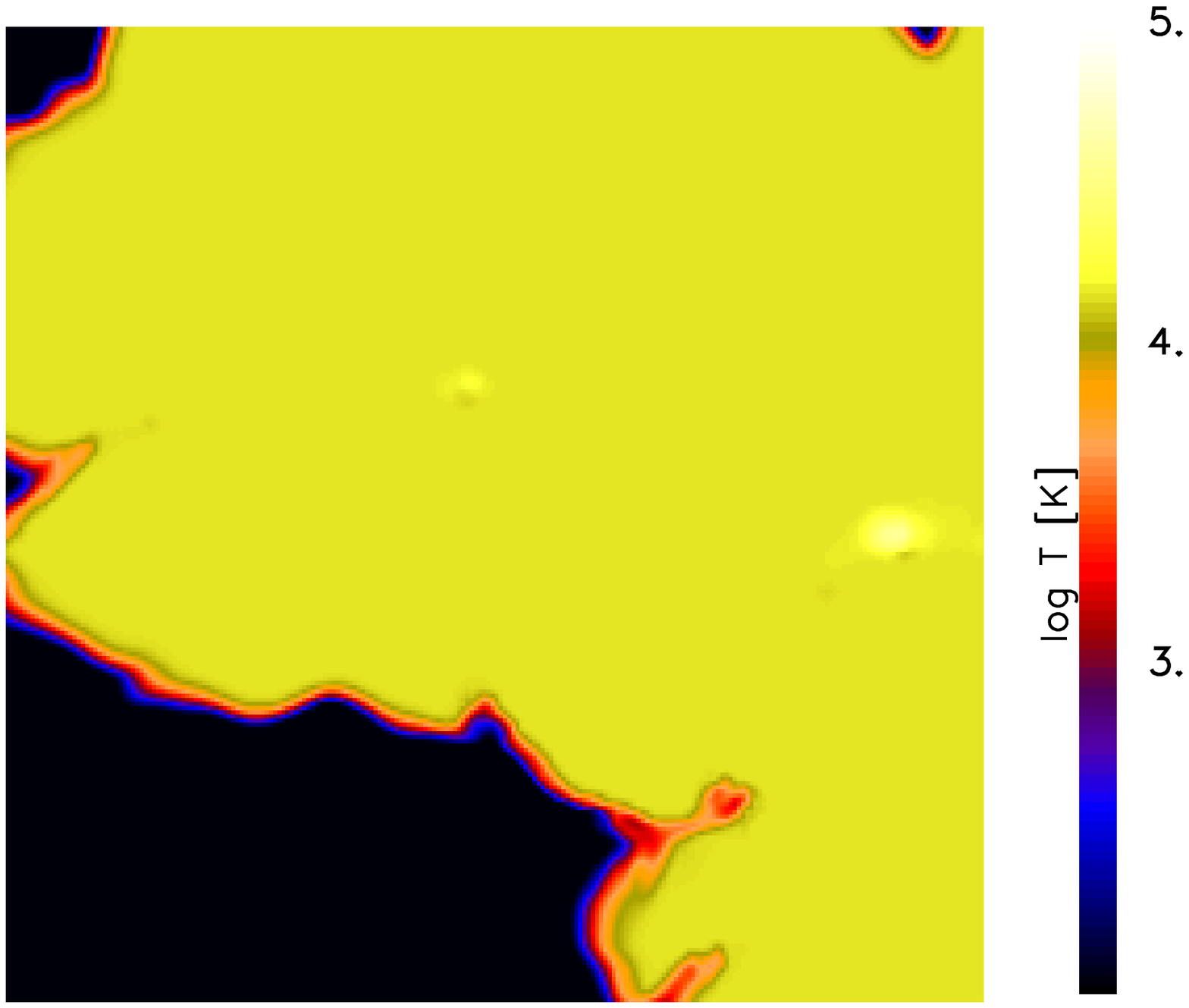}
\includegraphics[width=0.33\textwidth]{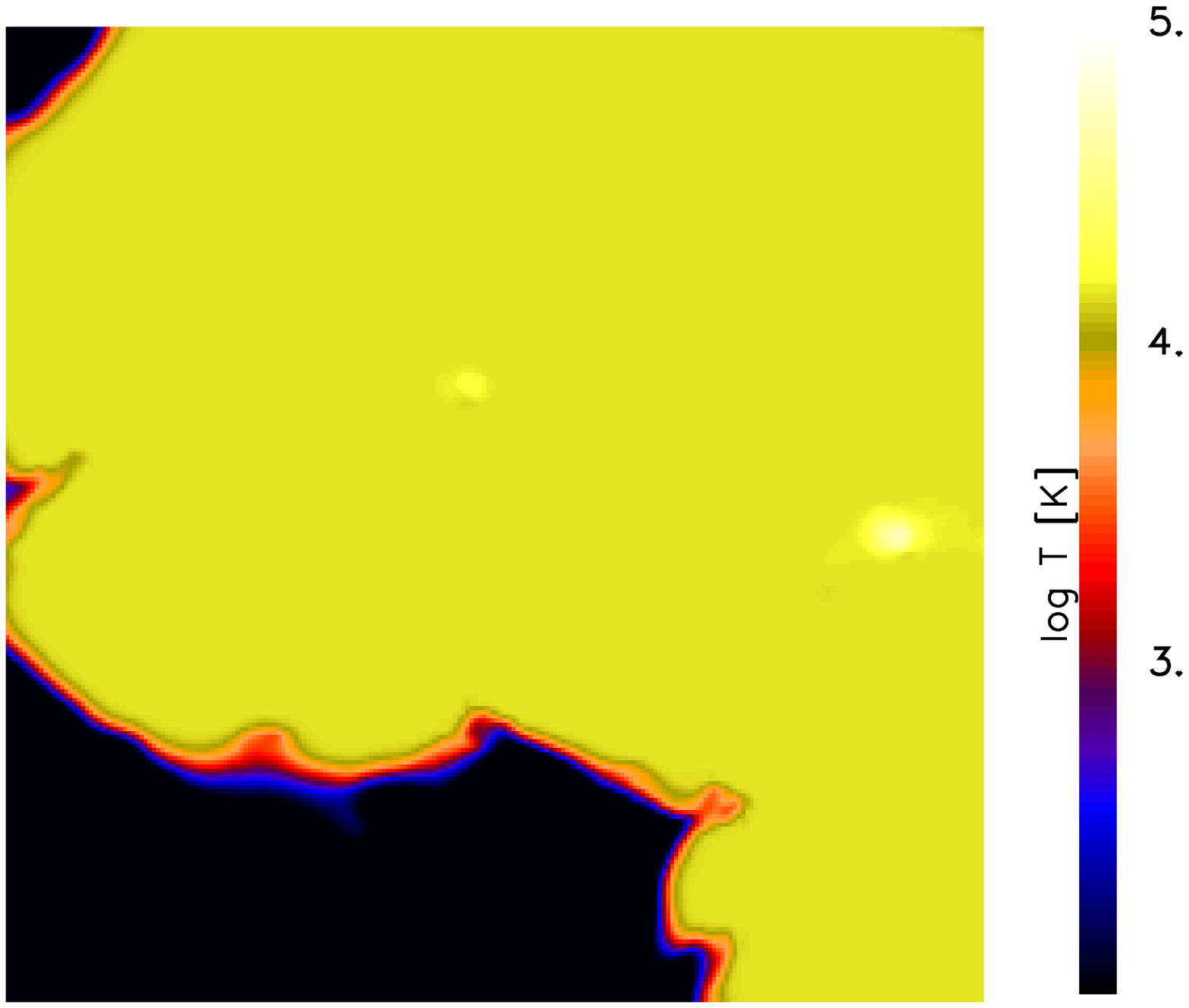}
\end{center}
\caption{Maps of the ionised fraction (top row) and temperature (bottom
  row) in a slice through the middle of the simulation volume at time
  $t=0.4 \, \rm Myr$ of our cosmological density field ionisation test.
  In the results shown in the left column, all sources have been treated
  independently as point sources. In the middle column, only the locally
  four brightest sources have been considered independently, while the
  remaining luminosity has been treated with radiation
  diffusion. Finally, the results in the right column are based on our
  cone transport algorithm with a division of the full solid angle into
  12 cones of equal size, corresponding to the coarsest {\small HEALPIX}
  resolution.
  \label{fig:ion_slice_cosmo}}
\efigs

In Figure~\ref{fig:clump}, we show the temperature and neutral
fraction as a function of distance from the geometric clump
center. The results are compared to those obtained in the comparison
study of \citet{Iliev2006comparison} for the Monte-Carlos transfer
code {\small CRASH}.  The position and shape of the ionising front
agree well with the results from the {\small CRASH} code, both at
times $t = 1\, {\rm Myr}$ and $15\, {\rm Myr}$.  The temperature
profile shows, however, some differences. This discrepancy can be
traced back to inaccuracies in {\small CRASH}, where lower energy
photons penetrate into the gas ahead of the ionising front and heat it
there.

\bfigs
\begin{center}
\includegraphics[width=0.45\textwidth]{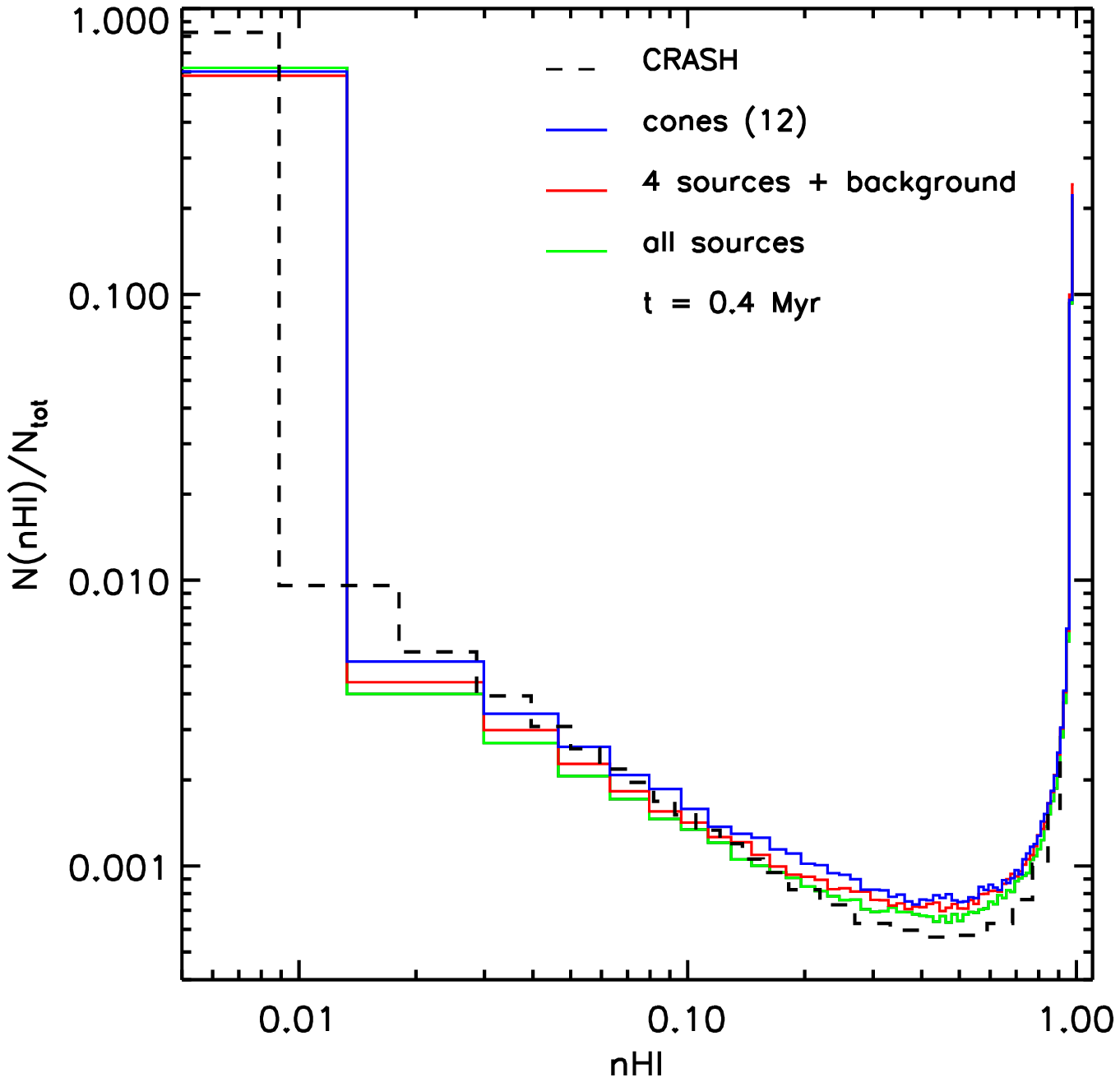}
\includegraphics[width=0.45\textwidth]{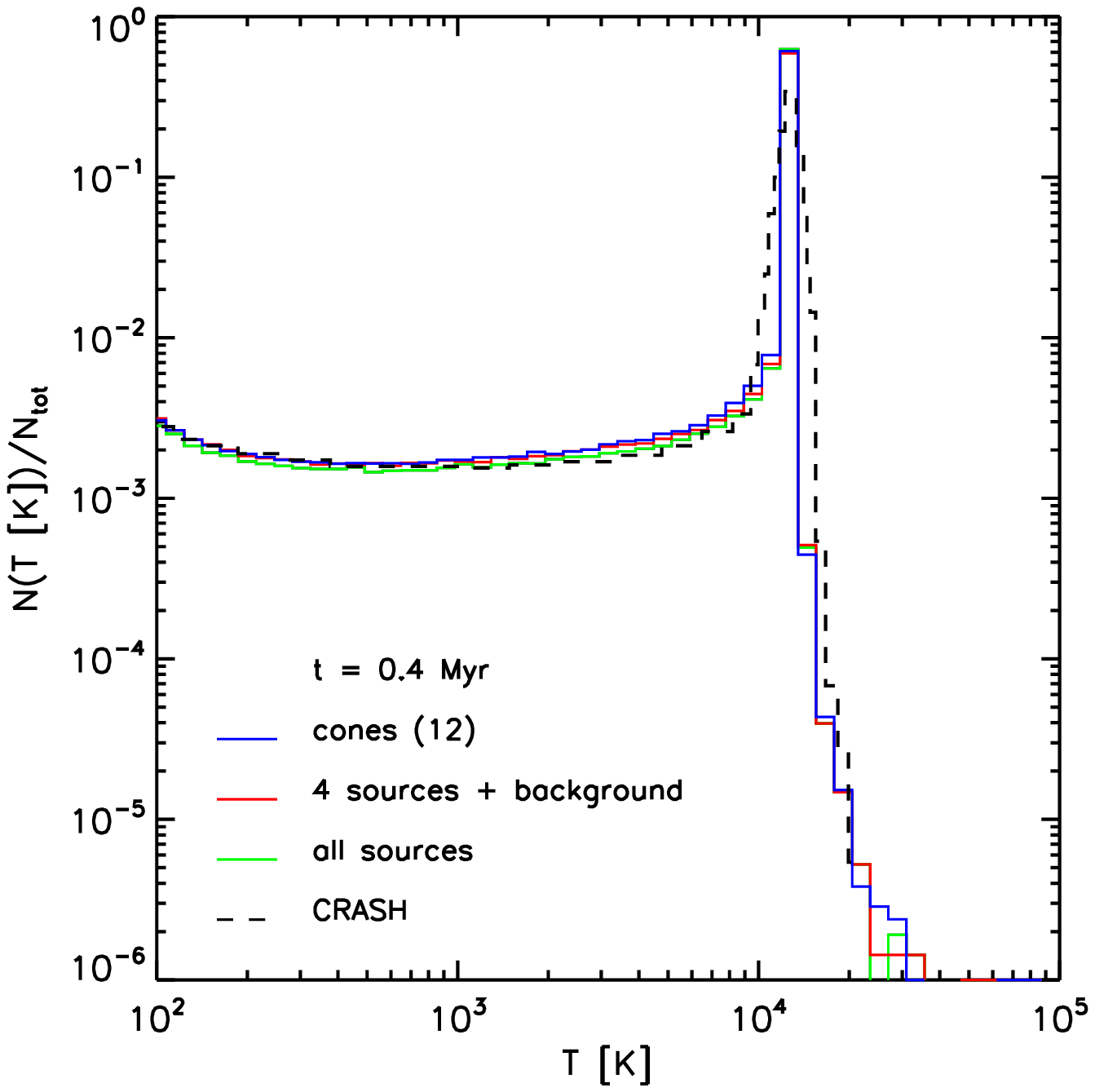}
\end{center}
\caption{Distribution functions of the neutral fraction (top panel)
  and the temperature (bottom panel) in the simulated cosmological
  volume at time $t=0.4\, \rm Myr$, for three different variants of
  our radiation transfer scheme, as labelled: (1) all sources are
  followed in a linearly independent fashion, (2) only the four
  locally brightest sources are followed as point sources with the
  rest treated through radiative diffusion, and (3) a cone transport
  approach based on a division of the unit sphere into 12 cones. For
  comparison, we also include results obtained with the code {\small
    CRASH} in the RT code comparison project of
  \citet{Iliev2006comparison}.
\label{fig:nHI_hist}}
\efigs

Finally, in Figure~\ref{fig:clump_time} we show the time evolution of
the temperature, ionised fraction and position of the I-front in
the clump, compared to the results obtained with {\small CRASH}. The
clump is 60\% ionised at the end, its average temperature increases to
several $10^4\rm K$ and the I-front becomes trapped around the
geometric center of the clump, which is all in good agreement with the
{\small CRASH} results. We hence conclude that our RT scheme yields
results of good accuracy for this test, which are comparable in
accuracy to those obtained with expensive yet accurate Monte-Carlo
treatments.

\subsection{Ionisation of a static cosmological density field} \label{sec:cosmo} 

In our most demanding test of pure RT we follow hydrogen ionisation in
a realistic cosmological density field, which is taken to be static
for simplicity. Again, in order to be able to compare our results with
those of the cosmological RT comparison project
\citep{Iliev2006comparison} we use the same density field, the same
cosmological box parameters, and assign sources in the same way.  The
test is based on a cosmological density field in a periodic box with
size $0.5\,h^{-1} {\rm comoving \, Mpc}$ that resulted from the
evolution of a standard $\Lambda$CDM model with the cosmological
parameters $\Omega_{\rm 0}=0.27$, $\Omega_{\rm b} = 0.043$, and $h=0.7$.
The gas density field at redshift $z=9$ is considered for further
analysis.

The source distribution is determined by finding halos within the
simulation box with a Friend-Of-Friends (FOF) algorithm and then
assigning sources to the $16$ 
most massive groups. The photon luminosity of these sources is taken to be
\be \dot N_\gamma = f_\gamma \frac{M\Omega_{\rm b}}{\Omega_{\rm 0}m_{\rm
    p}t_{\rm s}}, \ee where $M$ is the total halo mass, $t_{\rm s} = 3
\, \rm Myr$ is the assumed lifetime of each source, $m_{\rm p}$ is the
proton mass, and $f_\gamma = 250$ is the number of emitted photons per
atom during the lifetime of the source. For simplicity we also set the
initial temperature of the gas to $100\,{\rm K}$ throughout the whole
box.

\bfigs
\begin{center}
\includegraphics[width=0.3\textwidth]{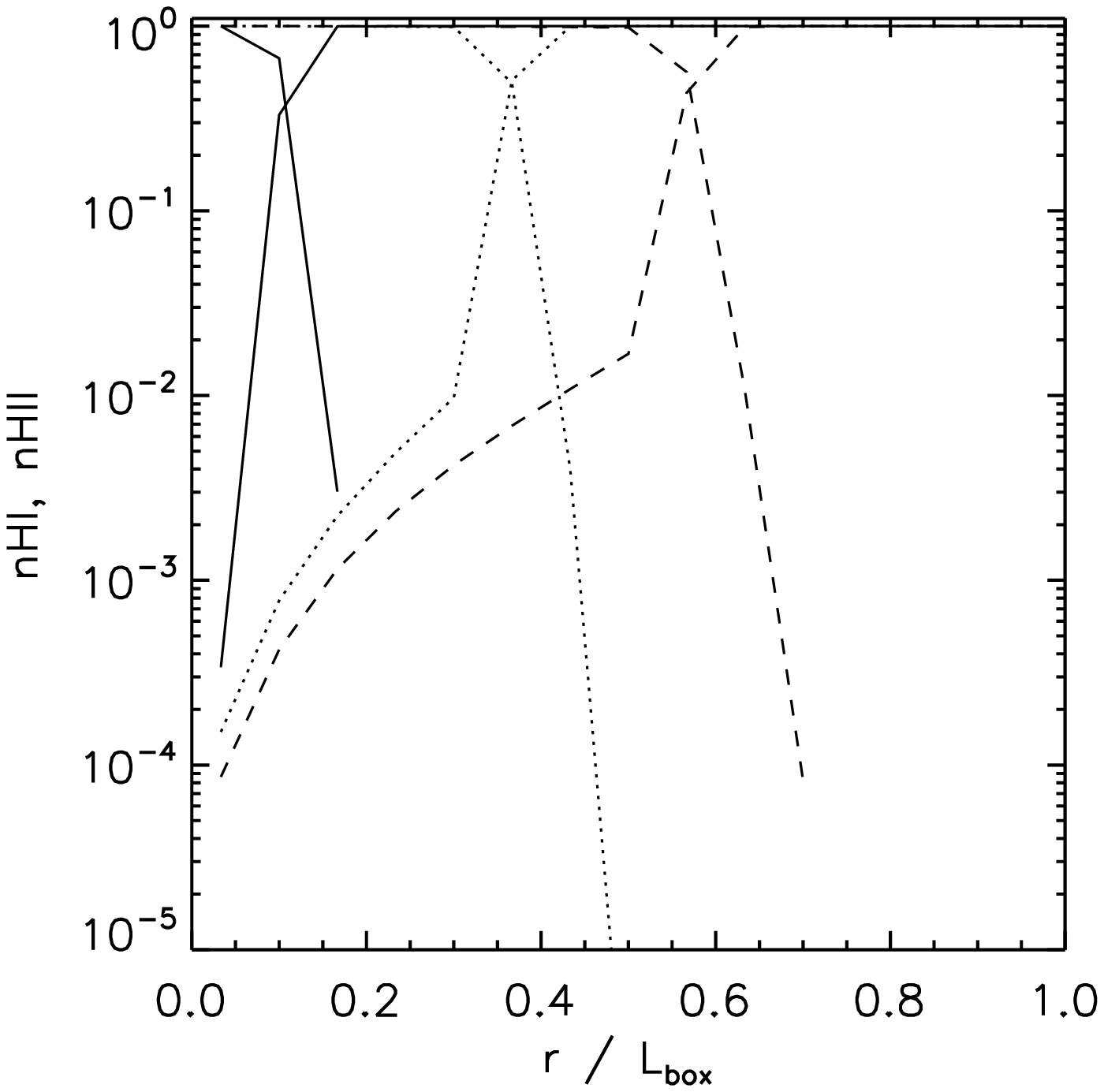}
\includegraphics[width=0.3\textwidth]{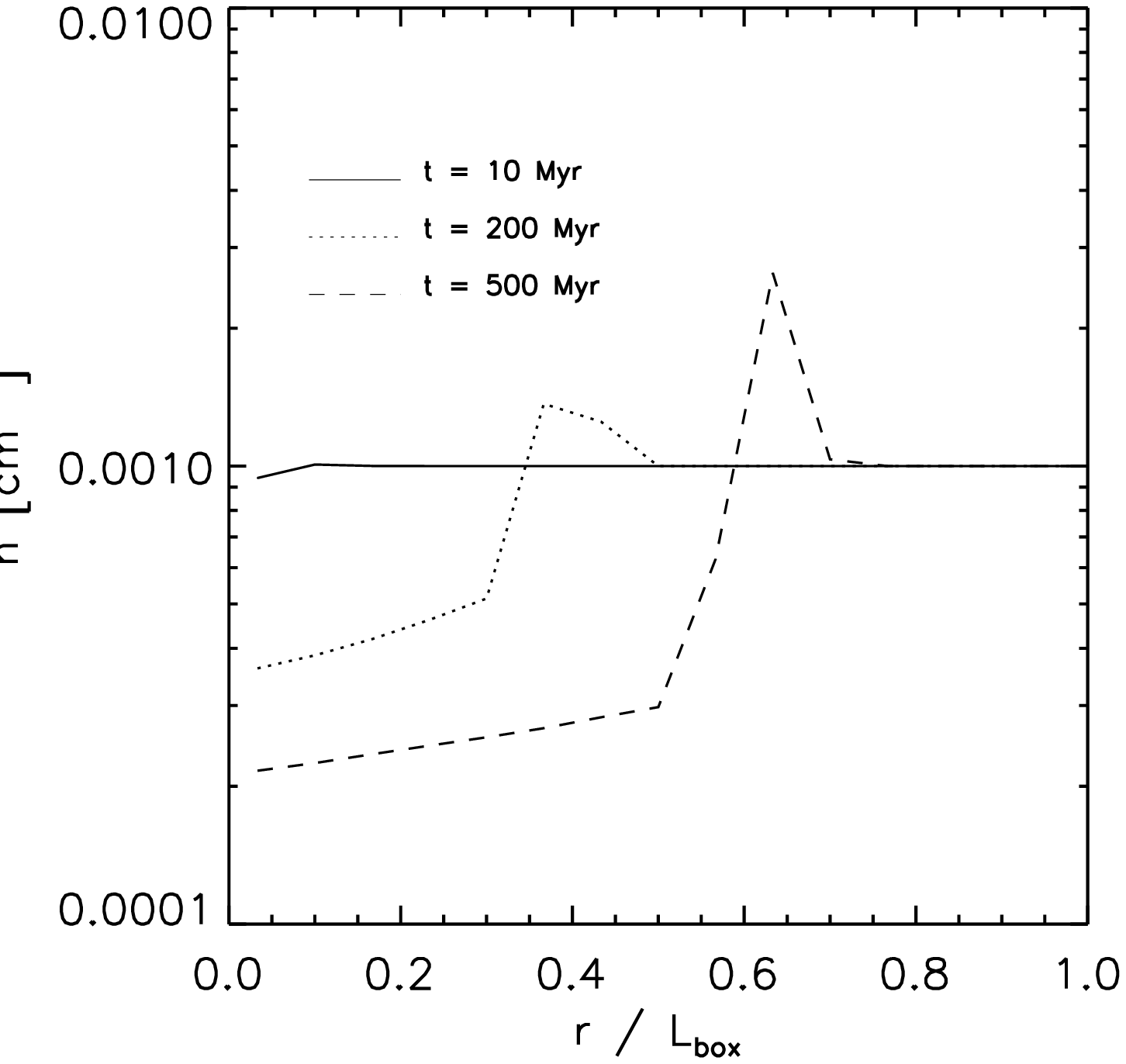}
\includegraphics[width=0.3\textwidth]{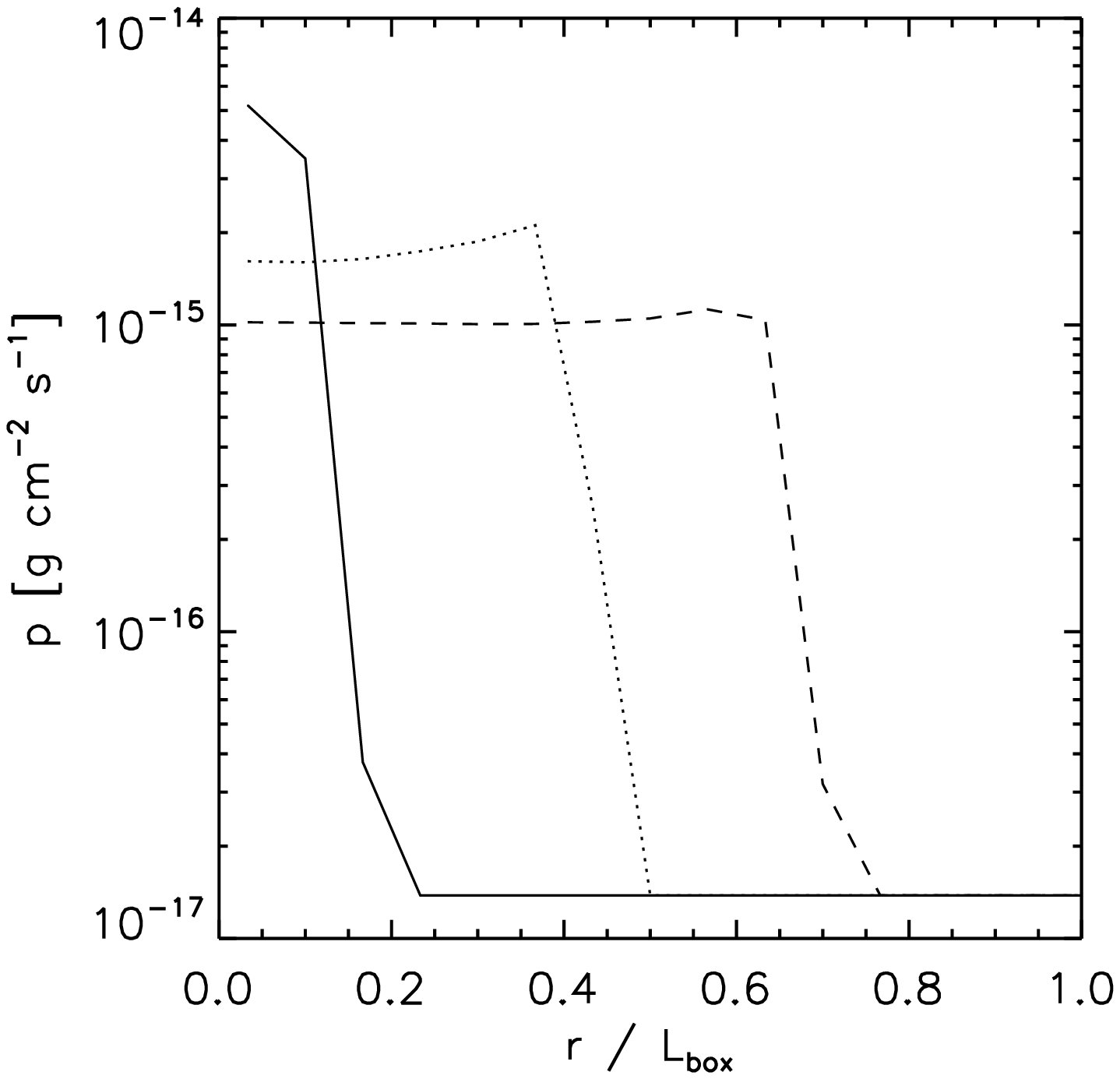}
\includegraphics[width=0.3\textwidth]{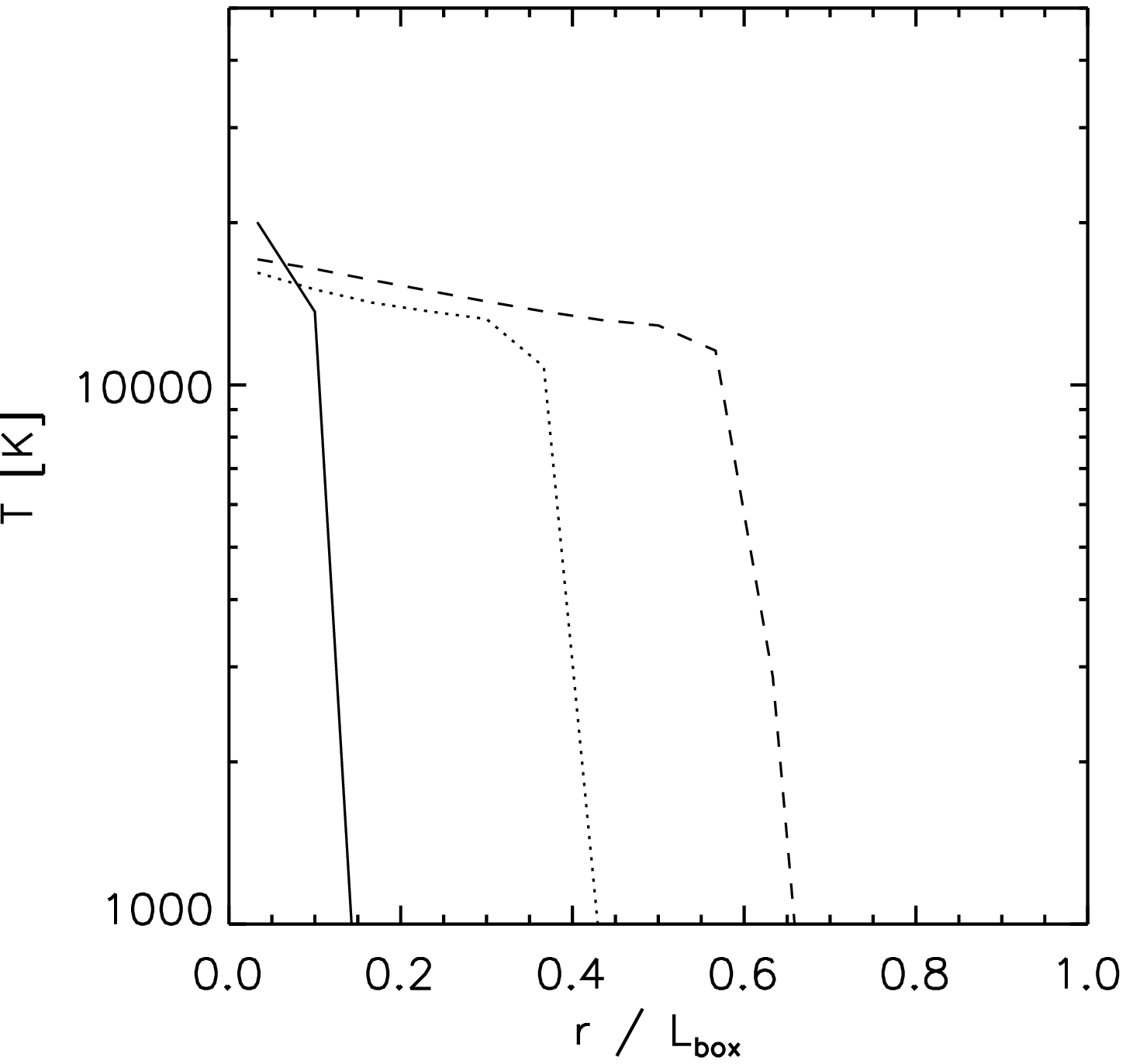}
\includegraphics[width=0.3\textwidth]{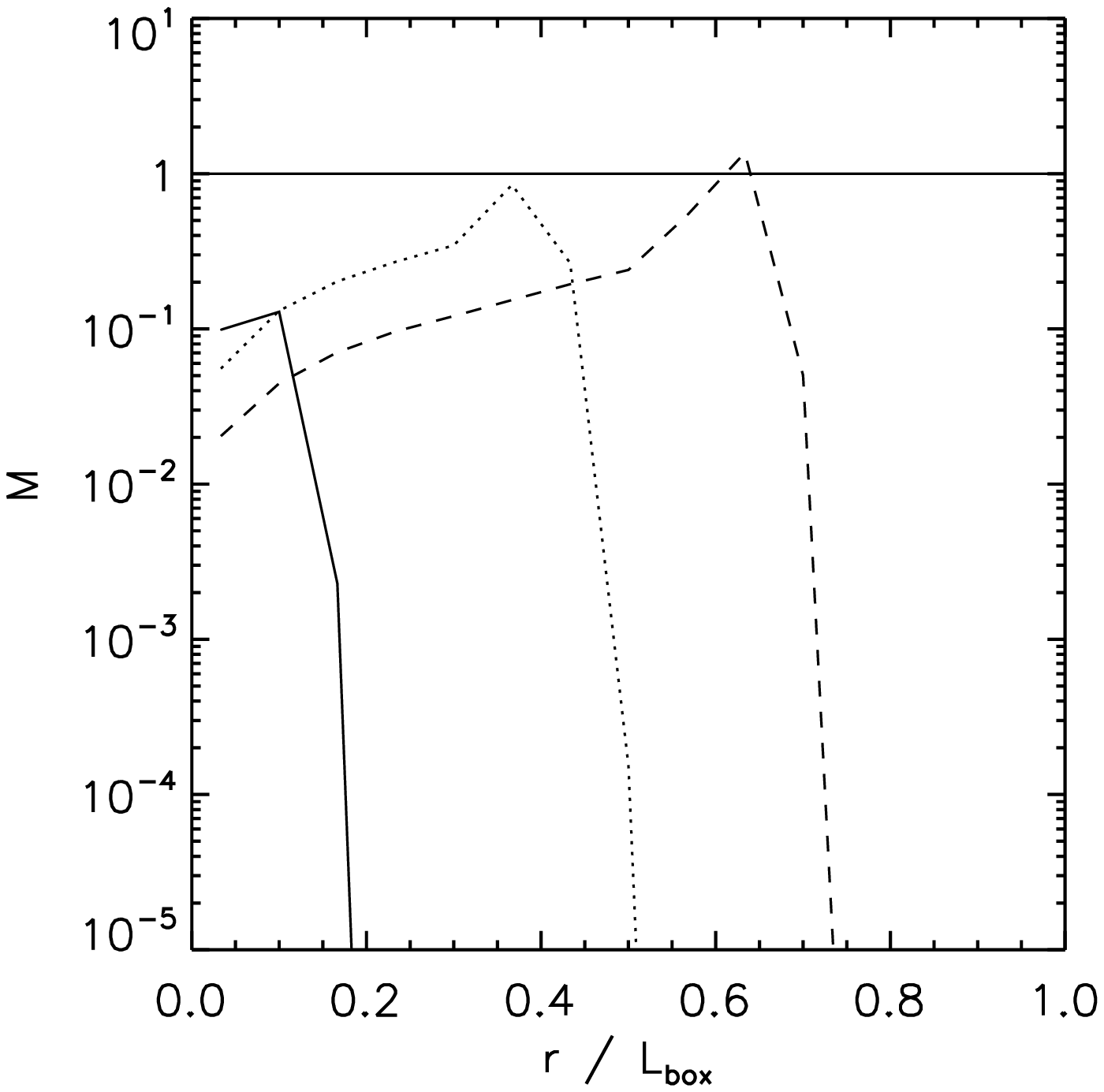}
\end{center}
\caption{Profiles of ionised fraction, hydrogen number density,
  pressure, temperature, and mach number at different times for a
  hydrodynamically coupled Str\"omgren sphere test.  The distance from
  the source is normalised by the box size $L_{\rm box} = 15 \, \rm
  kpc$. The three lines in each plot correspond to the times $t= 10 \,
  \rm Myr$ (solid), $t=200 \, \rm Myr$ (dashed), and $t = 500 \, \rm
  Myr$ (dotted). \label{fig:SST_dynamic}} \efigs

In Figure~\ref{fig:ion_slice_cosmo}, we show the neutral fraction and
the temperature in slices through the center of the simulated volume,
at the final evolution time of $t=0.4\,\rm Myr$. We have calculated
the radiation transfer in three different ways, corresponding to the
three variants of the radiation advection approach proposed in this
paper.  In the left panel, we show the results if all sources are
treated as linearly independent point sources. The middle panel shows
the result when only the $N_{\rm br} = 4$ brightest sources seen from a
given point are treated as point sources, while the remaining
luminosity is fed to a background radiation field which is treated
with radiative diffusion.  Finally, the right hand panel gives the
result when angular discretisation with 12 {\small HEALPIX} cones for
the full solid angle is applied.

Visually, based on Fig.~\ref{fig:ion_slice_cosmo}, all three results
agree very well with each other. However, there are some small
differences in the structure of the ionised regions and in the shape
of the I-fronts. For a better quantitative comparison we show
in the top panel of Figure~\ref{fig:nHI_hist} the volume filling
function of the neutral fractions for all three simulation methods,
where a comparison with {\small CRASH} results from the RT code
comparison study \citep{Iliev2006comparison} is also included. Our
results agree very well with the {\small CRASH} data, and we note that
there are also no substantial differences between the three different
approaches for dealing with multiple sources in our radiation
advection scheme. The same conclusion is also reached from a
comparison of the volume distribution function of the temperature,
which is shown in the bottom panel of Figure~\ref{fig:nHI_hist}. We
note that these results are considerably better than
those we obtained for the OTVET scheme implemented in the SPH code
{\small GADGET} \citep{Petkova2009}.

\subsection{Ionised sphere expansion in a dynamic density field}
\label{sec:tsphere}

As our final test, we again follow the expansion of an ionised sphere
in an initially homogeneous and isothermal medium, similar to
Section~\ref{sec:isphere}, but this time we allow the gas to be heated
up by the photons and to expand due to the raised pressure. This is
hence a radiation hydrodynamics test where both RT and hydrodynamics
are followed.  We design our test similar to the set-up studied in
\citet{Iliev2009comparison}. The source is at the center of the
simulation domain and emits at a luminosity of $\dot N_\gamma = 5 \times
10^{48} \,{\rm s}^{-1}$ . The surrounding hydrogen number density is
$n_{\rm H} = 10^{-3} \, \rm cm^{-3}$ at an initial temperature of $T =
10^2 \, \rm K$. The simulated box is $30 \, \rm kpc$ on a side, and is
resolved with $160^3$ cells. We evolve the system for $500 \, \rm
Myr$.

There are two critical gas velocities defined for such a set up
\citep{Spitzer1978}: the R-critical velocity $v_{\rm R} = 2c_{\rm
  s}^b$ and the D-critical velocity $v_{\rm D} = c_{\rm s}^b -
\sqrt{(c_{\rm s}^b)^2 - (c_{\rm s}^a)^2}$, where $c_{\rm s}^a$ and
$c_{\rm s}^b$ are the isothermal sound speeds ahead and behind the
I-front, respectively. If we assume the ionised gas has
temperature $10^4 \, \rm K$ and the neutral gas $10^2 \, \rm K$, than
we obtain $v_{\rm R} \approx \rm 25.70 \, km \, s^{-1}$ 
and $v_{\rm D} \approx 0.03 \, \rm km \,
s^{-1}$.  The I-front is called D-type when its speed is smaller than
the D-critical speed $v_{\rm I} \leq v_{\rm D}$. In this case it is
subsonic with respect to the neutral gas, which expands as the I-front
passes through it.  When $v_{\rm I} \geq v_{\rm R}$, the I-front
is called R-type. It is supersonic with respect to the neutral
gas ahead, and the gas does not expand as the I-front passes trough
it. When $v_{\rm D} < v_{\rm I} < v_{\rm R}$, there is a hydrodynamic shock wave
in front of the I-front. The position of the I-front in
this stage is given by \citep{Spitzer1978} \be r_{\rm I} = r_{\rm
  S,0}\left( 1+\frac{7c_{\rm s}t}{4r_{\rm
    S,0}}\right), \label{rI,d}\ee where $c_{\rm s}$ is the sound speed
of the ionised gas and $r_{\rm S,0}$ is the Str\"omgren radius given
by equation~(\ref{rs1}).

\bfig
\begin{center}
\includegraphics[width=0.4\textwidth]{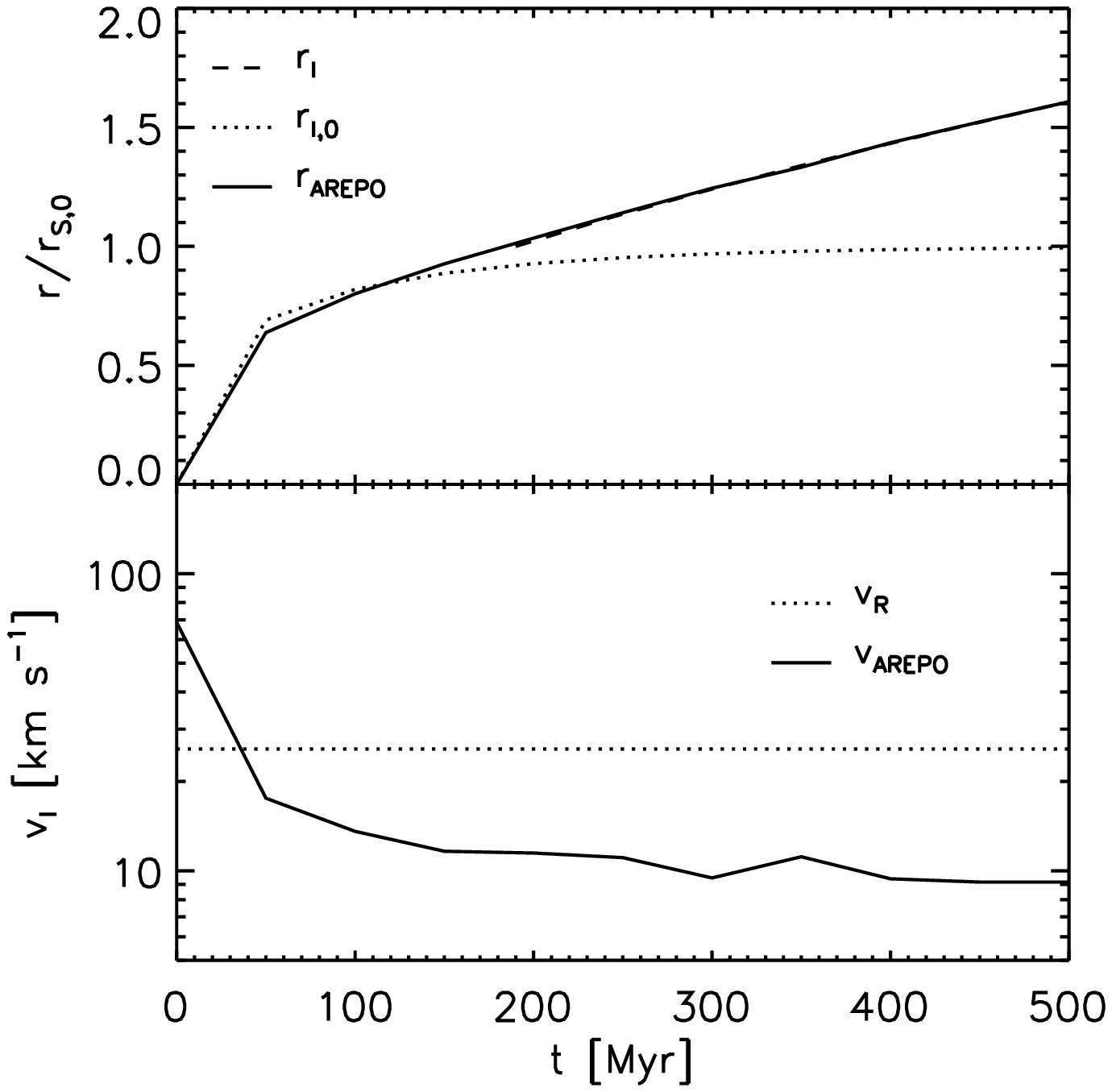}
\end{center}
\caption{Evolution of the position of the ionising front in a
  hydrodynamically coupled Str\"omgren sphere test. The distance is
  expressed in terms of the Str\"omgren radius $r_{\rm S,0}$ for
  the case of a static density field. The dotted line shows the
  analytic solution for the time evolution  in
  the static density case, while the dashed line gives the solution
   for the dynamic density case. The latter is well
  reproduced by our numerical {\small AREPO} calculation. In the
  bottom panel, we show the speed of the ionisation front. In the
  first $40 \, \rm Myr$ of the expansion, the front moves with a speed
  higher than the R-critical velocity (indicated by a dotted
  line). \label{fig:SST_dynamic_ifront}} \efig

We evolve our test setup for $500 \, \rm Myr$ and analyse the results
at three different times equal to $t = 10, \, 200$ and $500 \, {\rm
  Myr}$. Figure~\ref{fig:SST_dynamic} shows the time evolution of the
profiles of the ionised fraction, hydrogen number density, pressure,
temperature, and mach number profiles. At time $t=10 \, \rm Myr$, the
gas expands at subsonic speed. The pressure inside the ionised bubble
is very high as the density is still relatively close to $10^{-3} \,
\rm cm^{-3}$ and the temperature is several $10^4 \, \rm K$. At later
times, $t=200 \, \rm Myr$, there is a shock developing ahead of the
ionising front. The gas in this pseudo shock region is compressed,
leading to densities higher than $10^{-3} \, \rm cm^{-3}$ and an
increased pressure. At time $t=500 \, \rm Myr$, the dynamic situation
of the gas is similar, as there is still a shock ahead of the
ionising front, but the pressure in the ionised bubble has dropped
significantly due to the lowered density of the gas.

In Figure~\ref{fig:SST_dynamic_ifront}, we show the evolution of the radius of the
I-front. In the first $40 \, \rm Myr$, the ionising front moves
with a speed larger than the R-critical velocity: $v_{\rm I} > v_{\rm
  R}$. Its evolution corresponds to that of an I-front in a
static density field, and the position of the front follows the
analytical prediction from equation~(\ref{rI}). The gas does not expand
significantly in this stage. As the speed of the I-front drops below the
R-critical velocity, a shock develops ahead of it and the gas gets
compressed in these regions. Here the position of the front evolves
according to equation~(\ref{rI,d}).  

In general, our results for this test agree well with the other codes
that have been tested by \citet{Iliev2009comparison}. We find the best
agreement with the {\small ENZO-RT} results from the RT comparison
study, which is probably due to the specific monochromatic nature of
our code.

\section{Discussion and Conclusions} \label{sec:end}

In this study, we have proposed a novel implementation of radiative
transfer and implemented it in the moving-mesh code {\small
  AREPO}. The method differs substantially from commonly employed
ray-tracing or moment-based schemes in that it directly evolves a
discretised version of the Boltzmann equation describing the photon
distribution function. This is done in terms of an advection
treatment, where the photon transport is carried out with a
second-order accurate upwind scheme, based on methods that are
commonly employed in hydrodynamic mesh codes.  We have introduced
three different approaches to deal with multiple sources, either by
splitting up the radiation field into a linear sum 
of the partial fields created by all sources, by using a hybrid
approach consisting of an exact treatment of the locally brightest sources
combined with radiative diffusion, or by employing a direct
discretisation of angular space into a finite set of cones.  The
latter approach is the most general.  At a given angular resolution,
it can easily deal with an arbitrary number of sources as well as with
radiation scattering. Also, if the number of angular cones is
enlarged, its angular accuracy becomes progressively better, allowing
a simple way to test for convergence with angular resolution.

The radiation transport in our method is manifestly photon conserving.
Combined with a photon-conserving treatment of the source terms, this
yields a very robust description of the reionisation problem, ensuring
that ionisation fronts propagate at the correct speed.  If needed, our
code can employ a reduced speed-of-light approximation that avoids
overly small timesteps while not altering the growth of ionised
regions in any significant way.

We have presented tests of our new scheme in a variety of cases. Using
different photon transport tests in 2D, we have shown that our method
manages to accurately capture shadows, and to produce the correct
radiation fields independent of the mesh geometry. To test the
coupling of gas physics with the photon transport we have carried out
isothermal ionised sphere expansion tests and compared to the
analytical solutions. The results agree reassuringly well with
theoretical expectations, both for our linear summation method and the
cone transport approach. Furthermore, we have shown that our method
can treat multiple point sources in a highly accurate way, without the
problem of a detrimental mutual influence of the sources onto each
other, which is encountered in certain moment-based schemes
\citep{GA2001, Petkova2009}. We have also shown that our code performs
well on the problem of the ionisation of a static cosmological density
field, where we benchmarked our results against those obtained for the
same setup in the radiative transfer comparison study of
\citet{Iliev2006comparison}. Similarly, our results for I-front
trapping in a dense clump, and for a hydrodynamically coupled
Str\"omgren sphere agree well with those of other radiative transfer
codes included in \citet{Iliev2006comparison}.

Compared to other radiative transfer schemes, our new method based on
the cone transport features several interesting advantages. Unlike
long-characteristics or Monte Carlo schemes, it avoids any strong
sensitivity of the computational cost on the number of sources, and it
does not concentrate the computational effort in regions close to the
sources, which greatly helps in parallelising the calculations. Also,
the ability to easily treat time-dependent effects that are
consistently coupled to the hydrodynamic evolution is a substantial
asset. Compared to moment-based solvers, our method can cast sharp
shadows, and it performs accurately both in the optically thin and the
optically thick regimes.

Our cone-transport scheme bears some superficial resembles to the
{\small TRAPHIC} scheme of \citet{Pawlik2008}.  However, unlike their
approach, we do not rely on stochastic Monte Carlo techniques. Instead, we work
with an explicit spatial reconstruction of the radiation field and a
fixed set of angular cones. As a result, our radiation field is
essentially free of stochastic noise, which is a significant advantage
compared to  Monte Carlo approaches and offers much better convergence rate.

We hence think that our new method represents a promising treatment of
radiative transfer in cosmological simulations. Its simple and general
principles make it 
applicable in a wide variety of different problems in astrophysics. In
particular, it should not only be useful for studying reionisation of
the Universe, but also, for example, for studying star formation in
molecular clouds, where ionising radiation creates pillars or neutral
dense gas \citep[e.g.][]{Gritschneder2009}. Also, our RT solver
coupled to {\small AREPO} should allow self-consistent and more
accurate treatments of radiative feedback effects in hydrodynamic
simulations of star formation, something that we will study in
forthcoming work.

\section*{Acknowledgements}
This research was supported by the DFG cluster of excellence ``Origin and
Structure of the Universe'' (www.universe-cluster.de).

\bibliographystyle{mn2e}
\bibliography{paper}

\label{lastpage}

\bsp

\end{document}